\documentclass[12pt]{article}
\usepackage{tikz,pgfplots}
\usetikzlibrary{decorations.pathreplacing, arrows.meta,quotes,angles,patterns,arrows,calc,patterns.meta}
\tikzset{vert/.style = {circle, fill, inner sep = 0, minimum size = 5}}
\usepackage{amssymb}

\usepackage{amsfonts, bm}
\usepackage{amsfonts}
\usepackage{amsmath} 
\allowdisplaybreaks
\usepackage{amsthm}
\usepackage{amssymb,dsfont}
\usepackage{amssymb}
\usepackage[multiple,bottom]{footmisc}
\usepackage{enumerate}
\usepackage[textwidth=2cm]{todonotes}
\usepackage{epic,xcolor}
\usepackage{subcaption, graphicx}
\usepackage{booktabs}
\usepackage{natbib}
\usepackage{comment}
\usepackage{changepage}
\usepackage{setspace}
\usepackage{lscape}

\usepackage[unicode=true, pdfusetitle,
bookmarks=true,bookmarksnumbered=false,bookmarksopen=false,
breaklinks=false,pdfborder={0 0 1},backref=false,colorlinks=false]
{hyperref}
\hypersetup{
	colorlinks = true,
	urlcolor = chicago-maroon,
	linkcolor = chicago-maroon,
	citecolor = northwestern-purple,
}
\usepackage[capitalise,noabbrev]{cleveref}
\usepackage{multirow}
\usepackage{float}
\definecolor{clemson-orange}{RGB}{234,106,32}
\definecolor{chicago-maroon}{RGB}{128,0,0}
\definecolor{northwestern-purple}{RGB}{82,0,99}
\definecolor{cornell-red}{RGB}{179,27,27}
\definecolor{sauder-green}{RGB}{171,180,0}
\definecolor{gray}{RGB}{192,192,192}
\definecolor{lawngreen}{RGB}{0,250,154}
\definecolor{pink}{RGB}{255,0,128}

\usepackage{pdfpages}

\usepackage{microtype}
\usepackage{lipsum}

\makeatletter
\def\thm@space@setup{\thm@preskip=2pt
\thm@postskip=2pt}
\makeatother
\newtheoremstyle{newstyle}      
{} 
{} 
{\itshape} 
{} 
{\bfseries} 
{.~} 
{ } 
{} 
\theoremstyle{newstyle}

\newtheorem{fact}{\sc{Fact}}
\newtheorem{prop}{\sc{Proposition}}

\newtheorem{thm}{\sc{Theorem}}
\newtheorem{lem}{\sc{Lemma}}
\newtheorem{remark}{\sc{Remark}}

\newtheorem{cor}{\sc{Corollary}}
\newtheorem{as}{\sc{Assumption}}
\newtheorem{ex}{\sc{Example}}
\newtheorem{defn}{\sc{Definition}}

\newtheorem{innercustomthm}{\sc{Theorem}}
\newenvironment{customthm}[1]
  {\renewcommand\theinnercustomthm{#1}\innercustomthm}
  {\endinnercustomthm}

\newcommand{\bhat}{\hat{\beta}}

\newcommand{\ul}{\ensuremath{\underline{u}}}
\newcommand{\uh}{\ensuremath{\overline{u}}}
\def\D{\Delta}
\def\t{\theta}
\def\m{\mu}
\def\T{\Theta}
\def\e{\epsilon}

\newcommand{\MONTH}{%
	\ifcase\the\month
	\or January
	\or February
	\or March
	\or April
	\or May
	\or June
	\or July
	\or August
	\or September
	\or October
	\or November
	\or December
	\fi}

\usepackage{longtable}

\newenvironment{tabnotes}[2][1]{\begin{minipage}[t]{#1\textwidth}
\scriptsize{\emph{Note:} #2}}{\end{minipage}}

\newcommand\listappendixname{List of Appendices}
\newcommand\appcaption[1]{%
  \addcontentsline{app}{section}{#1}}
\makeatletter
\newcommand\listofappendices{%
  \section*{\listappendixname}\@starttoc{app}}
\makeatother

\newcommand{\reminder}[1]{\textcolor{red}{}} 

\usepackage{tabularx}
\newcolumntype{x}[1]{>{\centering\arraybackslash\hspace{0pt}}p{#1}}
\newcolumntype{H}{>{\setbox0=\hbox\bgroup}c<{\egroup}@{}}
\newcolumntype{P}[1]{>{\raggedright\arraybackslash}p{#1}}

\usepackage{enumitem}
\setlist{nosep}

\setlist[enumerate]{topsep=1ex,itemsep=-1ex,partopsep=0ex,parsep=1ex,leftmargin=4.5mm}

\setlist[itemize]{topsep=1ex,itemsep=-1ex,partopsep=0ex,parsep=1ex,leftmargin=4.5mm}

\usepackage[compact]{titlesec}
    \titlespacing{\section}{0pt}{-0.5ex}{-0.2ex}
    \titlespacing{\subsection}{0pt}{-0.5ex}{-0.2ex}
    \titlespacing{\subsubsection}{0pt}{0ex}{-0.2ex}

\renewenvironment{abstract}
{\small
	\begin{center}
		\bfseries \abstractname\vspace{-1em}\vspace{0pt}
	\end{center}
	\list{}{
		\setlength{\leftmargin}{1cm}%
		\setlength{\rightmargin}{\leftmargin}%
	}%
	\item\relax}
{\endlist}

\makeatletter
\renewcommand{\paragraph}{%
  \@startsection{paragraph}{4}%
  {\z@}{0.5ex \@plus 1ex \@minus .2ex}{-1em}%
  {\normalfont\normalsize\bfseries}%
}
\makeatother

\newcommand{\displaymargins}{
\setlength{\abovedisplayskip}{3pt}
\setlength{\belowdisplayskip}{3pt}
\setlength{\abovedisplayshortskip}{1pt}
\setlength{\belowdisplayshortskip}{1pt}}

\newcommand{\floatmargins}{
\setlength{\abovecaptionskip}{-0.1pt}
\setlength{\belowcaptionskip}{-0.1pt}
}

\usepackage{newtxtext}  

\usepackage[left=0.894in,right=0.894in,top=1.25in,bottom=1.25in]{geometry}

\newcommand{\fset}{B}

\newcommand{\assign}{\alpha}
\newcommand{\cU}{\mathcal{U}}
\newcommand{\R}{\mathcal{R}}
\newcommand{\ip}{t}

\newcommand{\PREF}{\mathcal{P}}

\newcommand{\pref}[1]{\PREF^{#1}}
\newcommand{\preff}{\widetilde\PREF}

\newcommand{\teps}[1]{TEPS$^{#1}$}

\newcommand{\uncelem}{\omega}
\newcommand{\unc}{W}
\newcommand{\uncset}{\Omega}

\newcommand{\TITLE}{Leveraging Uncertainties to Infer Preferences:\\Robust Analysis of School Choice}

\begin{document}
\title{\bf\Large \TITLE
\thanks{We are grateful to Nikhil Agarwal, Guy Aridor, Pierre-Andr\'e Chiappori, Chao Fu, Aram Grigoryan, Bernard Salanie, Xiaoxia Shi, Sukjoon Son, seminar participants at Boston College, Sciences Po, PSE, Columbia University, HKUST, USC, SNU, University of Tokyo, ASU, University of Toronto, UT San Antonio and participants of NBER IO Program meeting, ESWC, WEAI, AMES, NASM, Australian Education Markets conferences for valuable comments and suggestions. Thanks also go to the NYC DOE, particularly to Nadiya Chadha, Benjamin Cosman, Stewart Wade, and Lianna Wright. Che acknowledges funding from the Ministry of Education of the Republic of Korea and the National Research Foundation of Korea  (NRF-2020S1A5A2A03043516). Hahm acknowledges funding from the Program for Economic Research (PER) at Columbia University. All errors are our own.
}
}
		\author{Yeon-Koo Che\thanks{Department of Economics, Columbia University,  USA.  Email: {\href{mailto:yeonkooche@gmail.com}{\texttt{yeonkooche@gmail.com}}}.} \and Dong Woo Hahm\thanks{Department of Economics, University of Southern California,  USA.  Email: {\href{mailto:dongwooh@usc.edu}{\texttt{dongwooh@usc.edu}}}.} 
        \and YingHua He\thanks{Department of Economics, Rice University, USA. \ Email:
			{\href{mailto:yinghua.he@rice.edu}{\texttt{yinghua.he@rice.edu}}}.}
        }
		
\date{September 22, 2023}		

\maketitle
\vspace{-0.7cm}  
\begin{abstract}
\begin{spacing}{1.25}
{Inferring applicant preferences is fundamental in many analyses of school-choice data. Application mistakes make this task challenging. We propose a novel approach to deal with the mistakes in a deferred-acceptance matching environment. The key insight is that the uncertainties faced by applicants, e.g., due to tie-breaking lotteries, render some mistakes costly, allowing us to reliably infer relevant preferences. Our approach extracts all information on preferences robustly to payoff-insignificant mistakes. We apply it to school-choice data from Staten Island, NYC.  Counterfactual analysis suggests that we underestimate the effects of proposed desegregation reforms when applicants' mistakes are not accounted for in preference inference and estimation.}
\end{spacing}
		\end{abstract}

\displaymargins
\floatmargins
\begin{spacing}{1.33}
\newpage
\section{Introduction}
School choice, a popular framework for public school assignment in the US and elsewhere, is a widely debated subject among scholars and policymakers.  An important issue, still far from resolved, is how to reform an assignment mechanism to diversify and desegregate the student body.

For instance, nearly 80\% of NYC's Black or Hispanic public high school students are concentrated in just half of the high schools, and in response, various desegregation policies have been proposed including the elimination of admissions based on academic qualifications or residence locations \citep{shapiroNYT}.  The crucial first step for such an inquiry is to understand students' preferences. Only when the preferences are accurately inferred {from data} and estimated can one hope to {correctly} evaluate the effects of possible policy reforms.

In principle,  the widespread use of strategyproof mechanisms such as the deferred acceptance (DA) algorithm should make the inference straightforward as it gives applicants the dominant-strategy incentive to report their preferences truthfully. 
However, recent literature found evidence that a significant fraction of participants make strategic ``mistakes." That is, they do not report their preferences truthfully even in strategyproof environments.\footnote{We follow the existing literature and call such behaviors mistakes.} The vast majority of these documented mistakes are payoff-irrelevant: they typically involve applicants neglecting to rank, or ranking incorrectly, popular schools that are so out-of-reach that they would not have obtained them even if they had correctly ranked them.\footnote{For instance, \cite{ACHwp} find that even though 35\% of the applicants mistakenly forgo lower-tuition options in their application for colleges in Victoria, Australia, only 1-20\% of such mistakes had real consequences since such options were out-of-reach for them. Mistakes reported in \cite{shorrersovago2023} and \cite{hassidim2021limits} are of similar nature.  See Table 1 of \cite{ACHmistakes}.}   Yet, even such payoff-insignificant mistakes would lead to biased preference estimates and importantly, to misleading predictions of policy effects if applicants' rank-order lists (ROLs) are interpreted at face value.  For instance,  Weak Truth-Telling (WTT), a popular hypothesis that interprets students' ROLs as truthful representations of their most preferred choices, would incorrectly infer elite schools as unpopular to disadvantaged students when many of them find those schools out-of-reach and neglect to rank them on their ROLs.  Accordingly, WTT would predict those students would not apply to those out-of-reach elite schools even when they become within-reach after certain admissions reforms; such policies will then be incorrectly seen to have smaller desegregation effects than they truly do.

The stability hypothesis has been proposed as a way to deal with payoff-irrelevant mistakes in preference inference.
 \cite{Fack-Grenet-He(2015)} and \cite{ACHmistakes}  show that, as the economy becomes large, the aggregate uncertainty about schools' cutoffs vanishes, so that students, while making payoff-irrelevant mistakes, ``recognize'' which schools are feasible and are able to secure their {\it stable assignments}, i.e., they receive their favorite schools among those that are feasible for them given their priority standings.  Stability is clearly weaker than WTT and proves more robust to mistakes.\footnote{See \cite{Fack-Grenet-He(2015),ACHwp,ACHmistakes} for their Monte Carlo simulations and goodness-of-fit exercises.} By now, the method has been adopted by many papers \citep[e.g.,][]{otero2021affirmative,10.1093/restud/rdac002,combe2022market,HahmPark2022}.

However, the preceding argument for stability crucially rests on students' ability to recognize their feasible schools in a large market, which relies on the assumption that students face no uncertainties about their priorities.  In practice, students face priority uncertainties from a number of sources.
For example, the vast majority of the US public school districts use lotteries to break ties. Also, even when tie-breaking is not needed, the priorities may not be known in advance.\footnote{For example, admissions to NYC Specialized High Schools are based on the Specialized High School Admissions Test scores, but students must submit their ROLs before learning their scores. Similar uncertainty exists in China where students in some provinces apply to colleges before taking the entrance exam or after taking the exam but before learning their scores \citep{chen2017chinese}.} Finally, regardless of the need for tie-breaking, uncertainty arises from not knowing other students' priorities or preferences.  All these uncertainties imply that even in a large economy 
students no longer clearly recognize the schools that are feasible for them when they submit their applications---the argument underpinning the stability hypothesis.

Hence, it is unclear whether stability generalizes to environments with uncertainties and how it can lead to robust inference of preferences. In fact, it is not even clear that the types of payoff-irrelevant mistakes discussed earlier would arise in the presence of uncertainty. Without knowing which schools are feasible, mistakes generally become costly, and applicants would become more careful not to make them.  However, this logic does not mean uncertainties make {\it all} types of mistakes costly. For instance, some schools may still be out of reach for a student who does not live in the right geographic zone, regardless of her lottery draw. Moreover, the type of lotteries used in practice, such as Single-Tie-Breaking (STB), leaves room for payoff-irrelevant mistakes even when all schools may be within reach, as we illustrate in \Cref{ex:stb-mistakes}. Hence, even in the presence of uncertainty, payoff-irrelevant mistakes exist and preference must still be inferred robustly with respect to these mistakes.

At the same time, uncertainties do make \textit{some} types of mistakes costly, making applicants reveal certain preferences more reliably in order to avoid such mistakes.\footnote{\Cref{ex:stb-mistakes} also illustrates this point.}  For instance, the introduction of an assignment lottery could put previously out-of-reach elite schools within reach for a student. Neglecting these schools then becomes costly to the student, as it means forgoing real admission opportunities to them, and therefore, we can expect her to correctly reveal her preferences regarding these schools in the presence of a lottery. Therefore, such uncertainties present an opportunity to learn about student preferences for researchers.

 The question is: \emph{what} preferences can be revealed by the uncertainty present in any given environment?  
To answer this, we develop a novel method called \emph{Transitive Extension of Preferences from Stability}, or TEPS, which exploits the structure of uncertainty faced by applicants to infer their preferences robustly to payoff-insignificant mistakes within a DA matching environment.  We then apply TEPS to school choice data from Staten Island, NYC, and conduct counterfactual analyses of proposed desegregation policies. Our results highlight the importance of robustly inferring student preferences in predicting policy effects.

The theoretical foundation for our method rests on the concept of {\it robust equilibrium} developed by \cite{ACHmistakes}, which allows applicants to make payoff-insignificant mistakes.  Specifically, we consider applicants in a DA matching market and require them to employ strategies that arbitrarily closely attain their best response payoffs as the economy grows large.  While all applicants playing their dominant strategies of truth-telling is clearly a robust equilibrium, truth-telling need not be the only robust equilibrium behavior, as applicants may make payoff-insignificant mistakes---strategies that entail negligible payoff losses in a large market.\footnote{While we do not take a specific stance on why applicants make those mistakes, one way to understand the robustness concept is their lack of attention toward low probability options.} Hence, robust equilibrium is a relaxation of the dominant strategies equilibrium.  The main implication of this solution concept is  \Cref{thm:stability}, which establishes {\it asymptotic stability}: 
 as the economy grows large, with high probability, almost all students are assigned stably---namely, they are assigned to their favorite schools among those that are feasible given the realized uncertainty.
 
The asymptotic stability result justifies our TEPS method, described as follows.  The procedure first seeks to recover the underlying structure of priority uncertainty applicants face by simulating it. For instance, in a typical DA-STB environment, this means conducting the STB lottery and running the DA algorithm in the same manner as the official assignment algorithm.\footnote{The first step depends on the exact context of the setting. If the uncertainty arises from other sources, such as test scores that are yet to be realized at the time of application, then one may build an empirical model to simulate the distribution of test scores based on the observables (which may include various performance measures from preceding schools.)} We then invoke asymptotic stability to infer that the assigned school is preferred to all other feasible schools, given the lottery realization. We repeat this procedure for each lottery draw many times and obtain preference relations for each draw.  The final step then links together the preference relations obtained from the multiple realizations of uncertainty by invoking the transitivity axiom. Specifically, we take a transitive closure of all preference relations obtained from the pairs of assigned and feasible schools.

The prediction that stable assignments occur for {\it all} uncertainty realizations is valid only in a large limit economy. However, \Cref{thm:stability}  is consistent with violations of stability in \textit{finite} economies.\footnote{Payoff relevant mistakes are observed for a significant minority of applicants \citep[e.g.,][]{ACHwp}.}  Our general version of TEPS accommodates such possibilities.  Specifically, we define a family of \teps{\tau}, indexed by the attention parameter $\tau\in (0,100]$, such that only those uncertainties (or feasible sets) cumulatively most likely up to $\tau$\%  are considered for preference inference, meaning that the uncertainties belonging to the bottom $100-\tau$\% in likelihood are discarded for being unreliable. 

In principle, the level of inattention and tolerance for payoff-relevant mistakes is an empirical matter that depends on the characteristics of the participants. Accordingly, we develop a data-driven testing procedure to select from among alternative attention levels, ranging from WTT at one extreme to \teps{\tau} with varying attention levels $\tau$.  Monte Carlo simulations reveal that biased estimation arises when one ignores mistakes made by market participants and demonstrate that our selection procedure performs well.  

Finally, we apply our methods to public high school choice data in Staten Island, NYC.  We assess the performance of alternative hypotheses used for preference inference in describing the real data. In particular, we conduct counterfactual analyses to compare the predicted effects of several proposed desegregation policies based on these hypotheses. The goal of this analysis extends beyond the academic realm. There is a growing concern about the segregation of student bodies in large urban districts such as NYC, and proposals to remove priorities based on academic qualifications and/or residence locations have emerged as methods for desegregation. The effectiveness of these policies in achieving their stated goal is a matter of significant public interest, to which our methodological advance may take a surprising relevance.  The counterfactual analysis of such policy reforms relies on estimating the preferences of target students for schools that are currently inaccessible due to priority restrictions.  If any inattention these students exhibited toward these schools is not accounted for, the effectiveness of these reforms could well be understated.  

We do find that the effects of the desegregation policies are underestimated if one uses WTT, which neglects students' mistakes/inattention, as the hypothesis for preference inference.   
 For example, under the current regime, minority students are assigned to programs with an average proportion of Black or Hispanic students equal to 57 percent, which is 26 percentage points higher than the assigned programs of non-minority students. When all current priority rules are removed, our preferred TEPS-based estimates predict a 2 percentage point decrease in the racial gap, while WTT-based estimates predict that the decrease is only half of that. 
Intuitively, WTT incorrectly neglects mistakes and assumes that students do not prefer unranked out-of-reach schools, so they would not apply to them even when these schools become within reach under new policies.  
Indeed, our testing procedure selects TEPS with varying $\tau$ against WTT. 
Next, the removal of academic and geographic priorities has a rather modest effect on desegregation even under our selected TEPS-based estimates.  This suggests that forces other than school priorities, such as residential segregation, may significantly contribute to school segregation. This raises questions about the effectiveness of reforming priority rules as a tool to alleviate school segregation.

\paragraph{{Related Literature}.}
Strategic mistakes in a strategyproof environment were documented in lab settings \citep{CHEN2006202, li:17} and high-stake real-world contexts such as school applications \citep{larroucau2018short, ACHwp, hassidim2021limits, arteaga2022smart, shorrersovago2023}, and medical resident matching \citep{rees-jones:16, Rees-Jones_NRMP:2018}. For instance, \cite{hassidim2021limits} find in their analysis of Israeli postgraduate psychology program admissions that approximately 19\% of students either did not list a scholarship position for a program or ranked a non-scholarship position higher than the corresponding scholarship position against their interests. \cite{ACHwp} and \cite{shorrersovago2023} find similar findings in college admissions in Australia and Hungary, respectively. \cite{rees-jones:16}  reports that 17\% of the 579 surveyed US medical seniors indicate misrepresenting their preferences in the National Resident Matching Program. We propose a novel, theoretically-based method to infer student preferences robustly to those documented mistakes.

Our main contribution is to the literature on identifying and estimating student preferences in school choice. We focus on DA and allow students to deviate from truth-telling. Several papers have suggested alternative approaches to WTT. Most related to our proposed method, stability has been considered by \cite{Fack-Grenet-He(2015)} and \cite{ACHmistakes} (henceforth FGH and ACH, respectively). They argue that applicants make payoff-irrelevant mistakes in the presence of either application costs or inattention to (small) payoff losses, but are  stably assigned.
In particular, ACH uses robust equilibria to capture applicant inattention and justify asymptotic stability, just like the current paper. The crucial difference is that these papers assume that applicants do not face any uncertainties over the feasible set of schools when they apply.  As already argued, the ability to allow for uncertainties is important given the ubiquity of lotteries and the endemic nature of finite sample uncertainty. 
 Both the methodology we propose and our theoretical justification are nontrivial.\footnote{While the TEPS procedure can be seen as a generalization of the stability hypothesis, it is both novel and theoretically well-founded in a way a naive application of the stability hypothesis is not. For instance, one may directly ``implement'' the stability hypothesis in the lottery environment by applying it using the lottery that has been used in the actual assignment (if it is available in the data) or simply by simulating a lottery once. As we argue in \Cref{sec:method}, such a method is not well justified by our theory and is problematic in other respects.}  Our TEPS procedure of extracting and combining preferences using transitivity and the data-driven method for selecting the right version of TEPS are novel and theoretically well-justified, and they are computationally fast enough to be readily deployed in empirical research.  Likewise, our theoretical justification for TEPS, namely \Cref{thm:stability}, despite the resemblance to the analogous result in ACH, is not a straightforward extension of the latter; as we remark in \Cref{sec:theory}, the effect of deviating one's strategy is now stochastic, and  ``controlling'' the stochastic payoff effect presents a new challenge and requires an additional assumption on the structure of the underlying uncertainty. 

The current paper is also related to a strand of literature that views each applicant's submitted ROL as an optimal portfolio choice under non-strategyproof mechanisms \citep[see, as early examples,][]{,he:2017,agarwal2018demand,calsamiglia2020structural}. In particular, the approach in \cite{agarwal2018demand}, when applied to DA, would be robust to payoff-irrelevant mistakes. However, it would not allow for mistakes with small but positive payoff consequences.  Further, when the approach is taken to data, the curse of dimensionality becomes severe as the number of available schools grows.\footnote{To verify the optimality of a submitted ROL, one should compare it against all possible ROLs. However, for example, the number of possible ROLs with 10 available schools is nearly 9.9 million.}  By contrast, our proposed empirical method is computationally easy and robust to the strategical mistakes that are commonly observed in practice.

Some subsequent papers have attempted to deal with the computational issues associated with the portfolio choice approach. \cite{larroucau2018short} show that it is sufficient to consider a subset of ROLs (``one-shot swaps") to avoid the curse of dimensionality, but this result applies only in settings where the cutoffs, and hence the admission probabilities, are independent across schools (e.g., Multiple Tie-Breaking, or MTB).  \cite{Idoux_JMP} modifies \cite{agarwal2018demand} by assuming that applicants face application costs, are of limited rationality,\footnote{Relatedly, \cite{leeson2023} considers a model that allows the applicants to have wrong beliefs about admission chances and to consider only a limited set of schools when applying to schools.} and use a heuristic method to choose their ROLs. She thus avoids the curse of dimensionality and allows schools with small admission probabilities to be unranked, which makes her method robust to payoff-insignificant mistakes to some degree. However, her method does not allow for other common types of mistakes. For example, the ranked schools must be in the true preference order in her model, ruling out the possibility of a student ranking schools with small admission probabilities in a wrong order, a pattern documented for example in \cite{hassidim2021limits}.
Moreover, the method neglects that some ranked schools can be completely irrelevant under STB, including those with high admission probabilities (see \Cref{ex:stb-mistakes}). In contrast, \teps{} is robust to all such mistakes as long as the payoff consequence is not too large.

Stability has been a key identifying condition in the two-sided matching literature \citep[see][for a survey]{chiappori_econometrics_2016}. There, agents on both sides have preferences that are unknown to the researcher. This setting includes decentralized school choice, such as that in Chile \citep{he2021identification}, in which private schools can rank students freely. On the contrary, in our framework of centralized school choice, school preferences/priorities are known to some extent because schools rank students according to certain pre-specified rules. 

\section{Theoretical Analysis \label{sec:theory}}

\def\D{\Delta}
\def\t{\theta}
\def\m{\mu}
\def\T{\Theta}
\def\e{\epsilon}

\def \sl {\underline s}
\def \sh {\overline s}

We consider a model of a large matching market operated by DA. After introducing the primitives and the robust equilibrium solution concept, we will establish our main theorem. {We begin by motivating why payoff-irrelevant or -insignificant mistakes remain relevant even when the assignment is determined purely by lotteries and nevertheless how the presence of uncertainty helps to reveal students' preferences. }

\begin{ex}\label{ex:stb-mistakes} \rm  Consider three schools $a,b$, and $c$, which admit students exclusively based on STB lotteries: each student is assigned a single lottery number that applies to all schools.\footnote{Hence, all schools are `within reach', and the usual reason for mistakes---clearly out of reach schools---does not apply in this example.} Given DA, each student is assigned her highest-ranked school among those whose cutoffs are below her lottery number (i.e., feasible.) Suppose that the schools' cutoffs are always ordered such that $P_a>P_b>P_c$, where $P_j$ is school $j$'s cutoff, $j=a,b,c$. If a student happens to have preferences $b$-$a$-$c$, ranked in the descending order, then no matter her lottery draw, school $a$ is irrelevant for her since whenever $a$ is feasible to her, $b$ is also feasible to her, and she prefers $b$ over $a$.  Consequently, reporting ROLs, $b$-$c$ or $b$-$c$-$a$, entails no payoff loss compared with reporting truthfully, her dominant strategy.  In other words,  students need not report their ROLs truthfully. 
 The reason is that the uncertainty created by STB is not full support, so it leaves the scope for payoff-irrelevant mistakes. 
  
    Nevertheless, the uncertainty does help to reveal students' preferences through ROLs. Consider the same student with preference $b$-$a$-$c$.  Given the (STB-induced) uncertainty, the event that all schools are feasible and the event that {\it only} $b$ and $c$ are feasible have positive chances of being realized.\footnote{The lotteries yield the following four positive-probability events:  (i) all schools are feasible to her; (ii)  schools $b$ and $c$ are feasible to her; (iii) only school $c$ is feasible to her; and (iv) no school is feasible to her. Event (i) occurs when the student's score is higher than $P_a$, event (ii) occurs when her score is in between $P_b$ and $P_a$, event (iii) occurs when her score is between $P_c$ and $P_b$, and event (iv) occurs when her score is below $P_c$.} Hence, the student would exercise caution in ranking those schools to avoid payoff losses by considering those possibilities, and we can learn that $b\succ a,c$ by carefully examining the uncertainties.   
    This is much more than one could learn from invoking stability in an environment without uncertainty, mechanically applying ACH or FGH. Suppose there is no uncertainty and the student ``knows" that only $c$ is feasible. Then, no preference is revealed by invoking the stability hypothesis since \emph{any} ROL that lists $c$ will produce the same outcome i.e., $c$ is the only feasible school and she is assigned $c$.
\end{ex}

\subsection{Model Primitives}

\paragraph{Continuum Economy.}   We begin with  a \textit{continuum economy}, which will serve as a benchmark for finite economies---the object of our central interest.  The continuum economy $E$ consists of a finite set of \textit{schools}  $C=\{c_1, ..., c_{C}\}$  and a unit mass of students with types $\widetilde \t\in \widetilde \T$.  Note that, with slight abuse, we use $C$ to denote both the set and cardinality of the schools.  Let $\widetilde C:=C\cup \{\o\}$, where $\o$  corresponds to the outside option (which may involve attending a private school or an outside-district school).  
The schools have masses $S=(S_1, ..., S_C)$ of seats, where $S_j\in (0,1]$, and $S_{\o}=\infty$.  

Each student has an {\bf ex-ante type} $\widetilde\t=(u, t)$, where $u^{\widetilde\t}=(u_1^{\widetilde\t}, ..., u_C^{\widetilde\t})\in [\ul, \uh]^C$ are von-Neumann Morgenstern utilities of attending schools $C$, for some $\ul< 0<\uh$. By normalization, $u_{\o}^{\widetilde\t}=0$ for all ${\widetilde\t}$.  A student has ex-ante priority type $t$, interpreted as her ``intrinsic'' priority or belief on her merit.  Let $T$ be the set of all ex-ante priority types. Students observe their ex-ante type $(u,t)$ prior to application.  Let $\widetilde \eta$ denote the probability measure of $\widetilde\t\in \widetilde{\T}:= [\ul, \uh]^C\times T$.   

Given  $t\in T$, a student's  {\bf scores} $s\in [0,1]^{C}$ are drawn according   a distribution function $\Phi^t\in \Delta([0,1]^{C})$ that may depend on $t$.  The scores determine a student's ex-post priorities  used by schools for her assignment. Students do not observe these scores  when they submit applications.\footnote{In NYC school district, lottery draws have been recently disclosed to the families who request them prior to their application (see \href{https://www.schools.nyc.gov/enrollment/enroll-grade-by-grade/how-students-get-offers-to-doe-public-schools/random-numbers-in-admissions}{here}).  This is an exception than the norm.  Regardless, this revelation of scores does not invalidate our assumption since those revealed scores  can be treated as ex-ante priority type $t$.} Randomness of scores arises from the use of lotteries as tie-breakers or uncertainty about test scores.  Let $S^t:=\prod_c[\sl_c^t,\sh_c^t]$ be the support of the scores for type $t$, where   $\sl_c^t$ and $\sh_c^t$ are respectively the infimum and supremum  scores for type-$t$ student for school $c$.  
We call the pair $\theta=(u,s)$ for a student her  {\bf ex-post type}.  The measure $\widetilde \eta$ of ex-ante types, together with $(\Phi^t)_t$, induces a probability measure $\eta$ on the ex-post type $(u,s)\in \T:=[\ul, \uh]^C\times [0,1]^{C}$.  We assume that  $\eta$ is atomless.  (By contrast, $\tilde \eta$ can, and typically will, have atoms.)  For our purpose, the atomlessness assumption ensures that indifferences either in student preferences or in schools' scores arise only for a measure zero set of students. In summary, the continuum economy is summarized by $E=[\widetilde \eta, S, (\Phi^t)_t]$.
 
While we establish our main result under a general priority structure in Appendix~\ref{appendix:proof_theory}, for ease of exposition, here we focus on the following priority structure. The schools $C$ are partitioned into three types/subsets of schools, $C_1$, $C_2$, and $C_3$, so that schools of each subset use distinct priority rules. Specifically, we assume that
$T=T_1\times T_2\times T_3$ so that schools in $C_i$ apply priorities in $T_i, i=1, 2,3.$

  \begin{itemize}
 	
	\item [(a)] {\it Merit-based assignment with known scores}:  Each school in $C_1$ prioritizes students based on only priorities in $T_1=[0,1]^{|C_1|}$, where for each $t\in T_1$, $\phi^t$ is degenerate with $\sl_c^t=\sh_c^t=t_c$ for all $c$.  This corresponds to the case in which   priorities are given by some merit scores which are known to students when they apply.  Examples include Australian college admissions and Paris high school assignments.  This is the case studied by \citet{ACHmistakes}.
 	
    \item [(b)] {\it Coarse priorities}:   We assume that  $T_2$ is finite and ties are broken by lotteries.  More precisely, each school $c\in C_2$ has a finite number $n_c$ of intrinsic priorities  $\mathcal T_c=\{0, \frac{1}{n_c}, ..., \frac{n_c-1}{n_c}\}$. A student's ex-ante type consists of a profile $t=(t_c)_{c\in C_2} \in T_2=\prod_{c\in C_2} \mathcal T_c$. Then, for each school, $c\in C_2$, the score for a student with type $t$ is then a composite score, $s_c=t_c+\lambda_c\frac{1}{n_c}$, where $\lambda_c$ is the lottery score used for breaking a tie at school $c$.  The lottery scores, $(\lambda_c)_{c\in C_2}$, are either uniform on $[0,1]^{|C_2|}$ in the case of MTB rule, or $\lambda_{c}=\lambda$ for all $c\in C_2$ with $\lambda$ randomly drawn uniformly  from $[0,1]$ in the case of STB rule.  This feature corresponds to the common structure of priorities used in US public schools.  
 	
 	\item [(c)]  {\it Merit-based assignment with unknown scores}: We assume that $T_3=[0,1]$, and, for each $t\in T_3$, $\Phi^t$ is absolutely continuous with a strictly positive density over scores  $s\in [0,1]$. This corresponds to the case of NYC's Specialized High School assignment, which follows \'a la serial dictatorship based on the SHSAT scores. Students submit their ROLs at the time they take the SHSAT. The ex-ante type $t\in [0,1]$ corresponds to the student's belief about the score. 
 \end{itemize}
If $C=C_1$,   no students face any uncertainties, so the model specializes to that of ACH.

\paragraph{Finite Economies.} We are interested in a sequence of finite random economies converging to the continuum economy $E$ in the limit.  Specifically, let $F^k=[\widetilde\eta^k, S^k, (\Phi^{k,t})_t]$ be a \textit{$k$-economy} that consists of $k$ students each with type $\widetilde\t$ drawn independently according to $\widetilde\eta$, and the vector  $S^k=[k\cdot S]/k$  of capacity per student, where $[x]$ is the vector of integers nearest to $x$ (with a rounding down in case of a tie).  The distribution $\widetilde\eta^k$ is simply an empirical measure of ex-ante types $\widetilde \t$; likewise, we let $\eta^k$ denote an empirical measure of ex-post types $(u,s)$.

\paragraph{Matching and Mechanism.}    A \textit{matching} is a mapping  $\m: C\cup \T\to 2^{\T} \cup (C\cup \T)$ that describes how students are assigned to alternative schools based on their ex-post types. It satisfies the usual two-sidedness and consistency requirements as well as ``open on the right'' as  defined in \citet{azevedo/leshno:16} (henceforth AL, see p. 1241).  The key concept of our inquiry deals with the property known as stability.  To describe it, fix an economy, which can be either the continuum economy $E$ or a {\it realization} of a finite $k$-economy.  Fix any profile $p\in [0,1]^C$, interpreted as   {\bf cutoffs} of schools.   For each type $\t=(u,s)$, which consists of a realized utility vector $u$ as well as a realized score vector $s$, we say a school $c$ is {\bf ex-post feasible} if $s_c> p_c$.  Now let $D_c(p)$ denote the measure of students for whom $c$ is the best school among those that are feasible according to their true preferences, given $p$.  A stable matching is often defined in terms of two properties: {\it individual rationality} and {\it no blocking},\footnote{Individual rationality requires that no participant (a student or a school) is assigned a partner that is not acceptable. No blocking means that no student-school pair exists such that the student prefers the school over her assignment, and the school has either a positive measure of vacant positions or admits a positive measure of students whom the school ranks below that student.  In priority structure (b), the scores may be a result of tie-breaking, so no blocking, and therefore stability, should be defined based on ex-ante type $\widetilde \t$.  Nevertheless, we focus on ex-post stability since it, and not the stability, provides a foundation for our empirical method for inferring applicants' preferences.  Further, (asymptotic) ex-post stability implies (asymptotic) ex-ante  stability, so establishing the former will deliver the latter.} but  it is more convenient here to adopt the following characterization by AL as our definition.   A \textbf{stable matching} is a matching in which all students are assigned the best feasible schools at cutoffs $p$ which ``clear the market'' in the sense that $D_c(p)\le (=) S_c$ for all $c$ (if $p_c>0$).  
Importantly, in the $k$-economy, both the cutoffs and the resulting assignment are random, so the stability requires the stated condition to hold with probability one.  

Student-proposing deferred acceptance algorithm, or  DA in short,  takes students' reported ROLs and their scores  as input and outputs a stable matching that is Pareto-best for all students given the reported ROLs, or the matching that corresponds to the ``lowest'' market-clearing cutoffs.\footnote{By the well-known lattice property, the lowest market clearing cutoffs are well defined, given the responsive preferences on both sides.}  We call them {\bf DA cutoffs}.  As is well-known, DA makes it a weakly dominant strategy for each student to report a truthful ROL. In light of our main motivation, however, we shall allow dominated strategies to be used by students.  
 
\subsection{Robust Equilibria}

For each $k$-economy, $F^k$, we let $\{1, ..., k\}$ index the students present in that economy. We assume that each student observes her own ex-ante type $\widetilde \t$ but not the types of other students.  A (mixed) strategy  by each student $i\in \{1, ..., k\}$ is denoted by  a measurable function $\sigma_i: \widetilde \T \to \Delta(\mathcal{R})$, where $\mathcal{R}$ is the set of all possible ROLs of $\widetilde C$.  We use $\rho$ to denote a {\bf truthful reporting strategy (TRS)}: namely, $\rho(\widetilde \t)$ ranks school $a\in\widetilde C$ ahead of $b\in \widetilde C$ if and only if $u^{\widetilde \t}_a>u^{\widetilde \t}_b$.\footnote{How  $\rho$ breaks a tie is immaterial since $\eta$ is assumed to be atomless so that a tie will occur with zero probability. Hence, we leave this unspecified.}

In order to allow for the types of mistakes observed in  recent evidence, we focus on {\bf robust equilibria} developed by ACH.  This concept allows applicants to use dominated strategies but requires that the payoff loss resulting from such strategies be arbitrarily small in a large economy. In particular, this solution concept accommodates the commonly documented mistakes of not ranking, or ranking incorrectly, out-of-reach schools.

While we are agnostic about the sources of such mistakes, the current concept captures the decision-maker's inattention in a way similar to rational inattention \citep{matvejka2015rational} and quantile response equilibria \citep{GOEREE2002247}. Individuals facing decision costs would presumably rationally allocate more attention to choices that have high payoff consequences and less attention to those that have low payoff consequences, so they end up making more mistakes on the latter choices. Unlike these theories, however, such inattention becomes arbitrarily small as the economy grows large.\footnote{The idea of payoff-sensitive mistakes is also reminiscent of \cite{myerson1978refinements}'s proper equilibria. Note, however, the motivations are completely different: the goal here is to allow for such mistakes, whereas proper equilibria (and trembling-hand perfection) refine Nash equilibria by requiring robustness to  mistakes.}

To operationalize the concept, we need to consider a strategy profile along a sequence of finite economies.  Formally, we consider an infinite profile $\sigma:= (\sigma_{i})_{i \in \mathbb{N}}$ of strategies specified for each player $i\in \mathbb{N}$, with the interpretation that in each $k$-economy $F^k$, its  $k$-truncation $\sigma^k:=(\sigma_1, ..., \sigma_k)$ is employed by its participants as their strategies.  Note that we allow students to adopt asymmetric strategies, submitting different ROLs even when they have the same ex-ante type.  The solution concept follows:  

\begin{defn} An infinite profile $\sigma$ is a \textbf{robust equilibrium} if, for any $\e>0$, there exists $K\in \mathbb{N}$ such that for $k>K$, its $k$-truncation $\sigma^k$ is an  {interim $\e$-Bayes Nash equilibrium}---namely, for $i$, $\sigma_i$ gives student $i$ of each type $\t$ a payoff within $\e$ of the highest possible (i.e., supremum) payoff she can get by using any strategy when all the others employ $\sigma_{-i}^k$.
\end{defn}

The solution concept does not require exact optimality for the strategies but rather their near optimality in a sufficiently large economy.  
Although adding considerable demand on the proof of our main theorem, allowing for asymmetry makes the concept broadly applicable.  For our main result, we focus on a slight strengthening of this concept:

\begin{defn}  For any $\gamma\in (0,1)$, a strategy is   {\bf $\gamma$-regular} if it coincides with TRS with probability at least of $\gamma$.  A profile $\sigma$ is  a {\bf regular robust equilibrium} if it is robust and there exists a $\gamma\in (0,1)$ such that $\sigma_i$  is $\gamma$-regular for all $i\in \mathbb{N}$. 
\end{defn}

\subsection{Analysis of Robust Equilibria}

As illustrated in  \Cref{ex:stb-mistakes}, applicants may not use TRS in a regular robust equilibrium.\footnote{While the example in \Cref{ex:stb-mistakes} suggests only one preference type who may deviate from TRS, it is not difficult to imagine that there is a significant fraction of applicants with the same preference type.} Hence, one cannot  use truth-telling as a behavioral restriction.
Instead, we establish that the outcome of a regular robust equilibrium is asymptotically stable, which will give rise to a method for inferring preferences under the robust equilibrium.  

We begin with a few concepts.  For any potential cutoffs $p\in [0,1]^C$, we say a strategy $\sigma_i$ by student $i$ is {\bf {stable-response strategy (SRS)} against} $p$ if the student receives the best feasible school relative to the cutoffs with probability one;  $\sigma_i$ is {\bf non-SRS against $p$} if it is not an SRS against $p$.  In the context of DA, we say a strategy is {\bf stable} if it is stable against the prevailing DA cutoffs, given the strategies.  Certainly, TRS is SRS but an SRS need not be TRS, as illustrated in \Cref{ex:stb-mistakes}.  Next, we define \textbf{asymptotic stability} as:

 \begin{defn} A profile $\sigma$ is \textbf{asymptotically stable} if the fraction of students who employ SRS in each $\sigma^k$ converges  to 1 in probability as $k\to\infty$.
\end{defn}
 
 In words, asymptotic stability means that all except for a probabilistically vanishing fraction of students receive the stable assignment for almost all realizations of uncertainties they face.  This will provide a key method for inferring preferences from the choice data generated under the DA algorithm.  
ACH establishes this result when students face no uncertainty in $s$ (i.e., school types $C_1$) and under the assumption that $\eta$ has full support.  The full support assumption guarantees the uniqueness of stable matching, which plays a crucial part in their proof. However, it is  important to relax full support for our $\eta$, as simple STB with no intrinsic priorities and the more realistic scenario of coarse priorities with STB fail full support. This motivates us to weaken the condition:
\begin{as} [\textsc{Marginal Full Support}]
   For each $r\in \mathcal R$ and for each $c\in C$,  $\inf_{s_c\in [0,1]} \bar\eta(s_c)>0$, where
    \begin{align*}
        \bar\eta(s_c):=\int_{\rho(u)=r, s_{-c}}\eta(u, s_c, s_{-c}) du ds_{-c}
    \end{align*}
    is the marginal density for score $s_c$. 
\end{as}
This condition requires that any given ordinal preference type has full support for its \emph{marginal} distribution of score for each school.   If the scores have positive \emph{joint} density on the compact support $[0,1]^{C}$ as is assumed in AL, the condition will hold.  Importantly, however,  the condition allows for the support of scores to be low dimensional. In particular, Marginal Full Support holds in the case of STB  without any intrinsic priorities; in that case, the support of the uncertainty/lottery is diagonal and one-dimensional, yet the marginal of its lottery has full support.  With nontrivial intrinsic priorities, the condition states that each ordinal preference type has the full support of priorities.  We view the condition to be very mild.  We are able to generalize the uniqueness result by \cite{azevedo/leshno:16} and \cite{azevedo2014imperfect} under the weaker condition:\footnote{Relatedly, \cite{agarwal2018demand} and   \cite{grigoryan2022} provide other weak conditions for guaranteeing the uniqueness of stable matching in the continuum economy.}

\begin{lem}  \label{prop:uniquestable} If $\eta$ satisfies Marginal Full Support, then $E=[\eta,S]$ admits a unique stable matching. 
\end{lem}

We are now in a position to state the main result.

\begin{thm}
\label{thm:stability}  Suppose $\eta$ satisfies Marginal Full Support. Then, any regular robust equilibrium is asymptotically stable.
\end{thm}

This theorem states that in a large enough economy,    for almost every realization of uncertainty, virtually all students are assigned their most preferred schools among those that are feasible to them given the realized cutoffs, even though the feasible schools are unknown to them when they submit ROLs. This theorem implies that we can invoke stability for each realization of uncertainty; this is precisely what we will do in our TEPS procedure.

In light of the multiplicity in student behavior in robust equilibrium, another question is how to conduct counterfactual analysis, an exercise that is common in market design research.  The next corollary of \cref{thm:stability} justifies that, as far as predicting an assignment outcome is concerned, we can simply assume truth-telling among all students in a counterfactual regime. 
\begin{cor}\label{cor}  Suppose $\eta$ satisfies Marginal Full Support.  Then, in any regular robust equilibrium, the fraction of students whose assigned school is the one that would arise if everyone employs TRS converges to one in probability as $k\to \infty$. 
\end{cor}

\begin{remark} [Relationship with ACH's asymptotic stability result]
    Both ACH and the current paper build on the uniform convergence of cutoffs. If the priorities are strict and known in advance as in ACH, the cutoff convergence makes it clear that for most of the applicant types, playing a non-stable response entails a discrete loss, so this leads to no such strategies being used in that setting.  But, when priorities may be unknown and random, then the discreteness of the payoff loss from playing a non-stable response strategy is not easy to establish.  In fact, the nature of payoff loss depends crucially on the distribution of (unknown) priorities for the applicant.  Controlling this distribution requires the properties of the uncertainties embodied in the priorities assumed in (a), (b), and (c). In fact, \Cref{appendix:proof_theory} defines a general condition underlying the priority structure that guarantees the discreteness of the payoff loss. (Hence, the theorem is more general than the version stated in the text.)  The proof is highly nontrivial:  see \Cref{lem:assignment-gap} and its subsequent use in the proof of \Cref{thm:stability-general}.
\end{remark}

\section{Inferring Student Preferences from ROLs: Transitive Extension of Preferences from Stability\label{sec:method}}

Building upon \Cref{thm:stability}, we develop a procedure for inferring students' (ordinal) preferences from school choice data, called \textsf{Transitive Extension of Preferences from Stability (TEPS)}.  We show that it extracts the maximal information about participants' preferences based on our theory.  

  Suppose we have a typical dataset generated by a centralized DA assignment.
 Assume that the analyst observes the ROLs submitted by students/applicants and their (intrinsic) priorities, as well as the capacities of schools.  

\subsection{Preference Relations Inferred by Stability and Transitivity \label{sec:TE_0}}

Since we infer the preference of each applicant separately, it is without loss to focus on a single applicant.  Assume the applicant has submitted a ROL $R$ and has priorities $\ip$.  For notational simplicity, we shall drop the dependence on $R$ and $t$ whenever there is no ambiguity.  We assume there is no outside option although outside options can be easily accommodated.\footnote{Our procedure below can be extended straightforwardly with the set of schools redefined to $\widetilde{C}=C\cup \{\o\}$, where   $\o$ is an outside option representing an always-feasible   alternative, such as non-assignment.} Let $\uncset$ be the set of all assignment-relevant states, or more precisely, all possible profiles of priorities that our applicant, as well as all other applicants, may have at all schools.  In practice, each state $\uncelem\in \uncset$ corresponds to  the lottery draws for all students, the profile of their scores from the entrance exam, and other uncertainties each student faces on the other applicants' preferences, all of which affect the cutoffs of schools.

Given the data, the goal is to identify the set of   preference relations $\pref{}:=\{(x,y)\in C^2:x\in C\text{ is inferred   more preferred to }y\in C\}$ inferred from that data according to \Cref{thm:stability}.  The precise interpretation of \Cref{thm:stability}  may depend on the market size and the payoff relevance of mistakes.  For expositional ease, we begin with a rather strong interpretation of \Cref{thm:stability} that {\it the matching outcome must be stable for every realization $\uncelem$ of the uncertainty}; namely, no student makes any payoff-relevant mistakes.  While this implication is justified  only in the limit continuum economy, it serves as a good starting point for describing our method.  We will relax this assumption  later in \cref{sec:otherTEPS}.

The \teps{} procedure takes the following three steps to produce preference relations $\pref{}$.  To illustrate the procedure, consider the example.  
\begin{ex} \label{ex:TE} Suppose that there are six schools, $\{c_0,c_1,c_2,c_3,c_4,c_5\}$ and a student with intrinsic priorities $\ip$ and a submitted ROL $R= c_4$-$c_3$-$c_2$-$c_1$.  Depending on possible realizations of lottery draws, there are four possible events:  
1.  the set $\{c_3,c_4\}$ is feasible to the student with probability $0.4$;  
2. the set $\{c_0,c_1\}$  is feasible to the student with probability $0.3$; 
3. the $\{c_0,c_1,c_2\} $ is feasible to the student with probability $0.25$; 
4. the set  $\{c_1,c_4\}$ is feasible to the student with probability $0.05$.
\end{ex}

\noindent\textbf{Step 1.  Simulating uncertainties and compiling choice data:}  
 
 The first step simulates or ``traces out'' all possible profiles of priorities by simulating uncertainties.  For instance, lottery-driven uncertainties can be simulated easily by conducting the lotteries and running DA identically as the underlying process generating the data.\footnote{To simulate the student composition uncertainty,  we may bootstrap from the observed ROL to create multiple economies/applicant compositions.  If the uncertainty arises from other sources, such as test scores that are yet to be realized at the time of application, then one may build an empirical model to simulate the distribution of test scores say based on the observables (which may include various academic performance measures).}  Each simulated state $\uncelem\in \uncset$ leads the realized priorities of students, and given their ROLs, we can obtain the schools' cutoffs.  From this, we can identify for each student the set $\fset(\uncelem)$ of her {\it feasible schools} in state $\uncelem$, and her {\it assigned school}, $\assign(\uncelem)$---namely, the highest-ranked choice within $\fset(\uncelem)$ according to the student's ROL.   Naturally, there can be multiple $\uncelem$'s that lead to the same feasible set and, therefore, the same assignment.  Hence, we consider a partition $P_{\uncset}$ of $\uncset$ into equivalence classes, $\unc$, such that $\fset(\uncelem)=\fset(\uncelem')$ if and only if  $\uncelem, \uncelem'\in \unc$, for each $W\in P_{\uncset}$. In the example, we have four  equivalence classes of the uncertainties, $\unc_1,\unc_2,\unc_3,\unc_4$ in terms of distinct feasible sets,  
 $\fset_{\unc_1}=\{c_4,c_3\}$, 
 $\fset_{\unc_2}=\{c_1,c_0\}$, 
 $\fset_{\unc_3}=\{c_2,c_1,c_0\}$, and 
 $\fset_{\unc_4}=\{c_4,c_1\}$.  Given her ROL, she would be assigned 
 $\assign_{\unc_1}=c_4$, 
 $\assign_{\unc_2}=c_1$, 
 $\assign_{\unc_3}=c_2$, and 
 $\assign_{\unc_4}=c_4$ in the corresponding classes.

\begin{figure}[h]
    \centering
    \begin{subfigure}[h]{0.30\textwidth}
		\centering
        \vskip 0.2cm
		\begin{tikzpicture}

\draw (-1.95,1.6) circle (0.2 and 0.2) node {{$c_4$}};
\draw (-0.65,1.6) circle (0.2 and 0.2) node {{$c_1$}};
\draw (0.65,1.6) circle (0.2 and 0.2) node {{$c_2$}};
\draw (1.95,1.6) circle (0.2 and 0.2) node {{$c_4$}};



\draw  (-1.95,1.1) ellipse (0.5 and 0.8) node at (-1.95,0.85) {\begin{tabular}{c} $c_3$ \end{tabular}};
\draw  (-0.65,1.1) ellipse (0.5 and 0.8) node at (-0.65,0.85) {\begin{tabular}{c}  $c_0$ \end{tabular}};
\draw  (0.65,1.1) ellipse (0.5 and 0.8) node at (0.65,0.85) {\begin{tabular}{c}  $c_1~c_0$ \end{tabular}};
\draw  (1.95,1.1) ellipse (0.5 and 0.8) node at (1.95,0.85) {\begin{tabular}{c}  $c_1$ \end{tabular}};

\node at (-1.95,0) {$\fset_{\unc_1}$};
\node at (-0.65,0) {$\fset_{\unc_2}$};
\node at (0.65,0) {$\fset_{\unc_3}$};
\node at (1.95,0) {$\fset_{\unc_4}$};

\end{tikzpicture}
		\vskip -0.2cm
		\caption{\scriptsize Step 1}
	\end{subfigure}
    \hspace{0.5cm}
    \begin{subfigure}[h]{0.30\textwidth}
		\centering	
		\begin{tikzpicture}

\draw (1.2,1.6) -- (0.9,0.8) node [pos=0,above] {$c_4$} node [below] {$c_3$} ;
\draw (1.2,1.6) -- (1.5,0.8) node [below] {$\color{red}{c_1}$} ;

\draw (2.4,1.6) -- (2.1,0.8) node [pos=0,above] {$c_2$} node [below] {$\color{red}{c_1}$} ;
\draw (2.4,1.6) -- (2.7,0.8) node [below] {$c_0$} ;

\draw (3.6,1.6) -- (3.6,0.8) node [pos=0,above] {$\color{red}{c_1}$} node [below] {$\color{blue}{c_0}$} ;

\node at (1.2,0) {$\preff_1$};
\node at (2.4,0) {$\preff_2$};
\node at (3.6,0) {$\preff_3$};

\end{tikzpicture}
		\vskip -0.2cm
		\caption{\scriptsize Step 2}
	\end{subfigure}
	\begin{subfigure}[h]{0.30\textwidth}
		\centering	
		\begin{tikzpicture}

\draw (2.0,1.6) -- (1.7,1.4) node [pos=0,above] {$c_4$} node [below] {$c_3$} ;
\draw (2.3,1.0) -- (2.3,0.8) node [below] {$\color{blue}{c_0}$};
\draw (2.0,1.6) -- (2.3,1.4) node [below] {$\color{red}{c_1}$} ;

\draw (3.4,1.6) -- (3.4,1.4) node [pos=0,above] {$c_2$} node [below] {$\color{red}{c_1}$} ;
\draw (3.4,1.0) -- (3.4,0.8) node [below] {$\color{blue}{c_0}$} ;

\node at (2.7,0) {$\mathcal{P}$};
\end{tikzpicture}
		\vskip -0.2cm
		\caption{\scriptsize Step 3}
	\end{subfigure}
	\caption{An Example of TEPS\label{fig:transitiveextension}}
\end{figure}
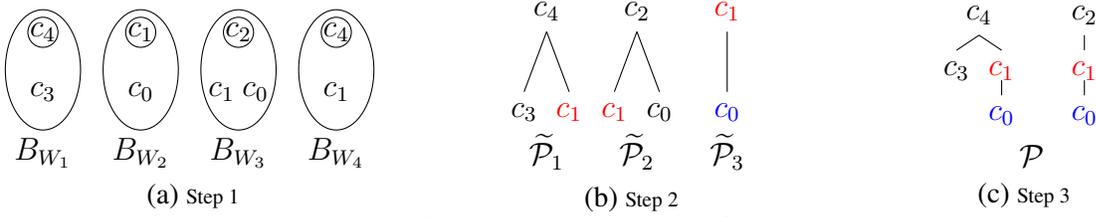

 We then compile  a {\it choice data} for the student---namely, a pair $(B_W, \alpha_W)$ of a feasible set of schools and an assigned school---for each equivalence class $\unc\in P_{\uncset}$.\footnote{Since we do not consider outside options, we omit $(\assign_\unc,\fset_\unc)$ if $\assign_\unc=\o$, a non-assignment. In case we include an outside option, we should also include $(\assign_\unc,\fset_\unc)$, where $\assign_\unc=\o$.} 
 The choice data for the example is depicted  in 
 Figure~\ref{fig:transitiveextension}(a).

\noindent\textbf{Step 2.  Inferring preference relations in each realized uncertainty:}

 \Cref{thm:stability} states that in the limit as the economy grows large, all but the vanishing fraction of students must be stably assigned for every state $\uncelem\in\uncset$.  This implies that, for any given student,  her assigned school $\alpha_\unc$ is her most preferred choice among the set of feasible schools $B_W$ for each $W\in P_{\uncset}$.  This yields preference relations $\preff_W\subset C^2$ for each  $W\in P_{\uncset}$.\footnote{Recall that each element in $\preff$ is an ordered pair of schools such that $c_1$ is inferred preferred to $c_2$ for each $(c_1,c_2)\in \preff$.}   Let $\preff :=\cup_{\unc\in P_{\uncset}}\preff_{\unc}$ denote the set of all preference relations obtained in this way.     We next partition $\preff$ into ordered sets of preference relations $\{\preff_1, \preff_2, ...., \preff_m\}$ such that 
 if $(c, d)\in \preff_i$ and $(c', d')\in \preff_j$, then $c$ is ranked ahead of $c'$ if $i<j$.  In our running example, this step produces ordered sets of preference relations, $\preff_1=\{ (c_4,c_3),(c_4,c_1)\}$, $\preff_2=\{ (c_2,c_1),(c_2,c_0)\}$, and $\preff_3=\{(c_1,c_0)\}$.  In Figure~\ref{fig:transitiveextension}(b), those are represented by three ordered trees with one length, where the roots correspond to the schools the student may be assigned in some realizations of the uncertainties.

\noindent\textbf{Step 3. Extending preference relations by transitivity axiom:}  

The last step links together the preference relations $\preff$ obtained from Step 2 via the axiom of transitivity.   Specifically, we use the following algorithm.\footnote{There are other computationally fast algorithms for finding a transitive closure of binary relations, based on shortest path search,  breadth- or depth-first search, that has polynomial time in the number of choices (see \url{https://en.wikipedia.org/wiki/Transitive_closure}.) In fact, the computational burden is further reduced by the length of one's ROL, which is often significantly smaller than by the number of all schools. This is because in order for a school to be involved in the transitive closure procedure, it must be  not only ranked on the ROL but also revealed preferred to some other schools that are ranked.}   Consider first   $\preff_m$ and $\preff_{m-1}$, where $m$ is the highest index.  The algorithm constructs a new set $\mathcal P_{m-1}$, by including  $\preff_m\cup\preff_{m-1}$ along
with preferences relations obtained by the transitive extension operation:  
If $(y,z)\in \preff_m$ and $(x,y)\in \preff_{m-1}$---i.e., if the assigned school under $\preff_{m}$ is  revealed inferior to the school that is assigned under $\preff_{m-1}$---then, we declare  $x$ to be revealed preferred to $z$ and add $(x,z)$ to $\pref{}_{m-1}$. We then iterate the same procedure on $\pref{}_{m-1}$ and $\preff_{m-2}$ to obtain $\pref{}_{m-2}$, and so on.  The repeated iteration returns the final output $\pref{} :=\mathcal P_1$. 
In the example,   $\pref{}=   \{(c_1,c_0),(c_2,c_1),(c_2,c_0),(c_4,c_1),(c_4,c_3),(c_4,c_0)\}$,  depicted in 
Figure~\ref{fig:transitiveextension}(c)  
as a collection of trees.  

\vspace{0.5cm}
Note that \teps{} procedure does not infer all preference ranks observed from a student's ROL.  In the example, \teps{} procedure does not infer any relations between $c_4$ and $c_2$ despite the ROL listing $c_4$-$c_3$-$c_2$-$c_1$.  The reason is that these two schools are never feasible simultaneously,\footnote{Such cases occur under MTB, which uses distinct lotteries to break ties at different schools.  Even with STB, such cases may occur as long as the cutoffs are not degenerate (i.e., in a finite economy.) This is because the relative ranking of schools' cutoffs may change under different realizations of the STB lotteries.} so their relative rankings are not directly revealed, nor are they inferred indirectly by transitivity.  Hence, \teps{} does not view this part of the ROL as payoff-relevant or reliable.  
Also, note that one of the unranked schools, $c_5$, is never feasible (i.e., out of reach) under any realization of the uncertainties, and as a result, ranking $c_5$ arbitrarily or even omitting it is payoff-irrelvant for the student.  Hence, \teps{} does not infer any preference relation regarding $c_5$. 
In short, \teps{} extracts only a subset of preference relations of what ROL (or WTT, as will be seen later) infers.  

Nevertheless, \teps{} procedure infers as many preference relations  as  possible consistent with \Cref{thm:stability}.  To state it formally, fix a student's ROL  $R$ and her observable priorities $\ip$.  For any $\uncset'\subset \uncset$, let $\pref{\text{ST}}(\uncset')$ denote the preference relations that can be inferred by stability on all possible uncertainties in $\uncset'$ and the  transitivity axiom.  Then, we have the following result:

\begin{prop}\label{prop:maximal}  Fix a student's ROL  $R$ and her observable priorities $\ip$.  If Step 1 of \teps{} simulates the uncertainties $\uncset'\subset \uncset$, then $\pref{}=\pref{\text{ST}}(\uncset')$.   That is,   the \teps{} procedure infers the preference relations if and only if they can be  inferred by the stability conditions on possible uncertainties $\uncset'\subset \uncset$ and transitivity. 
\end{prop}

\subsection{Allowing Violations of Stability in TEPS \label{sec:otherTEPS}}

In principle, Step 1 of the procedure may exhaustively simulate all uncertainties, i.e., by choosing $\uncset'=\uncset$.  Imposing stability for all uncertainties is tantamount to assuming that the applicant makes  no payoff-relevant mistakes.  By \Cref{thm:stability}, this  is an implication of a robust equilibrium only for the infinite economy.  For a finite economy, a robust equilibrium is consistent with violations of stability, or payoff-relevant mistakes, as long as the payoff consequence of the mistakes is sufficiently small.  To accommodate these more realistic scenarios and thus strengthen the robustness of preference inferences, a natural approach is to identify the set of states, or uncertainties $\uncset'\subset \uncset$, that are sufficiently likely and apply TEPS only on these $\uncset'$.  Specifically, we proceed as follows.  

Recall in Step 1, we partition $\uncset$ into equivalence classes in terms of the feasible sets $\fset_W$ the applicant faces.  We now order the feasible sets in descending order of their likelihoods.  In the example, the four feasible sets are indexed according to this order, with $\fset_{W_1}$ being most likely and $\fset_{W_4}$ being least likely.  We then consider the cumulative likelihood of the feasible set starting from the most likely feasible set.  

We then apply an ``attention parameter" $\tau\in [0,100]$ in terms of the cumulative likelihood (in percentage) for feasible sets to be included in our TEPS procedure.  In our example, applying the threshold of $\tau=95$ means that we stop at  $\fset_{W_3}$, as the cumulative likelihood  of $\fset_{\unc_1}, \fset_{\unc_2}$, and $\fset_{\unc_3}$ reach 95\%.  We then focus our TEPS inference only for $\uncset'=\{\fset_{\unc_1}, \fset_{\unc_2}, \fset_{\unc_3}\}$.\footnote{Formally, Step 1 of TEPS with threshold $\tau$ always includes the most likely feasible set $\fset_{\unc_1}$, and additionally includes up to $\fset_{\unc_{\bar{T}}}$ such that $\sum_{t=1}^{\bar{T}} Prob(\unc_t)\le \tau\%$ and $\sum_{t=1}^{\bar{T}+1} Prob(\unc_t)>\tau\%$.}

Let's call the resulting procedure \teps{\tau}, which produces preference relations $\pref{\tau}$.  We label \teps{100} and $\pref{100}$ as \teps{all} and $\pref{all}$ since they use \emph{all} realizations of the feasible sets.   Using this notation, the TEPS procedure and the resulting preference relations in Section~\ref{sec:TE_0}  are \teps{all} and $\pref{all}$, respectively.  Note that as we decrease the attention parameter $\tau$, \teps{\tau} tolerates more violations of stability.  When $\tau=0$, \teps{0} only infers a student's preferences based on her most likely feasible set, amounting to her  considering only the most likely feasible set.  Hence, \teps{0} is also called \teps{top}.  

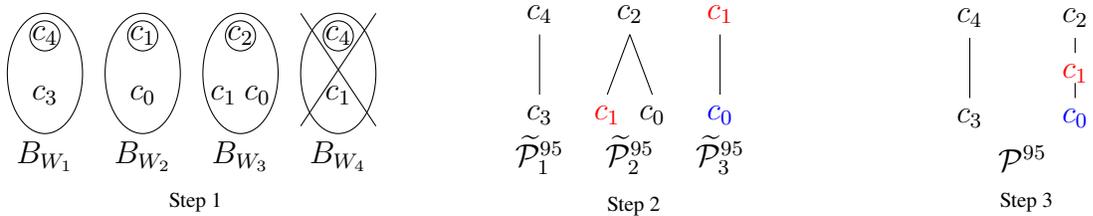
\begin{figure}[h]
    \centering
    \begin{subfigure}[h]{0.30\textwidth}
		\centering
        \vskip 0.2cm
		\begin{tikzpicture}

\draw (-1.95,1.6) circle (0.2 and 0.2) node {{$c_4$}};
\draw (-0.65,1.6) circle (0.2 and 0.2) node {{$c_1$}};
\draw (0.65,1.6) circle (0.2 and 0.2) node {{$c_2$}};
\draw (1.95,1.6) circle (0.2 and 0.2) node {{$c_4$}};



\draw  (-1.95,1.1) ellipse (0.5 and 0.8) node at (-1.95,0.85) {\begin{tabular}{c} $c_3$ \end{tabular}};
\draw  (-0.65,1.1) ellipse (0.5 and 0.8) node at (-0.65,0.85) {\begin{tabular}{c}  $c_0$ \end{tabular}};
\draw  (0.65,1.1) ellipse (0.5 and 0.8) node at (0.65,0.85) {\begin{tabular}{c}  $c_1~c_0$ \end{tabular}};
\draw  (1.95,1.1) ellipse (0.5 and 0.8) node at (1.95,0.85) {\begin{tabular}{c}  $c_1$ \end{tabular}};

\draw (1.45,1.9) -- (2.45,0.4);
\draw (2.45,1.9) -- (1.45,0.4);

\node at (-1.95,0) {$\fset_{\unc_1}$};
\node at (-0.65,0) {$\fset_{\unc_2}$};
\node at (0.65,0) {$\fset_{\unc_3}$};
\node at (1.95,0) {$\fset_{\unc_4}$};

\end{tikzpicture}
		\vskip -0.2cm
		\caption*{\scriptsize Step 1}
	\end{subfigure}
    \hspace{0.5cm}
    \begin{subfigure}[h]{0.30\textwidth}
		\centering	
		\begin{tikzpicture}

\draw (1.2,1.6) -- (1.2,0.8) node [pos=0,above] {$c_4$} node [below] {$c_3$} ;

\draw (2.4,1.6) -- (2.1,0.8) node [pos=0,above] {$c_2$} node [below] {$\color{red}{c_1}$} ;
\draw (2.4,1.6) -- (2.7,0.8) node [below] {$c_0$} ;

\draw (3.6,1.6) -- (3.6,0.8) node [pos=0,above] {$\color{red}{c_1}$} node [below] {$\color{blue}{c_0}$} ;

\node at (1.2,0) {$\preff_1^{95}$};
\node at (2.4,0) {$\preff_2^{95}$};
\node at (3.6,0) {$\preff_3^{95}$};

\end{tikzpicture}
		\vskip -0.2cm
		\caption*{\scriptsize Step 2}
	\end{subfigure}
	\begin{subfigure}[h]{0.30\textwidth}
		\centering	
		\begin{tikzpicture}

\draw (2.0,1.6) -- (2.0,0.8) node [pos=0,above] {$c_4$} node [below] {$c_3$} ;

\draw (3.4,1.6) -- (3.4,1.4) node [pos=0,above] {$c_2$} node [below] {$\color{red}{c_1}$} ;
\draw (3.4,1.0) -- (3.4,0.8) node [below] {$\color{blue}{c_0}$} ;

\node at (2.7,0) {$\pref{95}$};
\end{tikzpicture}
		\vskip -0.2cm
		\caption*{\scriptsize Step 3}
	\end{subfigure}
	\caption{An Example of Allowing Violations of Stability in TEPS: \teps{95}\label{fig:teps_95}}
\end{figure}

The logic behind the use of the attention parameter is intuitive and clear.  When a student submits her ROL, she may not pay sufficient attention to a low-probability feasible set such as $\fset_{\unc_4}$, which makes the preference ``inferred'' from such a feasible set to be unreliable.  Based on this, our \teps{\tau} simply focuses on the most likely feasible sets.   In our example, \teps{95} produces     preference relations $\pref{95}$, depicted in \Cref{fig:teps_95}, which is a subset of  $\pref{100}$ depicted in \Cref{fig:transitiveextension}(c).

\subsection{Comparison of TEPS with WTT and Stability\label{sec:wtt_stb}} 
The literature has traditionally used Weak Truth-Telling (WTT) to infer student preferences \citep[e.g.,][]{abdulkadiroglu/agarwal/pathak:15,laverde2022distance}.  WTT involves two assumptions: (a) the observed number of choices ranked in any ROL is exogenous to student preferences and (b) every student ranks her top preferred schools according to her preferences and may drop some least preferred schools.  Let $\pref{\text{WTT}}$ be the preferences of a student inferred by WTT.   
In \Cref{ex:TE}, for the student with $R=c_4$-$c_3$-$c_2$-$c_1$, WTT infers $u_{c_4} > u_{c_3} > u_{c_2} > u_{c_1} > u_{c_0}, u_{c_5}$, or $\pref{\text{WTT}}=\{(c_4,c_3), (c_4,c_2), (c_4,c_1), (c_4,c_0), (c_4,c_5), (c_3,c_2), (c_3,c_1), (c_3,c_0), (c_3,c_5), (c_2,c_1), (c_2,c_0),$\\$(c_2,c_5),(c_1,c_0), (c_1,c_5) \}$. Crucially, unranked schools, $c_0$ and $c_5$ in the example are inferred as less preferable than all ranked schools in $R$.

WTT would be justified by the strategyproofness if applicants  make no mistakes.  In the presence of mistakes, however, WTT may lead to biased estimates of preferences, as have been argued in \cite{Fack-Grenet-He(2015)} and \cite{ACHmistakes}. 
 \teps{\tau}, for $\tau\in[0,100]$, address the problem by offering a flexible menu of methods to identify preferences based on ROLs data in the face of uncertainties and (possible) mistakes.  It is important to note that TEPS does not necessarily contradict WTT; rather, it only focuses on the more robustly-inferred information from ROLs than WTT.  As one can see, WTT and \teps{\tau} with differing values of $\tau$'s form nesting relationships.

\begin{prop}\label{prop:nest}
Fix any student with a ROL $R$ and intrinsic priority $\ip$.  We have $\pref{top}
\subseteq
\pref{\tau} 
\subseteq
\pref{\tau'} 
\subseteq
\pref{all} 
=
\pref{\text{ST}}(\uncset) 
\subseteq
\pref{\text{WTT}}$ 
for any $0<\tau < \tau'<100$. 
\end{prop}

The existing literature, motivated by  mistakes mentioned earlier, has proposed  stability as an alternative to WTT for robustly inferring preferences from ROL data (\cite{Fack-Grenet-He(2015)}, \cite{ACHmistakes}) when there is no uncertainty in priorities. 
One can view the current method as a natural adaptation of the stability-based method to the environment involving uncertainties.  Evidently, the adaptation is nontrivial  both in terms of theoretical justification and empirical operationalization.  

To illustrate, suppose one adapts the stability restriction rather mechanically by applying it against single ex-post cutoffs obtained from an arbitrary lottery draw.  At first glance, this method seems analogous to  TEPS$^{top}$.  However, unlike TEPS$^{top}$, this method is problematic at least for two reasons.  First, 
this ad-hoc adaptation of stability is not well justified by \Cref{thm:stability}.  The ex-post cutoffs obtained from a single lottery draw could very well be an outlier that is a very low-probability event from the perspective of the applicant.  Hence, there is no guarantee that the preference ``inferred'' by this method reflects the true preference of the applicant.  
Second, in cases using a tie-breaking lottery, the realized lottery may not be observed, and the researcher may need to draw a lottery and simulate an observed outcome. 
 Since it is impossible to verify how many lottery draws the analysts have made, the method is not immune from an analyst  cherry-picking a lottery realization to suit his/her needs.  By contrast, TEPS$^{top}$ is both well-founded from theoretical justification (\Cref{thm:stability}) and is ex-post verifiable.

\subsection{Identification and Estimation}

The assumption of stability and transitivity, which provide the basis of \teps{}, does not uniquely pin down  a student's best response or the ROL she submits.  In fact, there are typically multiple best responses for a student given others' strategies. 
One may thus worry that this multiplicity may render our model incomplete 
in the sense of \cite{Tamer2003incomplete}. Incompleteness in our context would mean that the mapping from a student's type, ($u,t$), to \teps{} inferred preferences is a correspondence, which may cause the point identification of the distribution of $u$ to fail.  

These concerns do not apply here.  We provide some intuition about the completeness, which is formally established in Online Appendix~\ref{appendix:completeness}.  Consider a single student of type ($u,t$).  \teps{} assumes that she always receives her stable assignment in every realized uncertainty.  Integrating over all uncertainties, we obtain a distribution of her possible assignment outcomes that is unique in a large enough economy.  Because there are typically multiple ROLs that lead to the same outcome distribution, the mapping from ($u,t$) to best responses is indeed a correspondence.  Yet, \teps{} will infer the same set of preferences 
no matter which best-responding ROL is observed, because the \teps{} procedure only depends on the outcome distribution of the student.  Hence, the mapping from ($u,t$) to \teps{} inferred preferences is a function, implying that the  \teps{}  hypothesis is complete.

Once students' preferences are inferred, one may employ a parametric function of utilities to fit them with  parameter estimation.  As is popular in the literature, one may adopt a random utility model based on Extreme Value Type-1 (EVT1) or normal errors and estimate the parameters using maximum likelihood (MLE), Markov Chain Monte Carlo (MCMC) methods such as Gibbs sampler, or Expectation-Maximization (EM) algorithm.  We recommend adopting a multinomial probit model (either with or without random coefficients) and estimating it via Gibbs sampler due to its flexibility in drawing (random) cardinal utilities that are consistent with the inferred preferences.  For example, when the student's inferred preference is $c_1\succ c_2$ and $c_1\succ c_3\succ c_4$, it is not straightforward to write down the exact likelihood function even with an EVT1 error.  However, the inferred preferences provide restrictions (or bounds) on each school's utility, which can be easily incorporated into Gibbs sampler.\footnote{In the example, assuming no outside options, the bounds are given by $u_1\in(\max\{c_2,c_3\},\infty),u_2\in(-\infty,u_1),u_3\in(u_4,u_1),u_4\in(-\infty,u_3)$. For each student, one iteration of Gibbs sampler draws the utility of each school sequentially with truncation imposed by the bounds.  The procedure is given in the Online Appendix~\ref{appendix:MCMC}.}

\subsection{Selecting from among Alternative Preference Hypotheses \label{sec:tests}} 

\Cref{prop:nest} suggests a nested family of alternative preference inference hypotheses.  Each hypothesis in the spectrum offers a different way of trading off informational efficiency and robustness of inferences.  Which method is best depends on the extent of mistakes being made and their payoff significance.  Since the latter issue is empirical in nature, it would be ideal for the selection to be driven by the data.  

Specifically, we develop a sequential Wald test method, justified by the nesting structure in \Cref{prop:nest}. To this end, we assume that the preference estimates based on TEPS$^{top}$, the most robust inference, are consistent.  Using this as the alternative hypothesis, we test sequentially the consistency of more informationally demanding models, starting with WTT, and then TEPS$^{\tau}$ in the descending order of $\tau$'s.  

For concreteness, suppose we wish to estimate a parameter vector $\beta$ in a parametric model, say. 
Each method leads to an estimator for $\beta$.  We let them be the WTT-based estimator ($\widehat{\beta}^{\text{WTT}}$) and the TEPS$^{\tau}$-based estimator ($\widehat{\beta}^{\tau}$). As an example, let us consider TEPS$^{top}$ and 10 other possible values for $\tau$, $\tau_1=10,\tau_2=20,\ldots,\tau_9=90,\tau_{10}=100$. Note that \teps{\tau_{10}}=\teps{all}.
The following procedure selects the empirical method which is consistent and efficient among all consistent estimators that we consider due to the nesting structure in \Cref{prop:nest}.\footnote{One drawback of the testing procedure is that it relies on the parametric model and may suffer from model misspecification.  In that case, if one is willing to consider nonparametric models, the literature on nonparametric tests of nesting models may provide more general tests.}

\hangindent=1em
\hangafter=1
\noindent\textbf{Step 1: Test TEPS$^{top}$ vs.\ WTT}~~~We formulate two hypotheses.  $H_0^{\text{WTT}}$: $\widehat{\beta}^{\text{WTT}}$ and $\widehat{\beta}^{top}$ are both consistent, while $\widehat{\beta}^{\text{WTT}}$ is efficient; 
		$H_1^{\text{WTT}}$: only $\widehat{\beta}^{top}$ is consistent. Since WTT leads to an efficient estimator under the null, we conduct a Wald test based on the statistic,
		$$
		\left(\widehat{\beta}^{top} - \widehat{\beta}^{\text{WTT}}\right)' \left(V(\widehat{\beta}^{top})-V(\widehat{\beta}^{\text{WTT}})\right)^{-1} \left(\widehat{\beta}^{top} - \widehat{\beta}^{\text{WTT}}\right)
		$$
		where $V(\cdot)$ denotes the covariance matrix of its argument. 
		Under the null, the test statistic follows a $\chi^2_{|\beta|}$ distribution.  Justified by the nesting structure (\Cref{prop:nest}) and the maintained assumption on TEPS$^{top}$, if  $H_0^{\text{WTT}}$ is not rejected, we select $\widehat{\beta}^{\text{WTT}}$ and exit the procedure; otherwise, we perform the next test. 


\hangindent=1em
\hangafter=1
\noindent\textbf{Step 2: Test TEPS$^{top}$ vs.\ TEPS$^{\tau}$}~~~ We start with the largest $\tau$, $\tau_{10}=100$. $H_0^{\tau}$: $\widehat{\beta}^{\text{top}}$ and $\widehat{\beta}^{\tau}$ are both consistent, while $\widehat{\beta}^{\tau}$ is efficient; 
		$H_1^{\tau}$: only $\widehat{\beta}^{\text{top}}$ is consistent. We conduct a Wald test based on the statistic,
		$$
		\left(\widehat{\beta}^{\text{top}}-\widehat{\beta}^{\tau}\right)' \left(V(\widehat{\beta}^{top})-V(\widehat{\beta}^{\tau})\right)^{-1} \left(\widehat{\beta}^{\text{top}}-\widehat{\beta}^{\tau}\right).
		$$
		Under the null, the test statistic follows a $\chi^2_{|\beta|}$ distribution.  If $H_0^{\tau}$ is not rejected, we select $\widehat{\beta}^{\tau}$; otherwise, we proceed to test the next largest $\tau$, $\tau_9$. We continue until we reach a value of $\tau$ such that $H_0^{\tau}$ is not rejected, in which case we select $\widehat{\beta}^{\tau}$, or we reject $H_0^{\tau_1}$, in which case we select  $\widehat{\beta}^{top}$.

\vskip 0.2cm

\noindent The order of these steps guarantees the desired size of the tests and makes it appropriate to use the Hausman-type test statistics.

Besides formal statistical tests, some descriptive analyses can be informative as well.  For example, we may compare if there are any systematic differences among different groups of schools, e.g., schools ranked in a ROL, ever-feasible unranked schools, and never-feasible unranked schools.  WTT implies that there should not be any difference between ever-feasible unranked schools and never-feasible unranked schools, while both groups are worse than all ranked schools.  In contrast, TEPS distinguishes unranked schools by their feasibility status. 

\subsection{Performance of TEPS: Monte Carlo Simulations}
 
We analyze the performance of TEPS via Monte Carlo simulations; see Online Appendix~\ref{appendix:MC} for details.  Specifically, we consider 100 copies of finite economies in which 1,000 students apply and are assigned  12 schools through a student-proposing DA algorithm with coarse priorities and a single tie-breaking rule, similar to the NYC public school choice.  Each copy contains independent-randomly generated  priorities and preferences that follow a parametric random utility model. We consider three data generating processes (DGP); (i) \textsf{Truth-telling (TT)} in which each student submits a truthful ROL; (ii) \textsf{Payoff Irrelevant Mistakes (MIS-IRR)} in which students may skip or flip (i.e., arbitrarily rank) out-of-reach schools, and (iii) \textsf{Payoff Relevant Mistakes (MIS-REL)} in which  students may additionally skip schools with small but positive admission chances.\footnote{To be specific, we allow students to skip schools with admission probabilities lower than 10\%.} 
Table~\ref{table:skipsMC} summarizes the scenarios under each DGP. The fraction of students who make mistakes increases from 0\% in \textsf{TT} to 74.4\% in \textsf{MIS-IRR} and \textsf{MIS-REL}. As a result, the fraction of students reporting preferences consistent with WTT is only 27.0\% in \textsf{MIS-IRR} and 28.8\% in \textsf{MIS-REL}. Also, note that stability is satisfied for all students in \textsf{TT} and \textsf{MIS-IRR}, but not in \textsf{MIS-REL} since some students skip schools that are not completely out-of-reach for them.

\begin{table}[h]
	\centering
	\caption{Mistakes in Monte Carlo Simulations (\%)\label{table:skipsMC}}
	\resizebox{1.0\textwidth}{!}{
		\begin{tabular}{rccccc}
			\toprule
			\multicolumn{1}{l}{}                                                  & \multicolumn{5}{c}{Scenarios: DGP w/ Different Student Strategies}                                                                                  \\ \cline{2-6}
			\multicolumn{1}{l}{}                                                  & Truth-Telling  &  & \multicolumn{1}{c}{Payoff Irrelevant Mistakes} && \multicolumn{1}{c}{Payoff Relevant Mistakes} \\ \cline{2-2} \cline{4-4} \cline{6-6}
			\multicolumn{1}{l}{}                                                  & TT                                                                   &  &  MIS-IRR & & MIS-REL                                   \\ \midrule
            \multicolumn{1}{l}{Average length of submitted ROLs} & 12                                                                   &  & 6.1  & & 5.0                                 \\
			\multicolumn{1}{l}{WTT: \textit{Weak-Truth-Telling}} & 100                                                                   &  & 27.0  & & 28.8                                 \\
			\multicolumn{1}{l}{Matched w/ favorite feasible school}              & 100                                                                   &  &  100 & & 96.2                                    \\
			\multicolumn{1}{l}{Make Mistakes}                                          & 0                                                                     &  &  74.4    & & 74.4                              \\
			\bottomrule
	\end{tabular}}
    \begin{tabnotes}
        Each entry reported is a percentage that is averaged over the 100 estimation samples. A student is WTT if 1) ROL is in the true preference order, and 2) all ranked schools are more preferred to all unranked schools.
    \end{tabnotes}
\end{table}

For each of 300 scenarios (3 DGPs $\times$ 100 copies), we estimate the preference parameters using Gibbs sampler, based on WTT and \teps{} with attention parameters $\tau=0,20,40,60,80,100$. After the estimation, we perform the testing procedure in Section~\ref{sec:tests} to choose the optimal estimate and call it the ``selected'' estimate. 
\begin{figure}[h]
	\centering
	\begin{subfigure}{1\textwidth}
		\caption{Distribution of Estimates}
            \centering
		\includegraphics[width=0.33\textwidth]{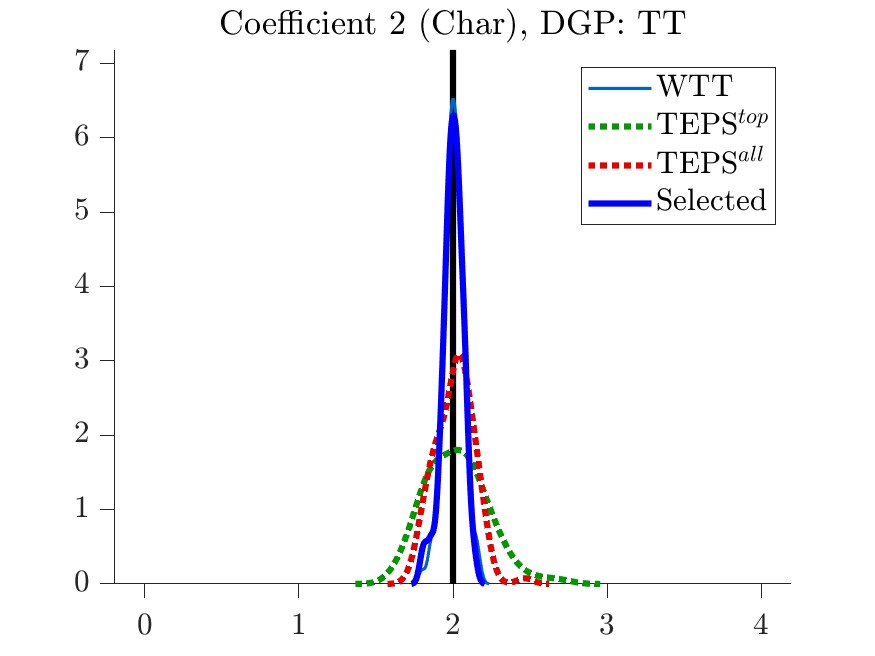}~
		\includegraphics[width=0.33\textwidth]{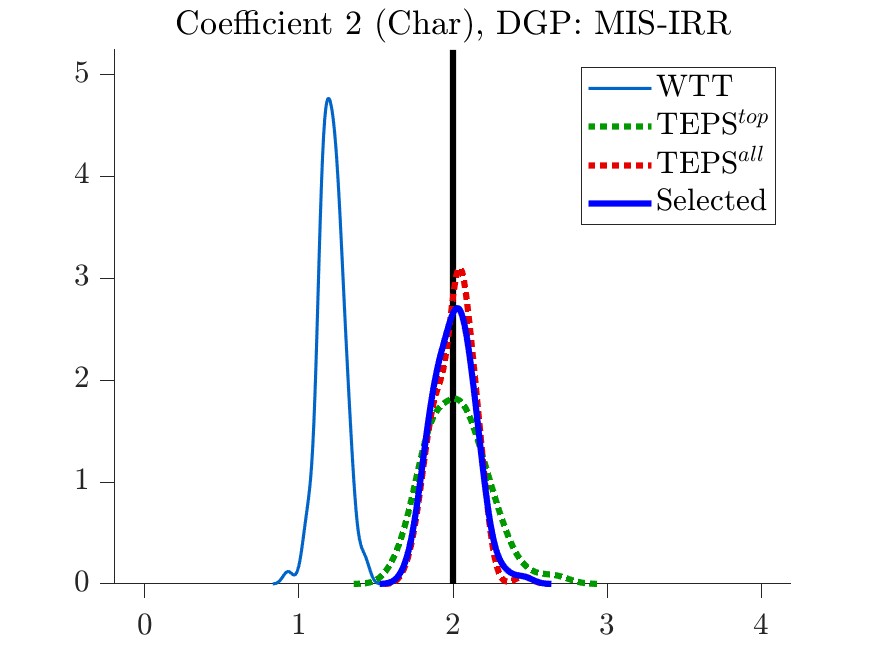}~
		\includegraphics[width=0.33\textwidth]{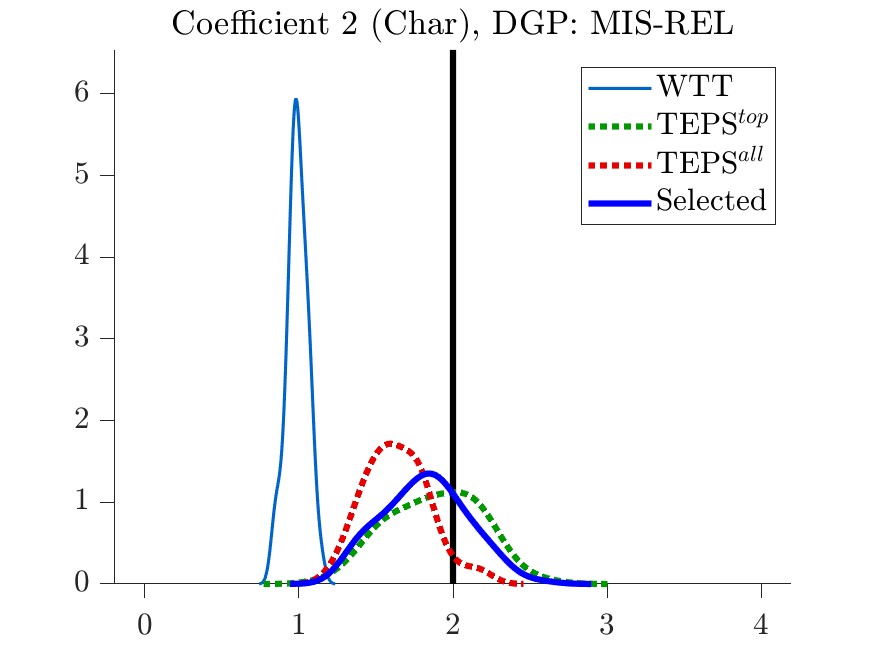}
	\end{subfigure}\\
	\begin{subfigure}{0.8\textwidth}\footnotesize
		\caption{Test Results: Fraction of Each Method Being Selected}
            \centering
		\begin{tabular}{crcccc}
		\toprule
		\multicolumn{2}{r}{{Data Generating Process: }}& & \textsf{TT}   & \textsf{MIS-IRR} & \textsf{MIS-REL}  \\ \midrule
		%
		
		
		\multirow{7}{*}{Estimation method} & &WTT                  			& \textbf{0.94} & 0 & 0   \\
        &&\teps{top}& 0.01 & 0 & 0\\
        &&\teps{20}& 0& 0 & 0\\
        &&\teps{40}& 0& 0 & 0.01\\
        &&\teps{60}& 0& 0.01 & 0.14\\
        &&\teps{80}& 0.03& 0.12 & \textbf{0.68}\\
        &&\teps{all}& 0.01&\textbf{0.87} & 0.17\\
		\bottomrule
	\end{tabular}
	\end{subfigure}
	\caption{Monte Carlo Simulations: Performance of TEPS and WTT\label{fig:mc_maintext}}
    \begin{tabnotes}
        See Online Appendix~\ref{appendix:MC} for the exact description of our Monte Carlo simulation.  In panel (a), we plot the kernel density plots of the estimates of $\beta_2$ from 100 Monte Carlo samples.  The left subfigure corresponds to \textsf{TT} DGP, the center subfigure corresponds to \textsf{MIS-IRR} DGP, and the right subfigure corresponds to \textsf{MIS-REL} DGP.  The black vertical line at 2 denotes the true value of the parameter.  We depict only WTT, \teps{top}, \teps{all}, and the Selected estimates for conciseness.  In panel (b), we report the fraction of each estimate being chosen as the `Selected' estimate among 100 Monte Carlo samples, according to the testing procedure in Section~\ref{sec:tests}.
    \end{tabnotes}
\end{figure}

The results are summarized in Figure~\ref{fig:mc_maintext}.  First, when the DGP is \textsf{TT}, all estimates are consistent, while the WTT-based estimator has the smallest variance since it uses the maximal, albeit possibly unreliable,  information that one can infer from observed ROLs. Second, when some students make payoff-irrelevant or -relevant mistakes (\textsf{MIS-IRR}, \textsf{MIS-REL}), the WTT-based estimator is susceptible to strategic mistakes with significant bias.  On the other hand, the estimators based on TEPS are robust to payoff-irrelevant mistakes.  Third, when some students make payoff-relevant mistakes (\textsf{MIS-REL}), \teps{all} is no longer consistent because \teps{all} does not take students' payoff-relevant mistakes into consideration.  On the other hand, \teps{top} remains consistent, so do \teps{\tau} with $\tau<100$ (see Online Appendix~\ref{appendix:MC}.)

Due to these bias-variance tradeoffs, our testing procedure selects WTT-based estimates 94\% out of 100 samples when the DGP is \textsf{TT}, but never selects WTT-based estimates when the DGP is \textsf{MIS-IRR} or \textsf{MIS-REL}.  Next, when students make no payoff-relevant mistakes (\textsf{MIS-IRR}), the procedure selects \teps{all} 87\% out of 100 samples since it uses the maximal information among all TEPS-based methods which are robust to payoff-relevant mistakes (\Cref{prop:nest}).  Finally, when students make payoff-relevant mistakes (\textsf{MIS-REL}), the procedure selects \teps{80} 68\% out of 100 samples since \teps{all} may no longer be robust to payoff-relevant mistakes.

\section{High School Choice in Staten Island, New York City\label{sec:nyc}}
We now turn to the study of high school choice in NYC. There are two purposes for this study.  First, we deploy and operationalize our framework to actual data and use it as proof of concept.  Specifically, we will put alternative assumptions of applicant inattention/mistakes to estimate student preferences and let the aforementioned testing procedure select the appropriate assumption. 
Second, and no less importantly, we highlight the importance of choosing the right assumption to infer student preferences by comparing the predicted effects of several proposed desegregation policies based on alternative assumptions on students' mistakes.\footnote{There is a substantive interest among the public and policymakers in diversifying and desegregating student bodies of NYC schools. Recently, the NYC DOE proposed to remove screening based on test scores for admissions and remove geographic preferences based on students' residence \citep{shapiroNYT}. Some versions of these proposals were implemented and subsequently reversed \citep{ClossonNYT}.}  
We show that alternative assumptions lead to different predictions, and ignoring students'  
mistakes in estimation may likely lead to an underestimation of the predicted effects.

NYC public high school admissions use the student-proposing DA \citep{abdulkadiroglu/pathak/roth:05}. Students apply to up to 12 school programs, and each program in a school has its own capacity and independently conducts its admissions. 

We focus on students and programs in Staten Island (SI, hereafter) that participated in Round~1, the main round, of NYC public high school choice in the 2016--17 academic year.
SI is one of the five boroughs of NYC and can be considered as an independent matching market.\footnote{SI is connected with the rest of the city only by the Staten Island Ferry or the Verrazzano-Narrow Bridge. There are pros and cons of analyzing SI instead of the entire NYC. While the direct implications for policies in NYC may be limited, SI's racial profile and size are closer to the medium-size cities in the rest of the United States and hence, the implications are more relevant for other US cities.}

Column~(1) of \cref{table:summ_stu} presents the summary statistics of the students, and \cref{table:summ_sch} describes the programs and schools. Most of the students participate in Round~1 and enroll in their assigned school from that round. In total, we have 3,731 students and 50 programs at nine schools. The length limitation of the ROL is not binding in SI. The average length was 4.05 and only about 3\% of the students in our sample exhausted the list.   
Online Appendix~\ref{appendix:data} gives more details on the data sources and sample restrictions. Almost all middle school students who reside in SI are enrolled in middle schools in SI (about 95\%) and apply to high school programs only in SI (see panel~D of \cref{table:summ_stu}). SI is on average, richer and has more white students relative to NYC, with a median household income \$74,580 vs.\ \$55,191 and the proportion of white students 53\% vs.\ 15\%.

\begin{table}[h!]
	\def\sym#1{\ifmmode^{#1}\else\(^{#1}\)\fi}
	\centering 	\footnotesize
	\caption{Summary Statistics: Sample Means across Students\label{table:summ_stu}}
		\resizebox{1\textwidth}{!}{
		\begin{tabular}{r*{9}{c}}
		\toprule
        & (1) & (2) & (3) & (4) & (5) & (6) & (7) & (8) & (9)\\
		&       Total&       Cell 1&       Cell 2&       Cell 3&       Cell 4&       Cell 5&       Cell 6&       Cell 7&       Cell 8\\
		\midrule
		\multicolumn{10}{l}{\it Panel A. Cell characteristics} \\
		Female              &       0.49&       no&       yes &       no&       yes &       no&       yes &       no&       yes \\
		Free or reduced-price lunch (FRPL)    &       0.53&       no&       no&       yes &       yes &       no&       no&       yes &       yes \\
		Black/Hispanic      &       0.36&       no&      no&       no&       no&       yes &       yes &       yes &       yes \\
		[0.5em]
		\multicolumn{10}{l}{\it Panel B. Student characteristics} \\
		Asian               &       0.10&       0.10&       0.09&       0.25&       0.22&       0.00&       0.00&       0.00&       0.00\\
		White               &       0.53&       0.90&       0.90&       0.73&       0.75&       0.00&       0.00&       0.00&       0.00\\
		Black               &       0.12&       0.00&       0.00&       0.00&       0.00&       0.30&       0.27&       0.33&       0.34\\
		Hispanic            &       0.25&       0.00&       0.00&       0.00&       0.00&       0.70&       0.74&       0.67&       0.66\\
		[0.25em]
		Median Income (\$1,000)     &      74.58&      85.21&      86.09&      77.72&      78.42&      65.97&      66.56&      59.55&      57.32\\
		Grade 7 ELA             &     310.87&     317.38&     326.83&     306.23&     316.91&     299.04&     319.93&     289.67&     301.48\\
		Grade 7 Math            &     310.21&     326.72&     326.11&     313.36&     315.34&     296.88&     308.43&     285.54&     287.27\\
		[0.5em]
			\multicolumn{10}{l}{\it Panel C. Admission outcomes} \\
		Matched with top choice     &       0.57&       0.61&       0.60&       0.56&       0.55&       0.55&       0.54&       0.54&       0.56\\
		Unassigned&       0.00&       0.00&       0.00&       0.01&       0.00&       0.00&       0.01&       0.00&       0.00\\
%
%
		[0.5em]
			\multicolumn{10}{l}{\it Panel D. Submitted ROLs} \\
		\# of SI Choices    &       4.05	 & 	4.03	 & 	3.97	 & 	4.01	 & 	4.06	 & 	4.14	 & 	3.69	 & 	4.11	 & 	4.20
		\\
		\# of Choices (all) &       4.06	 & 	4.05	 & 	3.98	 & 	4.02	 & 	4.07	 & 	4.16	 & 	3.70	 & 	4.12	 & 	4.21
		\\
%
		\midrule
		Observations        &        3731&         710&         743&         471&         455&         178&         136&         545&         493\\
		\bottomrule
	\end{tabular}
	}
\end{table}

\begin{table}[h!]
	\centering
	\footnotesize
	\caption{Summary Statistics: Sample Means across Programs/Schools\label{table:summ_sch}}
	\resizebox{0.7\textwidth}{!}{%
		\begin{tabular}{rcccc}
			\toprule
			& \multicolumn{2}{c}{Program} & \multicolumn{2}{c}{School} \\
			& Mean        & Std. Dev.           & Mean        & Std. Dev.          \\\midrule
			Capacity & 134.06 & (215.47) & 744.78 & (551.49)\\
			9th Grade Size           & 71.30       & (108.79)      & 396.78      & (293.48)     \\
			\% High Performer (ELA)  & 31.08       & (26.82)       & 34.12       & (15.01)      \\
			\% High Performer (Math) & 26.21       & (25.46)       & 25.83       & (12.19)      \\
			\% White                 & 36.32       & (22.14)       & 43.93       & (21.80)      \\
			\% Asian                 & 7.94        & (7.39)        & 7.61        & (3.40)       \\
			\% Black                 & 19.93       & (13.82)       & 17.51       & (12.79)      \\
			\% Hispanic              & 34.45       & (16.76)       & 29.57       & (12.72)      \\
			\% Free or Reduced-Price Lunch (FRPL)    & 63.40       & (18.08)       & 60.50       & (13.65)      \\
			$\mathds{1}$(STEM)                  & 0.20        & (0.40)        &             &              \\\midrule
			Observations             & \multicolumn{2}{c}{50}   & \multicolumn{2}{c}{9}      \\
			\bottomrule
		\end{tabular}%
	}
\begin{tabnotes}
    Programs that are `For Continuing 8th Graders' and zoned do not have capacity restrictions. We count the total number of students who are eligible for those programs and report that as their capacity.
\end{tabnotes}
\end{table}
	
\subsection{Preference Estimation}
To represent student preferences over programs, we adopt a random utility model.\footnote{We do not model outside options. Almost all students in our sample (3,724 out of 3,731) were assigned in the data (see \cref{table:summ_stu}), and less than 2\% of them chose not to enroll in their assigned school.} We divide the sample into cells and allow the parameters to vary freely across cells \`a la \cite{abdulkadiroglu2017parents}. Specifically, we construct eight (mutually exclusive and exhaustive) cells based on three binary student characteristics---gender, free or reduced-price lunch (FRPL) status, and Black/Hispanic. Columns~(2)--(9) of \cref{table:summ_stu} present the summary statistics of the students in each cell.

Let $u_{icps}$ denote the utility of student~$i$ in cell~$c$ when she is matched with program~$p$ at school~$s$. We parameterize the utility as follows:
\begin{align}
u_{icps} = \alpha_{cs}  + \beta_{1,c} D_{is} + \beta_{2,c} \textit{Nearest}_{is} + \beta_{3,c} z_p + \sum_l \gamma_{c}^l x_i^lZ_s^l + \sum_k \delta_{c}^k x_i^kz_p^k + \sigma_{\tau(p),c}\epsilon_{icps},\label{eqn:utility_SI}
\end{align}
where $\alpha_{cs}$'s are cell-specific school fixed effects;  $D_{is}$ is the distance from student $i$'s residence to school $s$;\footnote{We use the exact address for schools. For students, we use the centroid of the census tract that a student resides in as her residence. We calculate the line distance between the two using the Haversine formula.} $\textit{Nearest}_{is}$ is a binary variable which equals 1 if $D_{is}$ is the smallest for $i$ among all schools and 0 otherwise; $x_i$ is a vector of student characteristics to be interacted with program characteristics $z_p$ and school characteristics $Z_s$. Specifically, $x_i$ includes 7th grade standardized ELA and Math scores, their average, and median neighborhood income. $z_p$ and $Z_s$   
 include the proportions of high performers in 7th grade standardized ELA and Math (a high performer is defined as being above the 75th percentile), the size of the 9th grade, the proportion of each race/ethnicity group, the proportion of students on FRPL (calculated at program and school levels, respectively), and an indicator as to whether program $p$'s focus is in STEM fields. $\epsilon_{icps}$ is i.i.d.\ standard normal conditional on the above observables. $\sigma_{\tau(p),c}$ allows for heterogeneous variances of the unobserved idiosyncratic preference shocks based on program type $\tau(p)$ (STEM, or others), and we normalize $\sigma^2_{others,c}=1,\forall c$. 

We estimate equation~(\ref{eqn:utility_SI}) separately for the eight cells.  
In terms of observables, Table~\ref{table:summ_stu} shows that there is substantial heterogeneity across the cells. For example, the median neighborhood income of cell~1, which consists of male, non-FRPL, non-Black/Hispanic students, is \$85,207, much higher than \$57,321 for cell~8, consisting of female, FRPL, Black/Hispanic students. For each cell, we employ a Gibbs sampling procedure for the Hierarchical Bayesian estimation, which is described in detail in Online Appendix~\ref{appendix:MCMC}.

\subsubsection{Comparing Assumptions for Inferring Student Preferences \label{sec:compare}}
The full preference estimates are reported in Online Appendix~\ref{appendix:preferenceestimates}. 
For TEPS-based estimators, we consider \teps{top}, \teps{10}, \teps{20}, $\cdots$, \teps{90}, and \teps{all}.  Below, we present the test results (Section~\ref{sec:tests}) and descriptive statistics that are consistent with the test results. We then evaluate how well each set of estimates fits the data.

\paragraph{Test Results: Selected Estimates.}
The assumptions on student behavior \emph{selected} by our testing procedure in Section~\ref{sec:tests} are reported in Table~\ref{table:selected}. 
It shows that WTT is never selected, implying that students in our data across all the cells tend to make some, possibly payoff-insignificant, mistakes. Furthermore, the selected version ranges from \teps{20} to \teps{all} across the cells, indicating that there is some heterogeneity in the degree to which such mistakes are made in terms of students' observable characteristics.

\begin{table}[ht!]\centering\footnotesize
\caption{Test Results: Selected Estimates\label{table:selected}}
\begin{tabular}{lcccccccc}
\toprule
Cells              & 1         & 2          & 3         & 4         & 5         & 6         & 7         & 8         \\\midrule
Selected Estimates & \teps{70} & \teps{all} & \teps{40} & \teps{20} & \teps{70} & \teps{80} & \teps{50} & \teps{20}\\
\bottomrule
\end{tabular}%
\end{table}

\paragraph{Descriptive Analysis.}

\begin{figure}[h!]
	\centering
	\includegraphics[width=0.7\textwidth]{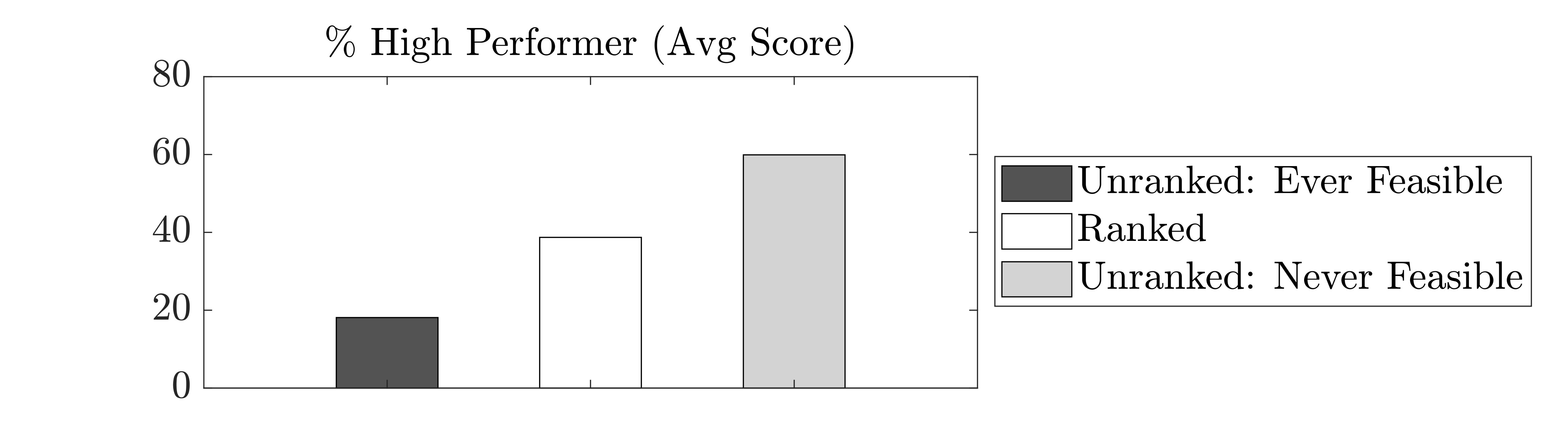}
	\vskip -0.3cm
	\caption{Characteristics of Ranked and Unranked Programs by Feasibility Status\label{fig:progcharbyfeasibility}}
	\begin{tabnotes}
		For each student, we classify the programs into three types---ever-feasible-unranked,  never-feasible-unranked, and ranked. Ranked programs are those included in the student's ROL, and Step 1 of \teps{all} procedure  determines the feasibility of each unranked program. We focus on the fraction of high-performing students (measured by the average of ELA and math scores) in a program.  In Online Appendix~\ref{app:add}, we report results on additional program characteristics.
        The figure reports the average across all students for each type of program. 
	\end{tabnotes}
\end{figure}

Recall \teps{} distinguishes unranked programs (in a student's ROL) depending on whether they are feasible or infeasible, whereas WTT makes no such distinction; i.e., it treats all unranked schools as if they are less preferred than ranked schools.  
To see which treatment makes more sense, it is instructive to inspect  whether the characteristics of programs vary systematically based on their feasibility status. Figure~\ref{fig:progcharbyfeasibility} reports the program's percentage of high-performing students in terms of the average score\footnote{A student is said to be a high performer if her average of ELA and math scores is above the 75th percentile.} by their feasibility status.

There is a clear order among the three types of programs:  ever-feasible unranked programs have the smallest percentages of high performers,  never-feasible unranked programs have the largest percentages, and ranked programs lie in the middle. The fact the ever-feasible unranked programs have a smaller percentage than the never-feasible unranked programs is likely mechanically driven  by the definition of feasibility, but the fact that the former has a higher percentage than ranked programs is noteworthy and is consistent with the selection of \teps{}, as opposed to WTT, in all cells.\footnote{We obtain the same conclusion when we look at other program characteristics such as the percentages of high performers in ELA and math separately and the proportion of non-FRPL students (see Online Appendix~\ref{app:add}.)}

\paragraph{Model Fit.}
To see how  alternative estimates based on the different assumptions fit the data, we predict the matchings and calculate the average characteristics of the assigned programs. Specifically, we use each set of estimates to simulate students' ordinal preferences, let them report truthfully to DA, and obtain a matching for each lottery draw. \cref{cor} validates the assumption that students are truth-telling in the analysis of matching \emph{outcomes}. 
For each program characteristic,  we calculate the average over 40,000 simulations for each set of estimates. \Cref{tab:modelfit} reports how close the average is to the data. Note that we use the estimates from all eight cells together for these exercises.

\begin{table}[h!]
\centering
	\caption{Model Fit: Average Characteristics of Assigned Programs}
	\label{tab:modelfit}
\resizebox{1\textwidth}{!}{%
\begin{tabular}{rccccccccccc}
\toprule
                         & \multicolumn{5}{c}{Black and Hispanic   Students}                          &  & \multicolumn{5}{c}{White, Asian, and   Other Students}                  \\
                         \cline{2-6} \cline{8-12}
                         & \multirow{2}{*}{Data} & \multicolumn{4}{c}{Difference b/t model prediction \& data} &  & \multirow{2}{*}{Data} & \multicolumn{4}{c}{Difference b/t model prediction \& data} \\
                         \cline{3-6} \cline{9-12}
                         &                       & WTT       & \teps{all}      & \teps{top}       & Selected    &  &                       & WTT       & \teps{all}      & \teps{top}       & Selected    \\\midrule
\% High Performer   (ELA) & 27.66  & -1.12  & 0.24   & 0.01   & 0.11   &  & 46.00  & -0.36  & -0.25  & -0.39  & -0.33  \\
                          &        & (0.46) & (0.49) & (0.51) & (0.52) &  &        & (0.26) & (0.24) & (0.27) & (0.28) \\
\% High Performer (Math)  & 20.66  & -0.76  & 0.30   & 0.09   & 0.21   &  & 39.68  & -0.56  & -0.49  & -0.44  & -0.51  \\
                          &        & (0.43) & (0.43) & (0.47) & (0.49) &  &        & (0.27) & (0.23) & (0.25) & (0.27) \\
\% FRPL (Program Level)   & 67.06  & 1.33   & 0.55   & 0.75   & 0.86   &  & 49.86  & 1.02   & 0.63   & 0.68   & 0.62   \\
                          &        & (0.31) & (0.36) & (0.43) & (0.40) &  &        & (0.23) & (0.17) & (0.21) & (0.19) \\
\% White                  & 34.19  & -2.58  & -0.88  & -1.27  & -1.30  &  & 58.41  & -1.06  & -0.58  & -0.89  & -0.65  \\
                          &        & (0.45) & (0.48) & (0.57) & (0.55) &  &        & (0.33) & (0.26) & (0.29) & (0.26) \\
\% Asian                  & 7.29   & 0.16   & 0.47   & 0.48   & 0.52   &  & 9.27   & -0.09  & -0.05  & -0.13  & -0.12  \\
                          &        & (0.15) & (0.17) & (0.17) & (0.16) &  &        & (0.08) & (0.09) & (0.09) & (0.09) \\
\% Black                  & 21.40  & 0.82   & -0.04  & 0.13   & 0.23   &  & 9.75   & 0.23   & 0.20   & 0.46   & 0.30   \\
                          &        & (0.29) & (0.32) & (0.36) & (0.34) &  &        & (0.16) & (0.17) & (0.18) & (0.16) \\
\% Hispanic               & 35.91  & 1.59   & 0.42   & 0.63   & 0.53   &  & 21.61  & 0.83   & 0.36   & 0.49   & 0.41   \\
                          &        & (0.33) & (0.34) & (0.39) & (0.38) &  &        & (0.22) & (0.17) & (0.19) & (0.18) \\
Size of 9th Grade         & 135.54 & -15.89 & -3.08  & 2.69   & 1.38   &  & 287.77 & -13.32 & -3.18  & -7.57  & -5.98  \\
                          &        & (4.50) & (4.75) & (5.16) & (5.02) &  &        & (3.96) & (2.70) & (2.93) & (2.79) \\
$\mathds{1}$(STEM)        & 15.67  & -0.89  & -1.89  & -2.81  & -1.73  &  & 14.56  & 0.77   & 0.84   & 1.13   & 0.84   \\
                          &        & (0.98) & (1.76) & (3.28) & (1.98) &  &        & (0.56) & (0.79) & (1.39) & (0.96)     \\\bottomrule
\end{tabular}}
\begin{flushleft}\scriptsize
\vskip -0.3cm
	Notes: We sample 200 draws from the posterior distribution of each parameter and, for each draw, draw 200 sets of lotteries and run DA $200\times200=40,000$ times. The mean and standard deviations across the preference estimates draws are reported. 
	\end{flushleft}
\end{table}
We report the model fit based on WTT,  \teps{all}, \teps{top}, and the selected {estimates} for conciseness. Two patterns are present. First, the WTT-based estimates do not fit the data well, while TEPS-based and the  selected estimates closely replicate the data. This reflects that WTT is vulnerable to students' payoff-insignificant mistakes and infers potentially unreliable information about their preferences. Second, {the predictions from \teps{top} tend to be less precise (i.e., have higher standard deviations) than those based on \teps{all} as well as the selected estimates (recall that \teps{\tau} with large enough attention parameter $\tau$ were selected in Table~\ref{table:selected}). This is because \teps{top} uses less information about student preferences than other TEPS-based estimates, resulting in less precise preference estimates (Tables~\ref{tab:SIpref1}-\ref{tab:SIpref4}).}

\subsection{Counterfactual Analysis}
We now proceed to evaluate the effects of desegregation policies in NYC based on alternative estimates. Motivated by the desegregation policies that were recently debated and partially implemented by the NYC DOE, 
we consider the following three counterfactual scenarios:\footnote{Similar policies are studied in \cite{Idoux_JMP} for middle schools and \cite{HahmPark2022} in a dynamic setting.} 

\begin{enumerate}[label = (\roman*)]
	\item \textsf{No screening}: no programs screen students based on their academic performances, {which means that a} lottery is the only tie-breaker to distinguish among applicants in the same priority group. 
	\item \textsf{No zoning}: no programs prioritize students based on their residence. This amounts to removing zone priorities at zoned programs in SI.
	\item \textsf{No priorities}: \textls[-25]{remove all priority rules so that admissions are based only on a single tie-breaking lottery.}
\end{enumerate}
We evaluate the desegregation effect of each scenario and compare it to the observed matching in the data given the current regime. To do so, for each set of estimates (WTT; \teps{\tau} with $\tau= top,10,\cdots,90,all$; or selected), we randomly select 200 draws of $u_{icps}$ for each $(i,c,p,s)$ from its respective posterior distribution and let each student submit a ROL representing her true ordinal preferences {per} \cref{cor}. We then draw 200 sets of random lotteries, construct students' priority scores under each scenario, and run DA to obtain counterfactual outcomes. We report the average of each measure we consider across 200$\times$200=40,000 simulation results for each scenario and each set of estimates. For conciseness, we only report the results from
{using the WTT estimates, \teps{top} estimates, and the selected estimates.}

\begin{figure}[h!]
	\centering
	\includegraphics[width=0.9\textwidth]{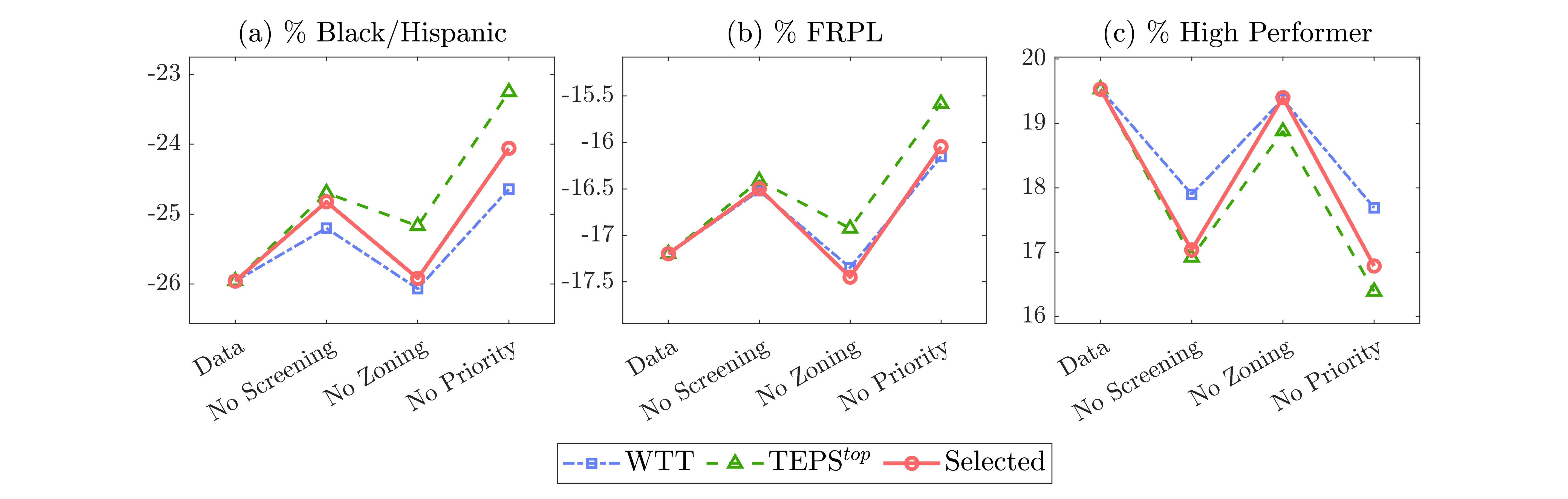}
	\caption{Racial Gap in Characteristics of Assigned Programs\label{fig:cf_chargap} }
\end{figure}

We evaluate the effects of the three policies on the racial gap between Black or Hispanic students and others. \cref{fig:cf_chargap} focuses on the average characteristics of the assigned program.\footnote{For each student in each racial group, we calculate three observable characteristics of the assigned program---proportion of Black and Hispanic students, proportion of FRPL students, and average 7th-grade standardized test score. We then calculate the mean across all students in each racial group and report it in \cref{fig:cf_chargap}.} The current regime produces a sizable racial gap. On average, a Black or Hispanic student is assigned to a program with 57\% Black or Hispanic students, while a student of other racial and ethnic groups is assigned to a program with only 31\% of Black or Hispanic, leading to a gap of $-26$ percentage points (panel~a). Similarly, significant gaps are present in terms of the proportion of FRPL students (panel~b) and high performers (panel~c). 

We find two prominent patterns in the predicted policy effects: 1) WTT-based estimates which neglect students' mistakes/inattention may underestimate the policy effects, and 2) 
the overall predicted effects are modest which suggests that forces other than priority rules may be more important in desegregating schools.

First, notably, the effects of the policies are  underestimated if one uses the WTT-based estimates for the counterfactual analysis.
  For example, WTT would predict that the decrease in the racial gap in the percentage of Black or Hispanic in the assigned program is only half (about 1 percentage point) compared to what the selected estimates predict.  This underestimation is expected. WTT neglects students' mistakes/inattention and assumes that students do not prefer unranked out-of-reach schools, so they would not apply to them even when these schools become within reach.

Second, considering the drastic nature of the contemplated policies, one may view the predicted effects, even under our selected estimates, to be rather small. While our selected estimates show that the no-priority policy reduces the racial gaps most significantly, the magnitude is modest.\footnote{For example, the racial gap in the percentage of Black or Hispanic in the assigned program is reduced by  2 percentage points. Dropping screening alone narrows the gap by only 1 percentage point, and dropping zoning alone reduces the gap only by a negligible amount.} This may be because there are other forces that may play an important role in school segregation, which raises a question on the effectiveness of removing priorities as a tool for achieving desegregation. 

\begin{figure}[h!]
    \centering
    \includegraphics[width=0.6\textwidth]{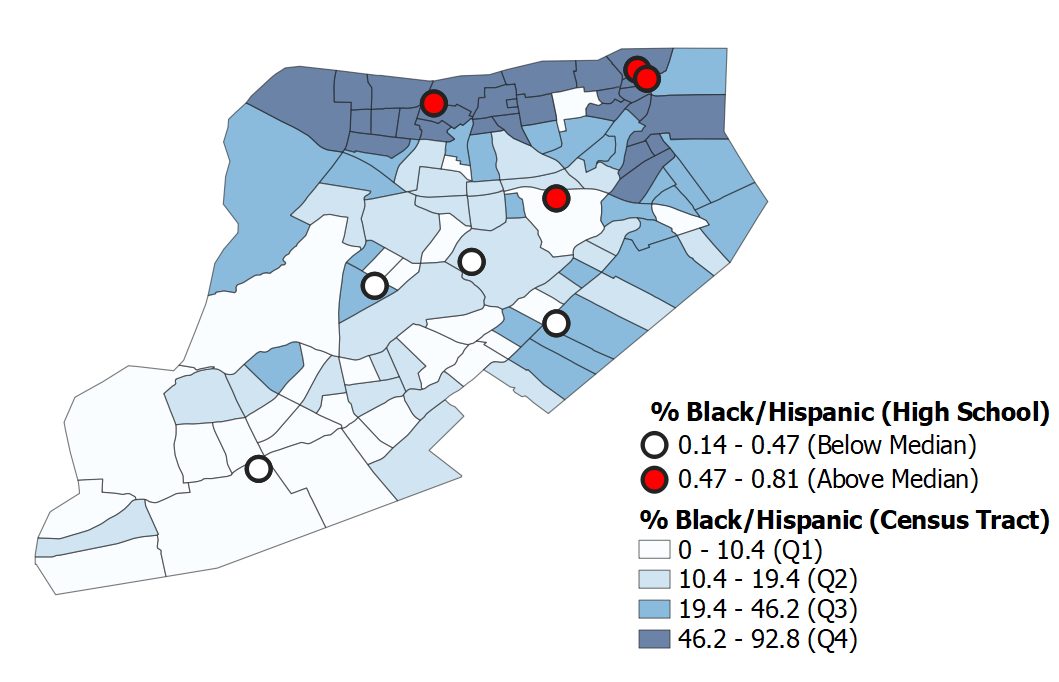}
    \caption{Percentage of Black or Hispanic: Census Tracts and High Schools\label{fig:SI_census}}
\begin{flushleft}\scriptsize
\vskip -0.3cm
	Source: 2017 American Community Survey: 5-Year Data, US Census TIGER/Line Shapefiles. 
	\end{flushleft}
\end{figure}

We provide descriptive evidence for one such explanation: residential segregation \citep{monarrez2020school,laverde2022distance,ParkHahm2022}. Figure~\ref{fig:SI_census} shows that SI has a high degree of residential segregation. Black and Hispanic students are mostly concentrated in the northern part of SI, which contains schools with more Black/Hispanic, FRPL, and low-performing students. Meanwhile, our preference estimates show that while students in general have a distaste for commuting, minority students have a particularly strong distaste for commuting. For instance, the selected estimates according to our testing procedure show that the willingness to travel for a 10 pp increase in the proportion of high performers in math is 0.45 miles for Cell 1 (male/non-FRPL/White or Asian) students, but is only 0.10 miles for Cell 7 (male/FRPL/Black or Hispanic) students (see Appendix Tables~\ref{tab:SIpref1}-\ref{tab:SIpref4}).\footnote{The average commuting distance of SI students is 2.3 miles. The willingness to travel for school/program characteristics $X$ can be calculated by dividing the coefficient on $X$---the marginal utility of $X$---by the negative of the coefficient on commuting distance---the marginal utility of traveling---in the estimated utility function.} Residential segregation, combined with such disparate distastes for commuting, means that school choice reforms equalizing school access alone may not significantly affect the enrollment outcome and thus may not significantly reduce the segregation of the student body, unless the residential segregation is also addressed. 
Our finding calls for a more comprehensive approach that studies school choice together with a family's residential choice;  see \cite{agostinelli2021spatial} and \cite{ParkHahm2022} that make progress in this direction.

\section{Conclusion}

This paper explores how we can leverage uncertainties faced by applicants to infer their preferences when they make payoff-insignificant mistakes. We study a general DA matching market in which students may face uncertainties about their priorities, and consider a robust equilibrium in which no applicant makes payoff-significant mistakes. We show that a robust equilibrium is asymptotically stable---the proportion of students assigned their favorite feasible schools converges to one in probability. Based on the theoretical finding, we develop a novel, three-step procedure called the \emph{Transitive Extension of Preferences from Stability} that exploits the structure of the uncertainties present
in a matching environment to robustly infer students' preferences. We show that it infers the maximal information on students' preferences justified by stability and transitivity of preferences. We then apply our method to estimate students' preferences using school choice data from Staten Island, NYC, and evaluate the  effects of several desegregation policies. The empirical results illustrate that the truth-telling assumption can lead to biased estimates and policy implications when students make payoff-insignificant mistakes, while our method is robust to such mistakes. 

 \bibliographystyle{economet}
\bibliography{bibmatching}

\begin{appendix}

\setcounter{footnote}{0}
\renewcommand{\thefootnote}{A-\arabic{footnote}}

\setcounter{equation}{0}
\renewcommand{\theequation}{\Alph{section}.\arabic{equation}}
\setcounter{table}{0}
\renewcommand{\thetable}{\Alph{section}.\arabic{table}}
\setcounter{figure}{0}
\renewcommand{\thefigure}{\Alph{section}.\arabic{figure}}
\setcounter{prop}{0}
\renewcommand{\theprop}{\Alph{section}.\arabic{prop}}

\section{Proofs from Sections~\ref{sec:theory} and \ref{sec:method} \label{appendix:proof_theory}}
\appcaption{Appendix \ref{appendix:proof_theory}: Proofs from Section~\ref{sec:theory}}

We first introduce a general priority structure that nests the priority structure assumed in the main text as a special case. We then establish asymptotic stability using that general priority structure in \Cref{thm:stability-general} below, which will in turn imply \Cref{thm:stability}.

\subsection{A General Priority Structure}

Here, we consider a more general priority structure than those consisting of (a), (b), and (c) assumed in the text.  Recall the ex-ante priority types $T$.  Let $T_d:=\{t\in T: \sl_c^t=\sh_c^t, \forall c\}$ be the types whose scores are all degenerate, and let $T_{n}:=T\setminus T_d$ denote types whose scores for some schools are non-degenerate.  We assume that $T$ is compact,  that $T_{n}$ is closed, and that there exists some $\kappa>0$ such that $\sh^t_c-\sl_c^t>\kappa$    if $\sh^t_c-\sl_c^t>0$.
 Further, for each $t\in T_n$,  $\Phi^t$ is absolutely continuous and admits density $\phi^t$ on $S^t$. As in the text, let $\widetilde \eta$ denote the probability measure of $\widetilde\t\in \widetilde{\T}:= [\ul, \uh]^C\times T$.   In summary, the continuum economy is summarized by $E=[\widetilde \eta, S, (\Phi^t)_t]$.  

In addition to Marginal Full Support, we introduce two additional conditions, which will be later shown to hold under the priority structure satisfying (b) and (c).  

\begin{as} [\textsc{Finite Atoms} \text{[}at Extremal ex-post Scores\text{]}]
The distribution of $(\sl_c^t,\sh_c^t)_t$, viewed as a random function of $t$, has at most a finite number of atoms.
\end{as}
 
This assumption can be seen as weakening the assumption of AL model.  First, for model (a), with no uncertainty on ex-post scores, we have $T=T_d$, with $\sl_c^t=\sh_c^t$ for all $t\in T$.  In this case, the atomlessness of ex-post measure $\eta$ (which is also assumed in AL) implies that the infimum and supremum scores have no point mass, so Finite Atoms will be trivially satisfied.  However, this condition does allow point mass for infimum and supremum scores but only finitely many of them; this is the case of coarse priorities in (b).  A mass of students belong to each of finitely many priorities, so there will be a finite number of atoms in the infima and suprema of their ex-post scores.  Last, in the priority structure (c), the infimum and supremum of $s^t_c$ are 0 and 1, respectively, so the finite atoms condition is clearly satisfied.

 For the last condition, for any $\delta\in (0,1)$, we say a school $c$ is {\bf $\delta$-feasible} for type $t$ given $p$ if $\sh^t_c-\delta>p_c$ and {\bf $\delta$-infeasible} for type $t$ given $p$ if $\sl^t_c+\delta<p_c$.   
 Plainly, $\delta$-feasibility and $\delta$-infeasibility mean feasibility and infeasibility, respectively, with probabilities that are bounded away from zero.

\begin{as} [\textsc{Rich Uncertainty}] \label{assm:richuncertainty}
 Fix any $p\in [0,1]^C$. Then, for any
		 $\delta<\bar\delta$, for some $\bar \delta>0$, there exists $\beta(\delta)>0$ such that, for any $t\in T$, for any $\delta$-feasible schools $a,b\in C$, and for any set $C^t \subset C\setminus \{a,b\}$ of $\delta$-infeasible schools,   \vspace{-0cm}
		\begin{equation} \label{eq:(iii)} \Pr\{s_c^t<p_c, \forall c\in C^t, 
		\mbox{ and }  s_a^t>p_a, s_b^t>p_b \}>\beta(\delta),\vspace{-0cm}
  \end{equation}
		whenever $\Pr\{s_c^t<p_c, \forall c\in C^t, 
	 s_a^t>p_a\}>0$ and $\Pr\{s_c^t<p_c, \forall c\in C^t, s_b^t>p_b \}>0.$\footnote{Rich Uncertainty is vacuously satisfied if schools with the specified restrictions do not exist. Also, if  $C^t=\emptyset$, the required condition reduces to  $\Pr\{s_a^t>p_a, s_b^t>p_b \}>\beta(\delta)$ whenever $\Pr\{s_a^t>p_a\}>0$ and $\Pr\{ s_b^t>p_b \}>0$. }
\end{as}

 The last condition, which we argue is quite weak, is most technical and thus requires unpacking.  
 In words, the condition states that whenever   school $a$ is feasible or $b$ is feasible, while all  $\delta$-infeasible schools in $C^t$ are infeasible, each with positive probability, then  $a$ and $b$ are {\it both simultaneously} feasible and $C^t$ are infeasible with  probability at least of $\beta(\delta)$, for some  $\beta(\delta)>0$.\footnote{ One non-obvious condition is the existence of a lower bound probability $\beta(\delta)$ that is independent of   $t$. The condition is still reasonable given the compactness of $T$ and $T_n$.}   Essentially, this rules out the case in which  a student's scores for $a$ and $b$ are extremely negatively correlated.  Hence, the condition is quite weak and is satisfied in all realistic environments in the main text.\footnote{See Online Appendix~\ref{app:lem-iii-proof} for the formal proof.}

\subsection{Proof of \Cref{prop:uniquestable}}
\label{sec:proof_uniquestable}
\begin{proof}  We first show that the following  condition, called {\it strict gross substitutes}, guarantees the uniqueness:
$\eta$ is such that for any $p, p'\in [0,1]^C$ with $p<p'$,\vspace{-0cm}
$$\sum_{c\in J} D_c(p')<\sum_{c\in J} D_c(p), \leqno{(SGS)}\vspace{-0cm}$$
where $J:=\{c\in C: p_c<p_c'\}$.  Suppose to the contrary there are two stable matchings characterized by two cutoffs $p$ and $p'$.  By the lattice property, we can assume without loss that  $p'>p$.  Since both $p$ and $p'$ clear the markets and since $p_c'>p_c\ge 0$ for each $c\in J$, for each $c\in J$, \vspace{-0cm}
$$D_c(p)\le S_c \mbox{ and } D_c(p')=S_c.\vspace{-0cm}$$
Summing across $c\in J$, we get
$$\sum_{c\in J} D_c(p')\ge \sum_{c\in J} D_c(p),\vspace{-0cm}$$
which contradicts $(SGS)$.

To prove the statement, it now suffices to show that $\eta$ satisfies $(SGS)$.  To this end, fix any $p, p'\in [0,1]^C$ with $p<p'$ and the corresponding set $J=\{c\in C: p_c<p_c'\}$. Observe that any ex-post type $\t$ who demands a school in $J$ at cutoffs $p'$ never switches its demand to a school in $C\setminus J$ or to the outside option $\o$, when the cutoffs shift to $p$. Hence, any such type continues to demand a (possibly  different) school in $J$ at cutoffs $p$.  Next, consider the type $(u,s)$ with $u_c>0> u_{c'}$, $c\in J$, $\forall c'\ne c$, and $s_c\in (p_c, p_c')$.  By Marginal Full Support, there is a positive mass of such types.  These types could not demand any school at $p'$ but now demand $c$ at $p$.  Collecting our observations, we conclude that $D_c(p')<D_c(p)$.  Since the same result holds for each $c\in J$,  $(SGS)$ holds. 
\end{proof}

\subsection{Proof of Asymptotic Stability of Robust Equilibria} 
\label{sec:proof_mainthm}

We now prove our main result under the general priority structure satisfying Marginal Full Support, Finite Atoms, and Rich Uncertainty:

\begin{customthm}{1$\bm ^\prime$}
\label{thm:stability-general}  Suppose $\eta$ satisfies Marginal Full Support, Finite Atoms, and Rich Uncertainty. Then, any regular robust equilibrium is asymptotically stable.
\end{customthm}

Note that \Cref{thm:stability} follows from \Cref{thm:stability-general} in light of the fact that the priority structures in the former satisfy Finite Atoms and Rich Uncertainty.

Before proceeding with the proof, we need to perform a few preliminary analyses.  Specifically, for each $k$-economy $F^k$, we study the strategy profile for that economy, or the $k$-truncation of $\sigma$, denoted by $\sigma^k:=(\sigma_1, ..., \sigma_k)$. A later analysis requires us to consider the consequence of an arbitrary student $i$ deviating to truthful reporting $\rho$. We denote the resulting profile by $\sigma_{(i)} :=(\rho, \sigma_{-i})$ which is obtained by replacing the $i$-th component of $\sigma$ with $\rho$.  Likewise, $\sigma_{(i)}^k$ denotes the $k$-truncation of $\sigma_{(i)}$. Note that if $i > k$, then $\sigma_{(i)}^k=\sigma^k$.  Let $P^k_{(i)}$ denote the cutoffs that would prevail if the students employ $\sigma_{(i)}^k$. Finally, let $\overline p$ be the unique, deterministic cutoff under the unique stable matching in the continuum economy $E$ (\Cref{prop:uniquestable}). We first establish a desirable limit behavior of $(P^k_{(i)})_{i\in \mathbb{N}_0}$ as $k\to \infty$, where $\mathbb{N}_0 := \mathbb{N} \cup \{0\}$. 

\begin{lem} \label{lem:uniform-convergence of P} 
 
	Let $\sigma$ be any $\gamma$-regular strategy profile. There exists a subsequence $\left\{F^{k_{\ell}}\right\}_{\ell}$ such that\vspace{-0cm}
	$$
	\sup_{0 \le i\le k_{\ell}} \lVert P^{k_{\ell}}_{(i)}-\overline p \rVert \overset{p}{\longrightarrow }0 \mbox{ as }\ell \to \infty.
 \vspace{-0cm}
	$$
\end{lem}

\begin{proof} The proof follows exactly the same argument as ACH, upon noting that the uniqueness of stable matching follows from (SGS) (instead of ACH's full support assumption).  
\end{proof}

		Next, define a set \vspace{-0.1cm}\\		
		\centerline{
		\begin{minipage}{\linewidth}
			\begin{align*}
		\tilde{\T}^{\delta}  :=   \left\{(u,t)\in \tilde{\T}: \inf_{c,c'\in\tilde C, c\ne c'} |u_c-u_{c'}|>\delta; \,   
		\forall c, \, \overline s_c^t=\underline s_c^t \mbox{ or } \overline s_c^t \ne \bar p_c \Rightarrow  |\overline s_c^t -\bar p_c| >\delta;\, \underline s_c^t \ne \bar p_c \Rightarrow  |\underline s_c^t -\bar p_c| >\delta\right\}.
		\end{align*}
		\end{minipage}
	}
 The Finite Atoms condition, together with atomlessness of $\eta$, implies that $\tilde \eta(\tilde{\T}^{\delta}) \to 1$ as $\delta\to 0$.\footnote{We note that the limit set $\tilde{\T}^{-}:=\tilde{\T}-\cup_{\delta>0}\tilde{\T}^{\delta}$ contains no mass point.  This is seen by the fact that the only possibility of a point mass in $\tilde{\T}^{-}$ may arise from a point mass occurring at some $\overline s_c^t=\underline s_c^t=p_c^t$ for some $c$. (Note that   any type $t$ with $\overline s_c^t>\underline s_c^t$ belongs to $\cup_{\delta>0}\tilde{\T}^{\delta}$, so it does not beling to $\tilde{\T}^{-}$.)   But if there were such a point mass, then there must be a positive mass at score $s_c=p_c^t$, which contradicts the atomlessness of $\eta$.} 

Further, we introduce a few notations.  Define $B_{\nu}(p):=\{p'\in [0,1]^C: ||p'-p||<\nu\}$.  Consider a student $\tilde \t$ and fix any arbitrary economy with other students playing some arbitrary reporting strategies.  If she adopts any arbitrary reporting strategy $\sigma_i'(\tilde{\t})$, this induces set of assignment probability 
$(X_c')_{c\in \tilde C}$, where $X_c'$ is the probability of the student being assigned to a school that she weakly prefers to $c$.  Suppose she switches to TRS which induces assignment probability $(X_c^*)_{c\in \tilde C}$, where $X_c^*$ represents the probability of the student being assigned a school that she weakly prefers to $c$.  The strategyproofness of DA implies that TRS yields better assignment than any other strategy in the sense of  first-order stochastic dominance:  	
\begin{fact} For each $c\in \tilde C$, $X_c^*-X_c'\ge 0$. 
\end{fact}	

Naturally, we define the {\bf probability gain} from the switch to be  $\max_{c\in \tilde C} (X_c^*-X_c')$.  The following preliminary result is useful.

\begin{lem}\label{lem:assignment-gap}  Fix any $\bar p\in [0,1]^C$ and any   $\delta\in (0,  \frac{1}{2} \min\{\kappa, \zeta\})$, where $\zeta:=  \inf_{c,c'\in C \cup\{x,y\}, \bar p_c\ne \bar p_{c'}}|\bar p_c-\bar  p_{c'}|$.  Then, there exist $\nu(\delta)>0$ and $\alpha(\delta)>0$ such that deviating from any non-SRS against $p$ to TRS yields a probability gain of at least $\alpha(\delta)$, and thus a payoff gain of at least $\delta\alpha(\delta)$, for any student with type $\tilde\t\in \T^{\delta}$, provided that non-SRS and TRS induce cutoffs $p'$ and $p''$ both in $B_{\nu(\delta)}(p)$. 
\end{lem}

\begin{proof} Let $\beta(\delta)$ be the probability lower bound defined in Rich Uncertainty.  Set $\nu(\delta)= \min \{\frac{\delta}{4 }, \frac{\beta(\delta)}{4 \upsilon } \}$, where $\upsilon:=\max_{t\in T_n,  s\in S^t} \phi^t(s)$.  Fix a student   with type $\tilde\t=(u,t)\in \tilde \T^{\delta}$.  Without loss, we index schools  $ \tilde C=\{1, ...,  {C+1}\}$ (including the outside option $\o$) so that $u_{j}>u_{j'}$ if and only if $j<j'$. 
	Let $C_j$ denote the set of schools that student $\tilde \t$ prefers to $j$.  Let $C^F\subset \tilde C$ be a set of  {\it feasible}  schools that the student would get with positive probabilities when she plays SRS against $\bar p$.  It means that for each $j\in C^F$, we have 
	\begin{equation} \label{eq:iii-hypothesis1}
	\Pr\{s_c < \bar p_c, \forall c\in C_j, s_j>\bar p_j  \}>0.
	\end{equation} 	
	There could be a feasible,  and less preferred, school outside $C^F$ that she is not assigned but would have been with positive probability if she had ranked it sufficiently favorably.
	Since $\tilde\t\in \tilde \T^{\delta}$, all these feasible schools are in fact $\delta$-feasible.  
	
	Now take any non-SRS $\sigma'$ against $p$ for that student. The strategy must rank some less-preferred, feasible---and thus $\delta$-feasible---school $b$ ahead of some school $a\in C^F$, where she prefers $a$ to $b$, and let $a$ be the most preferred school in $C^F$ that suffers from such a ranking-reversal by the non-SRS strategy. Let $C_a$ be the schools that the student prefers to $a$ among $\tilde C$. For the reversal to make a difference (which follows from $\sigma'$ being non-SRS), we must have \vspace{-0cm}
	\begin{equation} \label{eq:iii-hypothesis2}
	\Pr\{s_c < \bar p_c, \forall c\in C_a, s_b>\bar p_b  \}>0.\vspace{-0cm}
	\end{equation}

	Suppose as hypothesized that  $\sigma'$  in the hypothesized economy induces a cutoff  $p'\in B_{\nu(\delta)}$.  
	We wish to compute the probability $X'_a$ of the student receiving $a$ or better according to the true preferences. Let $C'\subset C_a$ be the set of schools among $C_a$ that $\sigma'$ ranks ahead of $a$.  Then,  \vspace{-0cm}
		\begin{align*}
	X'_a\le & \Pr\{ \exists c \in C' \text{ s.t. } s_c > p_c' \}+\Pr\{s_c < p_c', \forall c\in C', s_b<p_b',  s_a>p_a'   \} \\
	\le & \Pr\{ \exists c \in C_a \text{ s.t. } s_c > p_c' \}+\Pr\{s_c < p_c', \forall c\in C_a, s_b<p_b,  s_a>p_a'   \} \\
	\le & \Pr\{ \exists c \in C_a \text{ s.t. } s_c > p_c \}+\Pr\{s_c < \bar p_c, \forall c\in C_a, s_b<\bar p_b,  s_a>\bar p_a   \} + \upsilon \Vert \bar p'-\bar p\Vert\\
	\le & \Pr\{ \exists c \in C_a \text{ s.t. } s_c > \bar p_c \}+\Pr\{s_c <\bar  p_c, \forall c\in C_a, s_b<\bar p_b,  s_a>\bar p_a   \} + \frac{\beta(\delta)}{4},
    \vspace{-0cm}
	\end{align*}
	where the first inequality holds since it is possible that $\sigma'$ ranks some other less preferred school other than $b$ ahead of $a$, the second follows from the fact that $C'\subset C_a$, the third follows from the fact that for type $\tilde\t\in \tilde \T^{\delta}$, $B_{\delta/4}(\bar p)$ contains no point mass of scores and that $p',p\in B_{\nu(\delta)}(\bar p)\subset B_{\delta/4}(\bar p)$, and the last inequality follows from the fact that $\nu(\delta) \le \frac{\beta(\delta)}{4\upsilon}$ and $p' \in B_{\nu(\delta)}(\bar p)$.  
	
	Suppose next the student switches to TRS and as a result faces $p''\in B_{\nu(\delta)}(p)$ as cutoffs.  The probability of getting $a$ or better schools from the switch is given by:\vspace{-0cm}
	\begin{align*}
	X^*_a 
	= & \Pr\{ \exists c \in C_a \text{ s.t. } s_c > p_c'' \}+\Pr\{s_c < p_c'', \forall c\in C_a,   s_a>p_a''   \} \\
	\ge & \Pr\{ \exists c \in C_a \text{ s.t. } s_c > \bar p_c \}+\Pr\{s_c <\bar  p_c, \forall c\in C_a,   s_a>\bar p_a  \} - \upsilon \Vert p''-\bar p\Vert\\
	\ge & \Pr\{ \exists c \in C_a \text{ s.t. } s_c > \bar p_c \}+\Pr\{s_c <\bar  p_c, \forall c\in C_a,  s_a>\bar p_a   \} - \frac{\beta(\delta)}{4}.
 \vspace{-0cm}
	\end{align*}
	Again, the second inequality follows from the fact that for type $\tilde\t\in \tilde \T^{\delta}$, $B_{\delta/4}(\bar p)$ contains no point mass of scores and that $p'',p\in B_{\nu(\delta)}(\bar p)\subset B_{\delta/4}(\bar p)$, and the last inequality follows from the fact that $\nu(\delta) \le \frac{\beta(\delta)}{4\upsilon}$ and $p'' \in B_{\nu(\delta)}(\bar p)$.

	Consequently, the probability gain from switching from $\sigma'$ to TRS is at least: \vspace{-0cm}
	\begin{align*}
	\max_{c\in \tilde C}(X^*_c -X'_c) \ge &
	X^*_a -X'_a \\
	\ge & \Pr\{s_c <\bar  p_c, \forall c\in C_a, s_a>\bar p_a   \} -\Pr\{s_c < p_c, \forall c\in C_a, s_b<\bar p_b,  s_a>\bar p_a \} -  \frac{\beta(\delta)}{2}\\
	= & \Pr\{s_c < \bar p_c, \forall c\in C_a, s_a>\bar p_a, s_b>\bar p_b   \} -  \frac{\beta(\delta)}{2}\\
	\ge & \beta(\delta) -  \frac{\beta(\delta)}{2}=  \frac{\beta(\delta)}{2}.  
 \vspace{-0cm}
	\end{align*}
	The last inequality follows from Rich Uncertainty.  To see this, note first that each school in $C_a\setminus C^F$ is infeasible at $p$ to $\tilde \t=(u,t)$; then, since $\tilde \t\in \tilde \T^{\delta}$, it is $\delta$-infeasible at $p$ for $\tilde \t$. Consider next any $c\in C^F\cap C_a$.  The school is feasible given $p$ to $\tilde \t=(u,t)$, but we must have  $\sl_c^{t}< \bar p$; otherwise, the student would not have been assigned $a$ with positive probability under SRS given $p$. Then, since  $\tilde \t\in \tilde \T^{\delta}$, $\sl_c^{\delta}< \bar p-\delta$, so it is  $\delta$-infeasible. Finally, recall that both $a$ and $b$ are feasible for $\tilde \t$; again since $\tilde \t\in \tilde \T^{\delta}$, both $a$ and $b$ are $\delta$-feasible for  $\tilde \t$. Therefore,   (\ref{eq:iii-hypothesis1}) and (\ref{eq:iii-hypothesis2}) imply the inequality (\ref{eq:(iii)}), leading to that last inequality.

	Setting $\alpha(\delta)=\frac{\beta(\delta)}{2}$, we have established that the switch results in the probability gain of at least $\alpha$.  Since $\tilde\t\in \tilde \T^{\delta}_p$, the associated payoff gain is\vspace{-0cm}
	$$\sum_i ( X_i^*-X_{i-1}^*) u_i^{\tilde \t} -  \left[ \sum_i ( X_i'-X_{i-1}') u_i^{\tilde \t}\right]\ge \sum_i (X_i^*-X_i') (u_i^{\tilde \t}-u_{i+1}^{\tilde \t} )\ge (X_a^*-X_a') (u_a^{\tilde \t}-u_{a+1}^{\tilde \t})> \alpha \delta,\vspace{-0cm}$$
	where $ X_0^*= X_0'\equiv 0$. 
	We note that this bound does not depend on  $\tilde\t\in \tilde \T^{\delta}$.  
\end{proof}

\begin{proof}[Proof of \Cref{thm:stability}]  	
	For any sequence $\{F^k\}$ induced by $E$, fix any arbitrary regular robust equilibrium $\{(\sigma^k_{1\le i \le k})\}_k$. The strategies induce a random ROL, $R_i$, for each player $i$.
	We prove that the fraction of students who  play non-SRS in $\left\{
		\left(  \sigma_{i}^{k}\right)_{1\leq i\leq k}\right\}  _{k}$ converges in probability to zero as $k\to \infty$.	Suppose not.  Then, there exists $\varepsilon>0$ and a
		subsequence of finite economies $\left\{  F^{k_{j}}\right\}_{j}$ such that\vspace{-0cm}
		\[
		\Pr\left(  \text{The fraction of students playing SRS against } {p}^{k_{j}}%
		\text{ is no less than } 1-\varepsilon\right) < 1-\varepsilon.  \eqno{(*)}
  \vspace{-0cm}
		\]

		By \Cref{lem:uniform-convergence of P},  there exists a sub-subsequence of  economies $\left\{  F^{k_{j_{\ell}}}\right\}_{\ell}$,  
		such that the associated cutoffs
		$ {p}^{k_{j_{\ell}}}$ converge to $\bar{ {p}}$ in probability, where $\bar{ {p}}$ are the deterministic cutoffs stated in  \Cref{lem:uniform-convergence of P}.

		We choose $\delta>0$ small enough so that $\eta ( \tilde{\T}^{\delta}  )
		>\left(  1-\varepsilon\right)^{1/3}$ (this can be done since $\eta$ is
		absolutely continuous).  We then choose $\nu(\delta)$ and $\alpha(\delta)$ according to \Cref{lem:assignment-gap}.

		By WLLN, we know that $\eta^{k_{j_{\ell}}}( \tilde{\T}^{\delta}  )  $
		converges to $\eta ( \tilde{\T}^{\delta}  )  $ in probability, and therefore
		there exists $L_{1}$ such that for all $\ell>L_{1}$ we have\vspace{-0cm}
		\[
		\Pr\left(  \eta^{k_{j_{\ell}}} (  \tilde{\T}^{\delta} )  \geq\left(
		1-\varepsilon\right)  ^{1/2}\right)  \geq\left(  1-\varepsilon\right)  ^{1/2}.%
        \vspace{-0cm}
		\]

		For each economy $F^{k_{j_{\ell}}}$, consider the event \vspace{-0cm}
		\[
		A^{k_{j_{\ell}}}:=\left\{\sup_{0\le i\le k_{j_{\ell}}}   {||} P^{k_{j_{\ell}}}_{(i)}-\bar {p} {||}
		<\nu(\delta) \right\}.
        \vspace{-0cm}
		\]
		
		Since $P^{k_{j_{\ell}}}_{(i)}\overset{p}{ \rightarrow } \bar{p}$ uniformly over $i\in \mathbb{N}_0$, there exists
		$L_{2}$ such that, for all $\ell>L_{2}$ s.t. for all $\ell>L_{2}$ we have\vspace{-0cm}
		\[
		\Pr\left(  A^{k_{j_{\ell}}}\right)  \geq\max\left\{  \left(  1-\varepsilon
		\right)  ^{1/6},1-\left(  1-\varepsilon\right)  ^{1/2}\left[  \left(
		1-\varepsilon\right)  ^{1/3}-\left(  1-\varepsilon\right)  ^{1/2}\right]
		\right\}.  \eqno{(**)}
        \vspace{-0cm}
		\]

		Because $\left\{  \left(  \sigma_{i}^{k}\right)  _{1\leq i\leq k}\right\}
		_{k}$ is a robust equilibrium, there exists $L_{3}$ s.t. for all $\ell>L_{3}$ the
		strategy profile $(  \sigma_{i}^{k_{j_{\ell}}})  _{i=1}^{k_{j_{\ell}}}$
		is a $\delta\cdot \alpha(\delta)\left[  \left(  1-\varepsilon\right)^{1/6}-\left(
		1-\varepsilon\right)^{1/3}\right]  $-BNE for economy $F^{k_{j_{\ell}}}$.
		
		By WLLN, there exists a sufficiently large $\hat{L}$ such that  $\hat{L}$ i.i.d.\ Bernoulli random
		variables with probability $\left(  1-\varepsilon\right)^{1/3}$ have a sample mean
		greater than $\left(  1-\varepsilon\right)  ^{1/2}$ with probability no less
		than $\left(  1-\varepsilon\right)  ^{1/3}$. Define next $L_{4}$ such that  $\ell>L_{4}$ implies $\left(  1-\varepsilon\right)  ^{1/2}{k_{j_{\ell}}}>\hat{L}$.
		
		Now we fix an arbitrary $\ell>\max\left\{  L_{1},L_{2},L_{3},L_{4}\right\}$
		and wish to show that in economy $F^{k_{j_{\ell}}}$
        \vspace{-0cm}
		\[
		\Pr\left(  \text{The fraction of students playing SRS against }  {p}^{k_{j_{\ell}}}%
		\text{ is no less than } 1-\varepsilon\right)  \geq1-\varepsilon,
        \vspace{-0cm}
		\]
		which would contradict $(*)$ and complete the proof.
		
		We first prove that in economy $F^{k_{j_{\ell}}}$, a student with $ \tilde{\theta}\in
		 \tilde{\T}^{\delta} $ plays SRS against $\bar{ {p}}$ with probability no less than
		$\left(  1-\varepsilon\right) ^{1/3}$. To see this, suppose to the contrary that
		there exists some student $i$ and some type $ \tilde{\theta}\in \tilde{\T}^{\delta} $ such that 
        \vspace{-0cm}
		\[
		\Pr\left(  \sigma_{i}^{k_{j_{\ell}}} (   \tilde{\theta})  \text{ plays SRS
			against }\bar{ {p}}\right)  <\left(  1-\varepsilon\right)  ^{1/3}.%
        \vspace{-0cm}
		\]
		Then deviating to truthful reporting will give student $i$ with type $ \tilde{\theta}\in  \tilde{\T}^{\delta} $ at least a
		gain of
        \vspace{-0cm}
		\begin{align*}
			& \mathbb{E} \left[i\text{'s gain from deviation }\Big\vert \sigma_{i}^{k_{j_{l}}} \text{ plays
				non-SRS against } {p}^{k_{j_{\ell}}} \right]\Pr\{\sigma_{i}^{k_{j_{\ell}}} \text{ plays
				non-SRS against } {p}^{k_{j_{\ell}}}\}   \\
			\ge  & \mathbb{E} \left[i\text{'s gain from deviation }\Big\vert \{ \sigma_{i}^{k_{j_{l}}} \text{ plays
				non-SRS against } {p}^{k_{j_{\ell}}}\} \wedge A^{k_{j_{l}}} \right] \\
				& \cdot\Pr\left( 
				\sigma_{i}^{k_{j_{\ell}}} (   \tilde{\theta} )  \text{ plays non-SRS against			} {p}^{k_{j_{\ell}}}
			\text{ {\it and} event }A^{k_{j_{\ell}}}
			\right)  \\
			\ge	& \delta \alpha(\delta)\cdot\Pr\left(
				\sigma_{i}^{k_{j_{\ell}}} (   \tilde{\theta} )  \text{ plays non-SRS against
				} {p}^{k_{j_{\ell}}}
				\text{ {\it and} event }A^{k_{j_{\ell}}}
			\right)  \\
			\ge	&  \delta  \alpha(\delta)\cdot
			\Pr\left(
				\sigma_{i}^{k_{j_{\ell}}} (   \tilde{\theta} )  \text{ plays non-SRS against	}\bar{ {p}} 
				\text{ {\it and} event }A^{k_{j_{\ell}}}\right)   \\
			\ge	&  \delta  \alpha(\delta)\left[  \Pr \left(  A^{k_{j_{\ell}}}\right)  -\Pr\left(\sigma_{i}^{k_{j_{\ell}}}( \tilde{\theta} )  \text{ plays SRS against }\bar
			{ {p}}\right)  \right]  \\
			\ge	&  \delta  \alpha(\delta)\left[  \left(  1-\varepsilon\right)  ^{1/6}-\left(
			1-\varepsilon\right)  ^{1/3}\right],
        \vspace{-0cm}
		\end{align*}
		where the inequalities are explained as follows.  The first inequality holds since the gains from the deviation is nonnegative whether the event $A^{k_{j_{\ell}}}$ occurs or not.  The second inequality results from \Cref{lem:assignment-gap}, upon noting that by \Cref{lem:uniform-convergence of P} the cutoffs under the non-SRS and those that would prevail if $i$ deviates to TRS are within $\nu(\delta)$-distance from $\bar p$.
		The third inequality holds since, given our choice $\delta$, conditional on event $A^{k_{j_{\ell}}}$, a non-SRS against $\bar{ {p}}$ is a non-SRS against $ {p}^{k_{j_{\ell}}}$.    The above inequalities
		contradict   $\ell>L_{3}$, which implies that the strategy
		profile $\left(  \sigma_{i}^{k_{j_{\ell}}}\right)  _{i=1}^{k_{j_{\ell}}}$ is a
		$\delta  \alpha(\delta)\left[  \left(  1-\varepsilon\right)  ^{1/6}-\left(  1-\varepsilon
		\right)  ^{1/3}\right]$-BNE for the economy $F^{k_{j_{\ell}}}$.
		
		Therefore, in economy $F^{k_{j_{\ell}}}$, for each student $i=1,\ldots, {k_{j_{\ell}}}$
		and each $ \tilde{\theta}\in \tilde{\T}^{\delta} $, we have
        \vspace{-0cm}
		\begin{align*}
			& \Pr\left(  \left.  \sigma_{i}^{k_{j_{\ell}}} (   \tilde{\theta} )  \text{ plays
				SRS against }\bar{ {p}}\right\vert \eta^{k_{j_{\ell}}} ( \tilde{\T}^{\delta} )
			\geq\left(  1-\varepsilon\right)  ^{1/2}\right)  \\
			& =\Pr\left(  \sigma_{i}^{k_{j_{\ell}}} (   \tilde{\theta} )  \text{ plays SRS
				against }\bar{ {p}}\right)  \geq\left(  1-\varepsilon\right)  ^{1/3}, \tag{***}
        \vspace{-0cm}
		\end{align*}
		where the first equality holds because student $i$'s random report according
		to her mixed strategy is independent of random draws of the students' type.
		
		Then we have
        \vspace{-0cm}
		\begin{align}
			& \Pr\left(  \left.
			\begin{array}
				[c]{c}%
				\text{The fraction of students with } \tilde{\theta}\in \tilde{\T}^{\delta} \cr
				\text{playing SRS against }\bar{ {p}} \text{ is no less than } \left(  1-\varepsilon\right)  ^{1/2}%
			\end{array}
			\right\vert \eta^{k_{j_{\ell}}} (   \tilde{\T}^{\delta}  )  \geq\left(
			1-\varepsilon\right)  ^{1/2}\right)  \cr
			& \geq\Pr\left(  \left.
			\begin{array}
				[c]{c}%
				\eta^{k_{j_{\ell}}} (  \tilde{\T}^{\delta} )  \cdot  {k_{j_{\ell}}}\text{ i.i.d.\
					Bernoulli random variables with}\\
				\text{probability }\left(  1-\varepsilon\right)  ^{1/3} \text{have a sample mean no less than }\left(  1-\varepsilon\right)  ^{1/2}%
			\end{array}
			\right\vert \eta^{k_{j_{\ell}}} ( \tilde{\T}^{\delta} )  \geq\left(
			1-\varepsilon\right)  ^{1/2}\right)  \cr
			& \geq\Pr\left(
			\begin{array}
				[c]{c}%
				\hat{L}\text{ i.i.d.\ Bernoulli random variables with probability }\left(  1-\varepsilon
				\right)  ^{1/3}\cr
				\text{have a sample mean no less than }\left(  1-\varepsilon\right)  ^{1/2}%
			\end{array}
			\right)  \cr
			& \geq\left(  1-\varepsilon\right)  ^{1/3}, \label{****}	
        \vspace{-0cm}
        \end{align}
		where the first inequality follows from   (***)  and that
		$\sigma_{i}$'s are independent across students conditioning on the event
		$\eta^{k_{j_{\ell}}} ( \tilde{\T}^{\delta} )  \geq\left(  1-\varepsilon
		\right)  ^{1/2}$, and the second inequality holds since, for  $\ell>L_{4}$, 
		$\eta^{k_{j_{\ell}}} (  \tilde{\T}^{\delta} )  \geq\left(  1-\varepsilon
		\right)  ^{1/2}$ implies that $\eta^{k_{j_{\ell}}} (   \tilde{\T}^{\delta}  )  \cdot
		k_{j_{l}}>$ $\hat{L}$.
		
		Comparing the finite economy random cutoff $ {p}^{k_{j_{\ell}}}$ with the
		deterministic cutoff $\bar{ {p}}$, we have
        \vspace{-0cm}
		\begin{align*}
			& \Pr\left(  \left.
			\begin{array}
				[c]{c}%
				\text{The fraction of students with } \tilde{\theta}\in  \tilde{\T}^{\delta} \\
				\text{playing SRS against } {p}^{k_{j_{\ell}}} \text{ is no less than } \left(  1-\varepsilon\right)
				^{1/2}%
			\end{array}
			\right\vert \eta^{k_{j_{\ell}}} (   \tilde{\T}^{\delta} \ )  \geq\left(
			1-\varepsilon\right)  ^{1/2}\right)  \\
			& \geq\Pr\left(  \left.
			\begin{array}
				[c]{c}%
				\text{The fraction of students with } \tilde{\theta}\in \tilde{\T}^{\delta} \\
				\text{playing SRS against }\bar{ {p}} \text{ is no less than } \left(  1-\varepsilon\right)
				^{1/2}\text{,}\\
				\text{and event }A^{k_{j_{\ell}}}%
			\end{array}
			\right\vert \eta^{k_{j_{\ell}}} (   \tilde{\T}^{\delta}  )  \geq\left(
			1-\varepsilon\right)  ^{1/2}\right)  \\
			& \geq\Pr\left(  \left.
			\begin{array}
				[c]{c}%
				\text{The fraction of students with } \tilde{\theta}\in \tilde{\T}^{\delta} \\
				\text{playing SRS against }\bar{ {p}} \text{ is no less than } \left(  1-\varepsilon\right)  ^{1/2}%
			\end{array}
			\right\vert \eta^{k_{j_{\ell}}}( \tilde{\T}^{\delta} )  \geq\left(
			1-\varepsilon\right)  ^{1/2}\right)  \\
			& -\Pr\left(  \left. {A}^{k_{j_{\ell}}}  \mbox{ does not occur} \right\vert \eta^{k_{j_{\ell}}} (
			 \tilde{\T}^{\delta}  )  \ge \left(  1-\varepsilon\right)  ^{1/2}\right)  \\
			& \geq\left(  1-\varepsilon\right)  ^{1/3}-\frac{1-\Pr\left({A}^{k_{j_{\ell}}}\right)  }{\Pr\left(  \eta^{k_{j_{\ell}}} (   \tilde{\T}^{\delta}  )
				\geq\left(  1-\varepsilon\right)  ^{1/2}\right)  }\\
			&  \ge \left(  1-\varepsilon\right)  ^{1/3}-\frac{\left(  1-\varepsilon\right)
				^{1/2}\left[  \left(  1-\varepsilon\right)  ^{1/3}-\left(  1-\varepsilon
				\right)  ^{1/2}\right]  }{\left(  1-\varepsilon\right)  ^{1/2}}\\
			& =\left(  1-\varepsilon\right)^{1/2}.%
        \vspace{-0cm}
		\end{align*}
		The first inequality follows since in event ${A}^{k_{j_{\ell}}}$,  the strategy  $\sigma_i(\t)$ is SRS against $P^{k_{j_{\ell}}}$ if and only if $\sigma_i(\t)$ is SRS against $\overline p$ for type   $\t\in \tilde{\T}^{\delta} $. The third inequality follows from (\ref{****}).  The  fourth inequality follows from (**).
 		
		The construction of $L_{1}$ implies $\Pr\left(  \eta^{k_{j_{\ell}}} (
		 \tilde{\T}^{\delta}  )  \geq\left(  1-\varepsilon\right)  ^{1/2}\right)
		\geq\left(  1-\varepsilon\right)  ^{1/2}$,  so finally we have in economy
		$F^{k_{j_{\ell}}}$,
        \vspace{-0cm}
		\begin{align*}
			& \Pr\left(  \text{The fraction of students playing SRS against } {p}^{k_{j_{\ell}}}%
			\text{ is no less than } 1-\varepsilon\right)  \\
			& \geq\Pr\left(
			\begin{array}
				[c]{c}%
				\text{At least } \left(  1-\varepsilon\right)
				^{1/2} \text{ of students with } 	 \tilde{\theta}\in \tilde{\T}^{\delta} 
				\text{ play SRS against } {p}^{k_{j_{\ell}}}\\
				\text{\it and }\eta^{k_{j_{\ell}}} (  \tilde{\T}^{\delta}  )  \geq\left(
				1-\varepsilon\right)  ^{1/2}%
			\end{array}
			\right)  \\
			& =\Pr\left(  \eta^{k_{j_{\ell}}} (  \tilde{\T}^{\delta}  )  \geq\left(
			1-\varepsilon\right)^{1/2}\right)  \cdot\Pr\left(  \left.
			\begin{array}
				[c]{c}%
				\text{At least } \left(  1-\varepsilon\right)
				^{1/2} \text{ of students}	  \\ \text{with } \tilde{\theta}\in \tilde{\T}^{\delta}
				\text{ play SRS against } {p}^{k_{j_{\ell}}}\\
			\end{array}
			\right\vert \eta^{k_{j_{\ell}}} (  \tilde{\T}^{\delta}  )  \geq\left(1-\varepsilon\right)^{1/2}\right)  \\
			& \geq\left(  1-\varepsilon\right)^{1/2}\cdot\left(  1-\varepsilon\right)^{1/2}\\
			& =1-\varepsilon,
        \vspace{-0cm}
		\end{align*}
		where the last inequality follows from the above string of inequalities.  Therefore, we have obtained a contradiction to $(*)$, and the statement of \Cref{thm:stability} follows.  \end{proof}

\subsection{Proof of \Cref{prop:maximal}}
\begin{proof}
First, as TEPS only uses stability and transitivity to infer preferences, it  must be that $ \pref{} \subseteq \pref{\text{ST}}(\uncset') $.  
Second, we show that for any $(c,c')\in\pref{\text{ST}}(\uncset')$, $(c,c')\in \pref{}$.  
As $c'$ is inferred worse than $c$ by stability and transitivity,  there must be a sequence of schools, $c^1,\ldots,c^J$ for $1<J\leq C$ with $c^1=c$ and $c^J=c'$, such that 
$c^{j-1}$ and $c^{j}$, for $1<j\leq J$, are both feasible in some realized uncertainty in $\uncset'$ when $c^{j-1}$ is the assigned school.  
Since TEPS uses all uncertainties in $\uncset'$ in Stage~1,  
it must be that $(c_{j-1},c_j)\in\preff$ for $1<j\le J$ in Stage 2. 
By transitivity at Stage~3 of TEPS, $(c,c')=(c_1,c_J)\in\pref{}$.  
We thus have $\pref{} = \pref{\text{ST}}(\uncset') $. 
\end{proof}

\subsection{Proof of \Cref{prop:nest}}
\begin{proof}
First, for any $0<\tau < \tau'<100$, $\pref{top}  \subseteq \pref{\tau} \subseteq \pref{\tau'} \subseteq \pref{all}$ directly follows from Stage 1 of each \teps{\tau}, and $\pref{all}= \pref{\text{ST}}(\uncset)$ directly follows from \Cref{prop:maximal}
. We only need to prove that $ \pref{all} \subseteq \pref{\text{WTT}}$. Consider some $(c,c')\in \pref{all}$. We show that $(c,c')\in \pref{\text{WTT}}$. 

First, since \teps{all} infers only an assigned school in some realized uncertainty is preferred to other schools, and an unranked school can never be a student's assigned school, $c$ should be a ranked school in $R$. Next, notice that $\pref{all}$ can never include any $(c,c')$ such that $c'$ is ranked above $c$ in $R$. To see this, assume to the contrary that $(c,c')\in\pref{all}$ and that $c'$ is ranked above $c$ in $R$. Then at Stage 1 of \teps{all}, there must be a sequence of schools, $c^1,\ldots,c^J$ for $1<J\leq C$ with $c^1=c$ and $c^J=c'$, such that 
$c^{j-1}$ and $c^{j}$, for $1<j\leq J$, are both feasible in some realized uncertainty when $c^{j-1}$ is the assigned school. Then, for each $1<j\le J$, it must be that $c^{j-1}$ is ranked above $c^j$, which implies that $c^1=c$ is ranked above $c^J=c'$ in $R$, a contradiction.

Hence, $c'$ is either (i) a school ranked lower than $c$ on $R$, or (ii) an unranked school. Since WTT infers that $c$ is more preferable than \textit{every} school ranked below $c$ and \textit{every} unranked school, it should be that $(c,c')\in\pref{\text{WTT}}$. Therefore, $ \pref{all} \subseteq \pref{\text{WTT}}$. 
\end{proof}

\clearpage

\begin{center}
{(For Online Publication)}\\
[0.2cm]
{\Large Online Appendix to}\\
[0.3cm]
{\LARGE \TITLE}\\
[0.2cm]
{\large Yeon-Koo Che \hspace{1cm} Dong Woo Hahm \hspace{1cm} YingHua He} \\
[0.2cm]
{\large \MONTH ~\the\year}\\
\end{center}

\setcounter{section}{0}
\setcounter{page}{1}
\setcounter{footnote}{0}
\renewcommand{\thefootnote}{A-\arabic{footnote}}
\setcounter{equation}{0}
\renewcommand{\theequation}{\Alph{section}.\arabic{equation}}
\setcounter{table}{0}
\renewcommand{\thetable}{\Alph{section}.\arabic{table}}
\setcounter{figure}{0}
\renewcommand{\thefigure}{\Alph{section}.\arabic{figure}}
\setcounter{prop}{0}
\renewcommand{\theprop}{\Alph{section}.\arabic{prop}}
\setcounter{lem}{0}
\renewcommand{\thelem}{\Alph{section}.\arabic{lem}}

\section{Supplementary Materials for Section~\ref{sec:theory}}
\subsection{Rich Uncertainty Assumption\label{app:lem-iii-proof}}
We show that \Cref{assm:richuncertainty}, Rich Uncertainty, is satisfied in all realistic environments (a), (b), and (c) in the main text.

\begin{lem} Rich Uncertainty holds for priority structures, (a), (b), and (c). \label{lem:iii}
 \end{lem}
 
\begin{proof} For case (a), $\sl_c^t=\sh_c^t$ for all $t\in T$, so the probability in the LHS of \Cref{eq:(iii)} is equal to  one.  Hence, Rich Uncertainty holds trivially.  We thus focus on (b) and (c). 

\paragraph{Structure (b): STB version.}  The cutoff $p_c$ for each school $c$ corresponds to a lottery threshold $\ell_c^t\in [0,1]$, possibly dependent on the type $t$ of the student. To see this, the cutoff can be expressed as:  $p_c=\hat t_{c}+\frac{\lambda_c}{n_c}$, where the $\hat t_{c}$ is the intrinsic priority level that one needs for admission at $c$ and $\lambda_c$ is the  lottery cutoff score for $c$.  If $\sh^t_c-\delta>p_c$ for some $\delta>0$, either the student has an intrinsic priority level of  $\hat t_{c}$ or a higher priority level.  In the former, $s_c>p_c$ amounts to having a lottery draw above $\lambda_c$, so $\ell_c^t=\lambda_c$.  In the latter, $s_c>p_c$ regardless of the lottery draw, so $\ell_c^t=0$. Similarly, if $\sl^t_c+\delta<p_c$ for some $\delta>0$, either the student has an intrinsic priority level of  $\hat t_{c,i}$ or a lower priority level.  In the former, $s_c<p_c$ amounts to having a lottery draw below $\lambda_c$, in which case $\ell_c^t=\lambda_c$.  In the latter, $s_c<p_c$ regardless of the lottery draw, so $\ell_c^t=1$.

Given the STB structure,   the LHS of (\ref{eq:(iii)}) then reduces to\vspace{-0cm}
\begin{equation} \label{eq:(iii)-STB}
\Pr\left\{\max\{ \ell_a^t, \ell_b^t\}<\lambda^t< \min_{c\in C^t} \ell_c^t \right\},\vspace{-0cm}
\end{equation} 
	namely, the probability that the student draws an STB lottery number $\lambda^t\in (\max\{ \ell_a^t, \ell_b^t\}, \min_{c\in C^t} \ell_c^t)$. 
  Further, if $\Pr\{s_c^t<p_c, \forall c\in C^t, 
  s_a^t>p_a\}>0$ and $\Pr\{s_c^t<p_c, \forall c\in C^t, s_b^t>p_b \}>0$, it must be that  $\max\{ \ell_a^t, \ell_b^t\}<  \min_{c\in C^t} \ell_c^t$. Note that $\max\{ \ell_a^t, \ell_b^t\}=\lambda_i$ for some $i\in \{a,b,x\}$ and $\min_{c\in C^t} \ell_c^t= \lambda_j$ for some $j\in C^t \cup\{y\}$  such that $\lambda_i<\lambda_j$, where $\lambda_x:=0$ and  $\lambda_y:=1$.  
Hence,  (\ref{eq:(iii)-STB}) equals\vspace{-0cm}
\begin{align*}
\Pr\left\{\max\{ \ell_a^t, \ell_b^t\}<\lambda^t< \min_{c\in C^t} \ell_c^t \right\}
=&\Pr\left\{\lambda_i<U< \lambda_j \right\}\\
\ge &\min\left\{|\lambda_c-\lambda_{c'}|: c,c'\in \hat C, \lambda_c\ne \lambda_{c'} \right \}=: \beta(\delta),\vspace{-0cm}
\end{align*} 
where $U$ is uniform random variable on $[0,1]$ and $\hat C:=C\cup\{x,y\}$. Importantly, this lower bound $\beta(\delta)$ does not depend on $t$ or the particular pair $(a,b)$ or other $c\in C^t$. It only depends on $(\lambda_c)_c$, which is determined uniquely by $p$.
		
 \paragraph{Structure (b): MTB version.}  The approach is similar to that of STB.  In particular, the first part (mapping cutoffs $p$ to lottery cutoffs $\lambda=(\lambda_c)_c$) is exactly the same.  The difference is that the LHS of (\ref{eq:(iii)}) is now equal to\vspace{-0cm}
\begin{equation} \label{eq:(iii)-MTB}
\Pr\left\{ \lambda_c^t<\ell_c^t, \forall c\in C^t,  \ell_i^t<\lambda_i^t, \forall i=a,b\right\},\vspace{-0cm}
\end{equation} 
where $\lambda^t_i$ is $t$'s MTB draw for school $i$.
Clearly, given the hypothesis, this probability must be positive, which implies that
$\ell_i^t<1$ for $i=a,b$ and $\ell_c^t>0$ for  $c\in C^t$.  Since each $\ell_j^t=\lambda_j$ for some $j\in C\cup\{x,y\}$, we have \vspace{-0cm}
\begin{align*}
\Pr\left\{ \lambda_c^t<\ell_c^t, \forall c\in C^t,  \ell_i^t<\lambda_i^t, \forall i=a,b\right\}
\ge \Pr\left\{U^{(C-2: 1)}<\lambda_* \mbox{ and } U^{(2:2)}> \lambda^* \right\}=: \beta(\delta),\vspace{-0cm}
\end{align*} 
where $\lambda_*:=\min\{\lambda_i: i \in C\cup\{x,y\}, \lambda_i>0\}$,  $ \lambda^*:=\max\{\lambda_i: i \in C\cup\{x,y\}, \lambda_i<1\}$, and $U^{(n:m)}$ is the $m$-th highest value of $n$ independent draws of $U[0,1]$.  The proof is complete upon noting that the lower bound $\beta(\delta)$ is independent of $t$ and of particular pair $(a,b)$ or other $c\in C^t$. 
		
 \paragraph{Structure (c):}   
 The argument is similar to that of Structure (b).   
 
 We first consider a model  similar to the NYC Specialized High Schools, where the ex-post score is one-dimensional. The argument is similar to that of Structure (b): STB.  In particular, one can show that the LHS of (\ref{eq:(iii)}) is lower bounded by the probability that one's score lies within a certain interval:\vspace{-0cm}
 \begin{equation} \label{eq:(iii)-exam}
 \Pr\left\{\max\{p_a^t, p_b^t\}<s^t< \min_{c\in C^t} p_c^t \right\}.\vspace{-0cm}
 \end{equation} 
Similarly to (b), if $\Pr\{s_c^t<p_c, \forall c\in C^t, 
s_a^t>p_a\}>0$ and $\Pr\{s_c^t<p_c, \forall c\in C^t, s_b^t>p_b \}>0$, then we must have 
$\max\{p_a^t, p_b^t\}< \min_{c\in C^t} p_c^t$.  Hence,  (\ref{eq:(iii)-exam}) is lower bounded by\vspace{-0cm}
$$\beta(\delta):= \varrho \cdot \min\{\delta \wedge |p_c-p_{c'}|: c,c'\in \hat C, p_c\ne p_{c'}  \}>0,\vspace{-0cm}$$
where $\hat C:=C\cup\{x,y\}$ with $p_x:=0, p_y:=1$ and 
$\varrho:=\min_{t\in T}\min_{s\in [\sl_c^t,\sh^t]}\phi^t(s)>0.$
The positivity of $\varrho$ follows from the full and compact support assumption for each $t$ and the compactness of $T$.  In words, the probability lower bound is given by the shortest length of distinct score cutoffs or by $\delta$, whichever is smaller. The relevance of $\delta$ here comes from the fact that, due to the $\delta$-rejectability or $\delta$-acceptability of the schools,  the support of $t$'s score in the interval $(\max\{p_a^t, p_b^t\}, \min_{c\in C^t} p_c^t)$ spans at least the length of $\delta$, whenever that interval exceeds $\delta$ in length. The proof is complete upon noting that the lower bound $\beta(\delta)$ is independent of $t$ and of particular pair $(a,b)$ or other $c\in C^t$. 
 
 We next consider a model where ex-post scores are not perfectly correlated. The argument is similar to that of Structure (b): MTB.  Again, one can show that  \vspace{-0cm}
\begin{align*}
&\Pr\{s_c^t<p_c, \forall c\in C^t, 
		\mbox{ and }  s_a^t>p_a, s_b^t>p_b \} \\
\ge &\Pr\{s_c^t<p_*, \forall c\in C^t, 
		\mbox{ and }  s_a^t>p^*, s_b^t>p^* \} \\		
		\ge&  \hat\varrho^C \cdot (\delta \wedge p_*)^{C-2} (\delta \wedge(1- p^*))^2 =: \beta(\delta),\vspace{-0cm}
\end{align*} 
where  $p_*:=\min\{p_i: i \in C\cup\{x,y\}, p_i>0\}$,  $p^*:=\max\{p_i: i \in C\cup\{x,y\}, p_i<1\}$, and $\hat\varrho:=\min_{t\in T}\min_{s\in S^t}\phi^t(s)>0.$  Again, the appearance of $\delta$ in the lower bound follows from the fact that  the support of $t$'s score for $i=a,b$ in the interval $[0, p_i]$ and its scores for $c\in C^t$ in the interval $[p_c,1]$  each spans  $\delta$ in length, whenever each interval exceeds $\delta$ in length, again due to the $\delta$-acceptability  and $\delta$-rejectability of these schools.  The proof is complete upon noting that the lower bound $\beta(\delta)$ is independent of $t$ and of particular pair $(a,b)$ or other $c\in C^t$. 
\end{proof}

\section{Multiple Equilibria, Completeness, and Coherency \label{appendix:completeness}}

This appendix shows that our \teps{all} procedure does not suffer from incompleteness or incoherence in the sense of \cite{Tamer2003incomplete}. Incompleteness in our context would mean that the mapping from a student's type, ($u,t$), to \teps{} inferred preferences is a correspondence, which may cause the point identification of the distribution of $u$ to fail. Meanwhile, incoherency would imply that the model does not have a well-defined likelihood for \teps{} inferred preferences given exogenous variables, implying certain logical inconsistency.  

We start with some definitions. Recall that we consider a student with submitted ROL $R$ and intrinsic priorities $t$. We also make it explicit that the preferences inferred by \teps{all} depend on $R$ and $t$, i.e., $\pref{all}(R,\ip)$. 

We say ROL $R'$ is {\it consistent} with $\pref{all}(R,\ip)$ if $R'$ satisfies two conditions: (i) every ever-assigned school $c = \assign_\unc$ for some $\unc\subseteq\uncset$ is included in $R'$, and (ii) for any $c'$ ranked above $c$ in $R'$, $c'$ is {\it not} inferred worse than than $c$, i.e., $(c,c')\notin \pref{all}(R,\ip)$.  This implies that we may exclude a never-feasible school from $R'$ or insert it in $R'$ at any position; however, for an ever-feasible school that is included in $R'$, its position in $R'$ must respect the inferred preferences.  Let $\R^{*}(R,\ip)$ be the set of all ROLs that are consistent with ${\pref{all}(R,\ip)}$.   It can be verified that $\R^{*}(R,\ip)$ includes the student's true preference order given the assumption of stability and transitivity.  

Further, let $\cU^{*}(R,\ip) \subseteq [\underline{u},\overline{u}]^C$ be all the utility types that are consistent with ${\pref{all}(R,\ip)}$.  That is, if $u \in \cU^{*}(R,\ip) $,  the associated true preference order $\rho(u)$ is in $\R^{*}(R,\ip)$.

\begin{prop}\label{prop:coherent}
Assume that in every realized uncertainty, each student cannot change her own set of feasible schools and that the stable matching is unique and achieved. 
Suppose that a student of type $(u,\ip)$ submits a ROL $R$. 
We have the following results:

\noindent(i) {\textit{\textbf{Equivalent class of ROLs.}}} When her ROL $R$ is replaced by $R'$, the student receives her stable assignment in every realized uncertainty if and only if  $R'\in \R^{*}(R,\ip)$.  

\noindent(ii) {\textbf{\textit{Completeness.}}}  
$ \pref{all}(R,\ip) = \pref{all}(R',\ip) $, $\forall R' \in  \R^{*}(R,\ip)$.  That is, given stability, \teps{all} infers a unique set of preference relations for the student even if she submitted any ROL in $\R^{*}(R,\ip)$. 

\noindent(iii) {\textbf{\textit{Coherency.}}} \teps{all} infers $ \pref{all}(R,\ip) $ if and only if $u \in \cU^{*}(R,\ip)$. 

\end{prop}

\begin{proof}
	We prove the three statements one by one. 
	
	\noindent(i) We first prove sufficiency and then necessity. 
		 		 
		 		 \medskip
		 	
	\noindent {\bf Sufficiency.}  Suppose $R'\in \R^{*}(R,\ip)$.  In any given realized uncertainty, by the definition of $\R^{*}(R,t)$, $R'$ must rank the assigned school above any other {simultaneously} feasible schools, while the assigned school is included in $R'$. The student must be assigned the same school regardless of submitting $R$ or $R'$, because she cannot change her set of feasible schools in this realized uncertainty. Hence, she always obtains her stable assignment.

		 			 \medskip
		 			 
		 	\noindent {\bf Necessity.}  	  Suppose that $R$ and $R'$ give the student the same stable assignment in every realized uncertainty. We argue that  $R'$ is in $\R^{*}(R,\ip)$.  For any $c$ that is the assigned school in a realized uncertainty, $c$ must be ranked in $R'$.  Therefore, to show   $R'\in\R^{*}(R,\ip)$, we only need to prove that for any $c'$ ranked above $c$ in $R'$, $c'$ is {\it not} inferred worse than $c$, or $(c,c')\not\in\pref{all}(R,\ip)$. Suppose on the contrary that $c'$ ranked above $c$ in $R'$ and $(c,c')\in\pref{all}(R,\ip)$. For TEPS to infer that $c'$ is worse than $c$ given that $R$ is submitted, there must exist $c^1,\ldots, c^J$ for $1 < J \leq C$ with $c^1=c$ and $c^J=c'$ such that, for each $1<j \leq J$, there exists a realized uncertainty in which $c^{j-1}$ is the assigned school while $c^j$ is feasible.  This implies that $c^1,\ldots, c^{J-1}$ 
		 	must be included in $R'$.  Since $c^J=c'$ is ranked above $c^1=c$ in $R'$, $R'$ must rank $c^{j^*}$ above $c^{j^*-1}$ for some $1<j^*\leq J$.  Because the student cannot influence her own feasible set, there must exist a realized uncertainty in which the student's assignment when submitting $R'$ is $c^{j^*}$, in contrast to $c^{j^*-1}$ which is the assignment when submitting $R$. This contradiction lets us conclude that $R'$ must be consistent with  $\pref{all}(R,\ip)$, or equivalently, $R'\in \R^{*}(R,\ip)$. 		 

		 \medskip
		 
\noindent(ii)  
By part~(i), the set of ever-assigned schools is the same under $R$ or $R'$. Hence, for any $c$ which is never assigned to the student, there does not exist $(c,c'),\forall c'\in C$ such that $(c,c')\in\pref{all}(R,\ip)$ or $(c,c')\in\pref{all}(R',\ip)$. 

Consider $c'$ such that $(c,c') \in \pref{all}(R,\ip)$ for some $c$ that is assigned to the student in some realization of uncertainty. When $R$ is submitted,  there must exist $c^1,\ldots, c^J$ for $1< J \leq C$ with $c^1=c$ and $c^J=c'$ such that, for each $1<j \leq J$, there exists a realized uncertainty in which $c^{j-1}$ is the assigned school while $c^j$ is feasible.  Together with the assumption that $i$ cannot change her own feasible set, part~(i) implies that in each of those same realized uncertainties associated with $R$, $c^{j-1}$ remains the assigned school while $c^j$ is still feasible for each $1<j \leq J$ even when $R'$ is submitted. Hence, $(c,c')\in \pref{all}(R',\ip)$. 
Similarly, we can show that for any $c'$ such that $(c,c')\in\pref{all}(R',\ip)$ for some $c$ that is assigned to the student in
some realization of uncertainty, $(c,c') \in \pref{all}(R,\ip)$.  

Taken together, $ \pref{all}(R,\ip) = \pref{all}(R',t) $. 
		 \medskip
		 
\noindent(iii) We first prove sufficiency and then necessity. 

\medskip

\noindent {\bf Sufficiency.}  For $u \in \cU^{*}(R,\ip)$, to guarantee the unique stable matching in each realized uncertainty, part~(i) implies that the student must submit a ROL in $\R^{*}(R,\ip)$.  By part~(ii), TEPS infers $ \pref{all}(R,\ip) $ for the student. 

\medskip

\noindent {\bf Necessity.}   Suppose that TEPS infers $ \pref{all}(R,\ip) $ for the student while $u \notin \cU^{*}(R,\ip)$. By the definition of $\cU^{*}(R,\ip)$, the student's true preference order, $\rho(u)$, is not in  $\R^{*}(R,\ip)$. Part~(i) implies that the student's assignment from submitting $\rho(u)$ is not the same as her assignment from submitting $R$, while $R$ gives her the unique stable assignment. This implies that reporting truthfully leads to an unstable assignment for her, contradicting the property of the DA mechanism. 

\end{proof}

The assumptions in \Cref{prop:coherent} are justified by our \Cref{thm:stability} in large markets.  When the market becomes large, a student's impact on cutoffs and thus her own feasible sets becomes negligible and the matching in any realized uncertainty is virtually stable and unique. 

\begin{figure}[h!]
    \centering
    \includegraphics[width=0.6\textwidth]{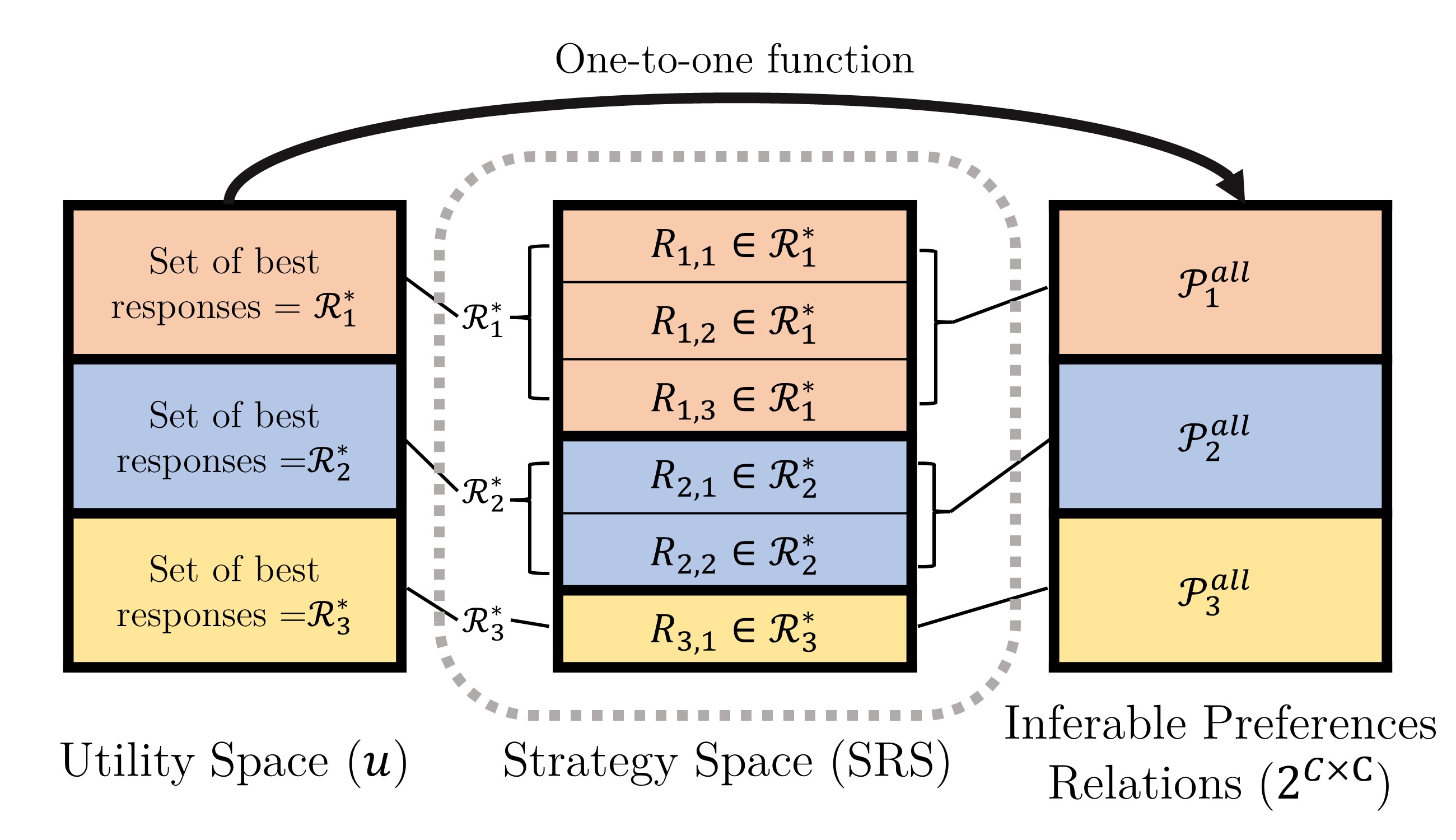}
    \caption{Completeness and Coherency \label{fig:completeness}}
\end{figure}

Part~(i) implies that, although stability does not predict a unique ROL for every student, it predicts a unique class of outcome-equivalent ROLs, $\R^{*}(R,t)$.  Further, part~(ii) shows that \teps{all} maps a student's true preferences into a unique set of inferred preferences, regardless of which ROL from $\R^{*}(R,t)$ the student submits.  Finally, part~(iii) indicates that our model is coherent and that the likelihood of a set of \teps{all} inferred preferences can be written as the likelihood of the student's cardinal preferences satisfying a set of conditions. 

\Cref{fig:completeness} illustrates the logic behind the proposition. Following part (i) of \Cref{prop:coherent}, we divide the action space into equivalent classes of ROLs that lead to the same distribution of stable assignments (middle panel). As part~(ii) implies, each equivalence class of ROLs uniquely maps into a set of preference relations inferred by \teps{} (right panel). 
Finally, although the mapping between the utility space and the action space is a correspondence, the mapping between the utility space and the space of possible sets \teps{} inferred preferences is a function as implied by part (iii).

\section{Performance of Transitive Extension of Preferences from Stability: Monte Carlo Simulation\label{appendix:MC}}
\appcaption{Appendix \ref{appendix:MC}: Performance of Transitive Extension of Preferences from Stability: Monte Carlo Simulation}
This section describes the Monte Carlo simulations that we perform to analyze the implications of our theoretical results.

\subsection{Model Specification}
Consider a finite economy in which $k=1000$ students apply to $C=12$ schools for admission. The vector of school capacities is specified as follows:
\begin{align*}
	\{S_c\}_{c=1}^{12} =\{110,50,100,100,50,100,100,50,100,100,50,100 \}
    \vspace{-0cm}
\end{align*}
The total capacity is set to be larger than the total number of students by 10 in order to ensure all students are matched to some school in the simulations.

The economy is located in an area within a circle of radius 1. Students are uniformly distributed in the circle and schools are evenly located on another circle of radius 1/2 around the center. Denote the Euclidean distance between student $i$ and school $c$ as $d_{i,c}$.

Students are matched with schools through a student-proposing DA algorithm with single tie-breaking (DA-STB), similar to the NYC high school choice. Students submit a ROL of schools that can include any number of available schools. We assume all schools are acceptable to all students; hence, student~$i$ submits a ROL including all 12 schools if she truthfully reports her preferences.

School priorities over students are coarse. Each school has 4 categories of school-specific (intrinsic) priority groups 0, 1, 2, and 3, with a larger number indicating a higher priority. Denote the priority group that student~$i$ belongs at school~$c$ as $t_{i,c}\in\{0,1,2,3\}$. Therefore, school~$c$ prioritizes student~$i$ over $i'$ if $t_{i,c}>t_{i',c}$. A student's priority at a school is drawn independently and uniformly from the four groups.  All students are eligible/acceptable at each school.  Each student knows her priority group at each school at the time of submitting ROL. 

To break ties in priorities, every student is assigned a random lottery number drawn from $\textit{Uniform}[0,1]$, $l_i$ for all $i$. Lottery numbers are not known at the time of submitting ROL. The score of student~$i$ at school~$c$ is $s_{i,c} = \frac{t_{i,c}+l_i}{4} \in [0,1]$. School~$c$ prioritizes student~$i$ over $i'$ if and only if $s_{i,c}>s_{i',c}$.

Student preferences over schools follow a random utility model without an outside option. Student~$i$'s utility from being matched with school $c$ is specified as follows:
\begin{align}\label{eqn:utility}
	u_{i,c}=\beta_1 \times c + \beta_2  (D_i \times A_c) + \beta_3 d_{i,c} + \beta_4 Small_c + \epsilon_{i,c},\quad\forall i,c
    \vspace{-0cm}
\end{align}
where $\beta_1 \times c$ is school $c$'s baseline quality; $d_{i,c}$ is the distance between student~$i$'s and school~$c$; $D_i=1\text{ or }0$ is student~$i$'s type (e.g., disadvantaged or not); $A_c=1\text{ or }0$ is school $c$'s type (e.g., known for resources for disadvantaged students); $Small_c=1$ if $S_c=50$, 0 otherwise; and $\epsilon_{i,c}$ is distributed as $N(0,\sigma_c^2)$ where $\sigma_c^2=1$ for $c=1,\cdots,6$ and $\sigma_c^2=2$ for $c=7,\cdots,12$. $\epsilon_{i,c}$ are independent across all $i$ and $c$.

The type of school $c$, $A_c$, equals 1 if $c$ is an odd number and otherwise 0. The type of student~$i$, $D_i$, is 1 with probability 2/3 among the lowest priority group of school~1 ($t_{i,1}=0$); and $D_i=0$ for all students in highest three priority groups ($t_{i,1}\in\{1,2,3\}$). 

The coefficients of interest are $(\beta_1,\beta_2,\beta_3,\beta_4)$ which are fixed at $(0.3,2,-1,0)$ in the simulations. By this specification, schools with larger indices are of higher quality, and $Small_c$ does not play a role in student preferences over schools. The purpose of estimation is to recover these coefficients and therefore the distribution of preferences.

\subsection{Data Generating Process}
Each simulation sample contains an independent preference profile obtained by randomly drawing $D_i$ and $\{t_{i,c},d_{i,c},\epsilon_{i,c}\}_c$  for all $i$ from the distributions specified above. In all samples, school capacities and school types ($A_c$ and $Small_c$) are kept constant.

\begin{figure}[htbp]
	\centering
	\includegraphics[width=0.4\textwidth]{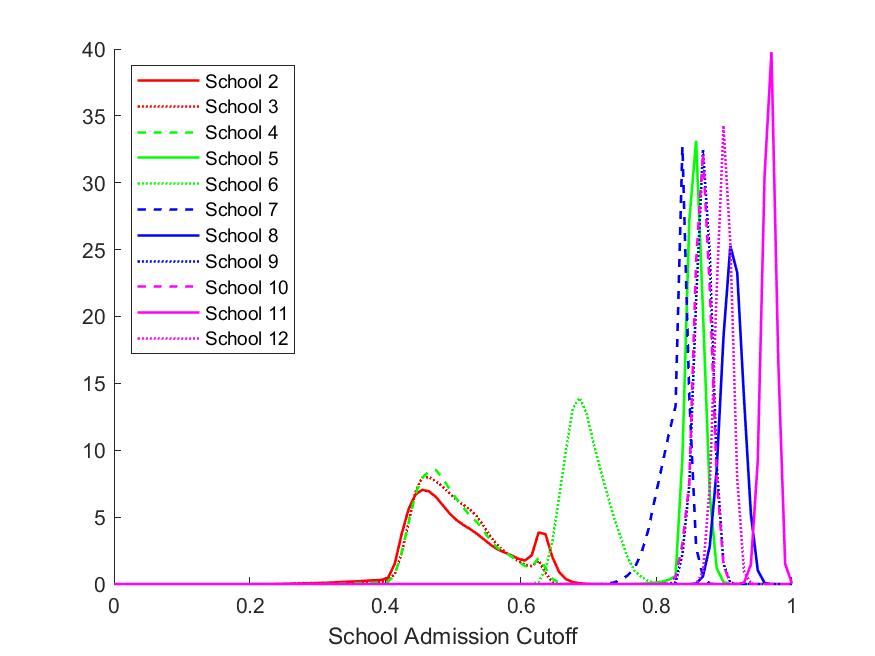}
	\caption{Simulated Cutoff Distribution\label{fig:cutoffs}}
\end{figure}

The first set of simulation samples, \textit{cutoff} samples, are used to simulate the joint distribution of the 12 schools' cutoffs (in terms of score, $s_{i,c}$) by letting every student submit a ROL ranking all schools according to her true preferences. To do so, we simulate 100 samples each consisting of 12 schools and 1,000 students. Each sample contains 1,000 sets of independent draws of tie-breaking lotteries $l_i$. After running the DA algorithm, we calculate the cutoffs in each simulation sample with each draw of the lottery. Figure~\ref{fig:cutoffs} shows the marginal distribution of each school's cutoff from $100\times 1000 = 100,000$ simulated realizations. Note that schools with smaller capacities tend to have higher cutoffs. For example, school 11 with 50 seats often has the highest cutoff, although school 12 with 100 seats has the highest baseline quality. Since every student is guaranteed a seat at some school, school 1 which has the lowest baseline quality has a cutoff equal to 0 (not depicted in the graph).

To generate data on student behavior and admission outcomes for preference estimation, we simulate another $100$ samples, the \textit{estimation} samples, with new independent draws of $T_i$ and $\{p_{i,c},d_{i,c},\epsilon_{i,c}\}_c$.  
For each of the 100 estimation samples, we calculate the distribution of cutoffs which represents the uncertainties in cutoffs that students face. By drawing a set of tie-breaking lotteries, we simulate the uncertainty due to random tie-breaking.\footnote{One may additionally simulate uncertainties in cutoffs due to the finiteness of the economy. To do so, we can randomly generate economies in the spirit of the bootstrap by randomly resampling a set of students and then use a set of tie-breaking lotteries for each resample.}   
The resulting cutoff distributions are used in the Transitive Extension of Preferences from Stability procedure to determine the feasible sets for each realization of the uncertainty. The \textit{estimation} samples are used for the estimation, and in each of them, we consider three types of data-generating processes (DGPs) with different student behaviors. 

\begin{enumerate}[label=(\roman*)]
	\item \textbf{TT (Truth-Telling)} : Every student submits a ROL ranking all 12 schools according to her true preferences.
	
	\item \textbf{MIS-IRR ((Almost) Payoff Irrelevant Mistakes)}: A fraction of students skip schools with which they are never matched according to the simulated distribution of cutoffs. For a given student,	a skipped school can have a high (expected) cutoff and thus be ``out of reach" (i.e., never feasible.) Alternatively, the school may also have a low cutoff, but the student is always accepted by one of her more preferred schools. To specify the fraction of skippers, we first randomly choose about 25.6\% of the students to be never-skippers who always rank all schools truthfully. All other students are potential skippers, and we make all of them skip. Students with $T_i = 1$ are more likely to skip than those with $T_i = 0$, as their scores tend to be lower: 95.6 percent of $T_i = 1$ are potential skippers, compared to 70.1 percent of $T_i = 0$. Finally, we introduce flips by adding back the most preferred school (according to each skipper's true preference) at the end of the ROL if it is never-feasible and thus skipped.  It is important to note that some of the mistakes may turn out to be payoff-relevant, because a never-matched school is determined by the simulated cutoff distributions (which may not exhaust all uncertainties in cutoffs).  \reminder{Students who are never matched to any school may skip all schools, in which case we randomly choose a school for such students so that they submit one-school ROLs. Is this still true?} 

	\item \textbf{MIS-REL (Payoff Relevant Mistakes)}: In addition to MIS-IRR, i.e., given all the potential skippers have skipped the never-matched schools, we now let them make payoff-relevant mistakes. That is, students skip some of the schools with which they have a small chance of being matched according to the simulated distribution of cutoffs. Recall that the joint distribution of cutoffs is only simulated once under the assumption that everyone is truth-telling. We specify a threshold and make the skippers omit the schools at which they have an admission probability lower than the threshold, where the threshold is equal to 10 percent. We allow for flips in the same fashion as in MIS-IRR.
\end{enumerate}
In summary, for each of the 100 estimation samples, we simulate the matching games 3 times: TT, MIS-IRR, and MIS-REL. Table~\ref{table:skipsMC_app} summarizes the scenarios under each DGP. The fraction of students who make mistakes increases from 0\% in \textsf{TT} to 74.4\% in \textsf{MIS-IRR} and \textsf{MIS-REL}. As a result, the fraction of students reporting preferences consistent with WTT is only 27.0\% in \textsf{MIS-IRR} and 28.8\% in \textsf{MIS-REL}. Also, note that stability is satisfied for all students in \textsf{TT} and \textsf{MIS-IRR}, but not in \textsf{MIS-REL} since some students skip schools that are not completely out-of-reach for them.

\begin{table}[h]
	\centering
	\caption{Mistakes in Monte Carlo Simulations (\%)\label{table:skipsMC_app}}
	\resizebox{1.0\textwidth}{!}{
		\begin{tabular}{rccccc}
			\toprule
			\multicolumn{1}{l}{}                                                  & \multicolumn{5}{c}{Scenarios: DGP w/ Different Student Strategies}                                                                                  \\ \cline{2-6}
			\multicolumn{1}{l}{}                                                  & Truth-Telling  &  & \multicolumn{1}{c}{Payoff Irrelevant Mistakes} && \multicolumn{1}{c}{Payoff Relevant Mistakes} \\ \cline{2-2} \cline{4-4} \cline{6-6}
			\multicolumn{1}{l}{}                                                  & TT                                                                   &  &  MIS-IRR & & MIS-REL                                   \\ \midrule
            \multicolumn{1}{l}{Average length of submitted ROLs} & 12                                                                   &  & 6.1  & & 5.0                                 \\
			\multicolumn{1}{l}{WTT: \textit{Weak-Truth-Telling}} & 100                                                                   &  & 27.0  & & 28.8                                 \\
			\multicolumn{1}{l}{Matched w/ favorite feasible school}              & 100                                                                   &  &  100 & & 96.2                                    \\
			\multicolumn{1}{l}{Make Mistakes}                                          & 0                                                                     &  &  74.4    & & 74.4                              \\
			\bottomrule
	\end{tabular}}
    \begin{tabnotes}
        Each entry reported is a percentage that is averaged over the 100 estimation samples. A student is WTT if 1) ROL is in the true preference order, and 2) all ranked schools are more preferred to all unranked schools.
    \end{tabnotes}
\end{table}

\subsection{Estimation and Results}
The random utility model described by Equation~\eqref{eqn:utility} is estimated under two methods, WTT and TEPS 
using Gibbs Sampler, where the procedure is described in the Appendix \ref{appendix:MCMC}. For TEPS estimators, we use \teps{top}, \teps{20}, \teps{40}, \teps{60}, \teps{80}, and \teps{all}. 
Table \ref{table:MCresults} presents the mean and standard deviation of the posterior mean of each parameter across the 100 samples.\footnote{We iterate through the MCMC 100,000 times and discard the first 75,000 for mixing. We calculated the Potential Scale Reduction Factor (PSRF) (\cite{gelman1992inference}) to ensure enough convergence of the posterior distributions.}

We evaluate the performance of the two sets of estimators along two dimensions. The first is the bias-variance tradeoff, focusing on  $\beta_2$ which measures how students of type $T_i=1$ value schools of type $A_c=1$. The second dimension is the performance of the TEPS estimators relative to the TEPS estimators that use smaller sets of inferred preferences, which are both robust to (some) strategic mistakes. 

\begin{table}[h!]
	\centering
	\caption{Estimation with Different Identifying Assumptions: Monte Carlo Results\label{table:MCresults}}
	\resizebox{\textwidth}{!}{
		\begin{tabular}{crcccccccccccccccc}
			\toprule
			\multirow{2}{*}{DGPs}                    & \multicolumn{1}{c}{\multirow{2}{*}{\begin{tabular}[c]{@{}c@{}}Identifying\\ Condition\end{tabular}}} &                               & \multicolumn{3}{c}{Quality ($\beta_1=0.3$)}   &           & \multicolumn{3}{c}{Interaction ($\beta_2=2$)} &           & \multicolumn{3}{c}{Distance ($\beta_3=-1$)}    &           & \multicolumn{3}{c}{Small ($\beta_4=0$)} \\ \cline{4-6} \cline{8-10} \cline{12-14} \cline{16-18} 
			& \multicolumn{1}{c}{}                                                                                 &                               & mean          & s.d.          & $\sqrt{MSE}$  &           & mean          & s.d.          & $\sqrt{MSE}$  &           & mean           & s.d.          & $\sqrt{MSE}$  &           & mean           & s.d.           & $\sqrt{MSE}$  \\ \hline
			\multirow{8}{*}{TT}  & WTT         &  & 0.30 & 0.00 & 0.00 &  & 2.00 & 0.06 & 0.06 &  & -1.00 & 0.03 & 0.03 &  & 0.00  & 0.02 & 0.02 \\
                     & \teps{top}   &  & 0.30 & 0.02 & 0.02 &  & 2.02 & 0.20 & 0.20 &  & -1.03 & 0.11 & 0.12 &  & 0.00  & 0.07 & 0.07 \\
                     & \teps{20} &  & 0.30 & 0.02 & 0.02 &  & 2.02 & 0.20 & 0.20 &  & -1.02 & 0.11 & 0.12 &  & 0.00  & 0.07 & 0.07 \\
                     & \teps{40} &  & 0.30 & 0.02 & 0.02 &  & 2.02 & 0.20 & 0.20 &  & -1.03 & 0.12 & 0.12 &  & 0.00  & 0.07 & 0.07 \\
                     & \teps{60} &  & 0.30 & 0.02 & 0.02 &  & 2.03 & 0.20 & 0.20 &  & -1.02 & 0.11 & 0.11 &  & -0.01 & 0.07 & 0.07 \\
                     & \teps{80} &  & 0.30 & 0.01 & 0.01 &  & 2.02 & 0.17 & 0.17 &  & -1.02 & 0.10 & 0.10 &  & 0.00  & 0.06 & 0.06 \\
                     & \teps{all}        &  & 0.30 & 0.01 & 0.01 &  & 2.01 & 0.12 & 0.12 &  & -1.01 & 0.07 & 0.07 &  & 0.00  & 0.04 & 0.04 \\
                     & Selected    &  & 0.30 & 0.01 & 0.01 &  & 2.00 & 0.07 & 0.07 &  & -1.00 & 0.04 & 0.04 &  & 0.00  & 0.03 & 0.03 \\ 
                     &&&&&&&&&&&&&&&&&\\
\multirow{8}{*}{MIS-IRR} & WTT         &  & 0.08 & 0.00 & 0.22 &  & 1.21 & 0.08 & 0.79 &  & -0.49 & 0.04 & 0.51 &  & -0.19 & 0.02 & 0.19 \\
                     & \teps{top}   &  & 0.30 & 0.02 & 0.02 &  & 2.02 & 0.20 & 0.20 &  & -1.03 & 0.11 & 0.12 &  & 0.00  & 0.07 & 0.07 \\
                     & \teps{20} &  & 0.30 & 0.02 & 0.02 &  & 2.02 & 0.20 & 0.20 &  & -1.03 & 0.11 & 0.12 &  & 0.00  & 0.07 & 0.07 \\
                     & \teps{40} &  & 0.30 & 0.02 & 0.02 &  & 2.02 & 0.20 & 0.20 &  & -1.03 & 0.11 & 0.12 &  & 0.00  & 0.07 & 0.07 \\
                     & \teps{60} &  & 0.30 & 0.02 & 0.02 &  & 2.03 & 0.20 & 0.20 &  & -1.02 & 0.11 & 0.11 &  & -0.01 & 0.07 & 0.07 \\
                     & \teps{80} &  & 0.30 & 0.01 & 0.01 &  & 2.02 & 0.17 & 0.17 &  & -1.02 & 0.09 & 0.10 &  & 0.00  & 0.06 & 0.06 \\
                     & \teps{all}        &  & 0.30 & 0.01 & 0.01 &  & 2.01 & 0.13 & 0.13 &  & -1.01 & 0.07 & 0.07 &  & 0.00  & 0.04 & 0.04 \\
                     & Selected    &  & 0.30 & 0.01 & 0.01 &  & 2.01 & 0.13 & 0.13 &  & -1.01 & 0.08 & 0.08 &  & 0.00  & 0.04 & 0.04\\
                     &&&&&&&&&&&&&&&&&\\
\multirow{8}{*}{MIS-REL}  & WTT        &  & 0.13 & 0.00 & 0.17 &  & 0.99 & 0.07 & 1.01 &  & -0.44 & 0.03 & 0.56 &  & -0.11 & 0.02 & 0.11 \\
 & \teps{top} &  & 0.29 & 0.02 & 0.02 &  & 1.88 & 0.29 & 0.31 &  & -0.98 & 0.14 & 0.14 &  & 0.01  & 0.09 & 0.09 \\
 & \teps{20}  &  & 0.29 & 0.02 & 0.02 &  & 1.88 & 0.29 & 0.31 &  & -0.97 & 0.14 & 0.14 &  & 0.01  & 0.09 & 0.09 \\
 & \teps{40}  &  & 0.29 & 0.02 & 0.02 &  & 1.88 & 0.29 & 0.31 &  & -0.97 & 0.14 & 0.14 &  & 0.01  & 0.09 & 0.09 \\
 & \teps{60}  &  & 0.29 & 0.02 & 0.02 &  & 1.87 & 0.28 & 0.31 &  & -0.96 & 0.13 & 0.14 &  & 0.01  & 0.09 & 0.09 \\
 & \teps{80}  &  & 0.28 & 0.02 & 0.03 &  & 1.82 & 0.25 & 0.30 &  & -0.92 & 0.12 & 0.14 &  & 0.00  & 0.10 & 0.10 \\
 & \teps{all} &  & 0.24 & 0.01 & 0.06 &  & 1.64 & 0.21 & 0.42 &  & -0.77 & 0.08 & 0.24 &  & -0.07 & 0.10 & 0.12 \\
 & Selected   &  & 0.28 & 0.03 & 0.04 &  & 1.82 & 0.27 & 0.32 &  & -0.91 & 0.13 & 0.16 &  & -0.01 & 0.09 & 0.09 \\
			\bottomrule
	\end{tabular}}
	\begin{tabnotes}
		The results are from the 100 Monte Carlo samples.
	\end{tabnotes}
\end{table}

\begin{figure}[ht]
	\centering
	\begin{subfigure}{0.32\textwidth}
		\caption{DGP: Truth-telling}
		\includegraphics[width=\textwidth]{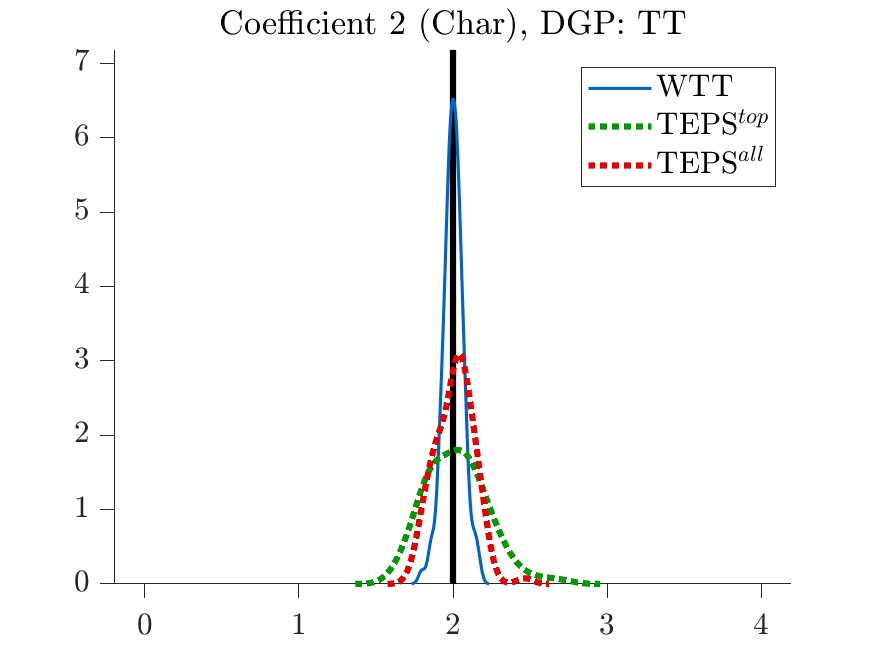}\\
		\includegraphics[width=\textwidth]{figures/TT_coeff_2_select.jpg}
	\end{subfigure}
	\begin{subfigure}{0.32\textwidth}
		\caption{DGP: MIS-IRR}
		\includegraphics[width=\textwidth]{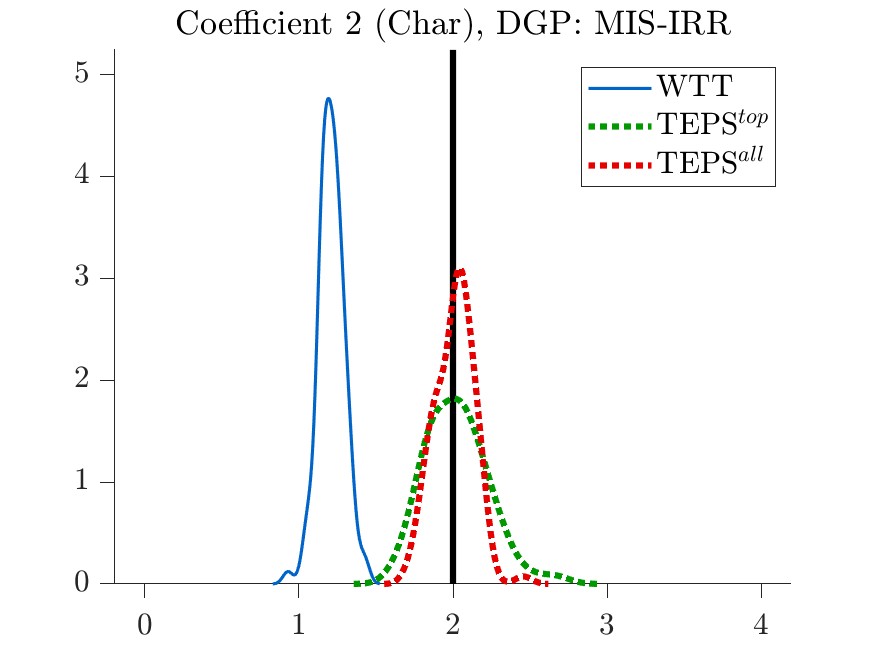}\\
		\includegraphics[width=\textwidth]{figures/MIS-IRR_coeff_2_select.jpg}
	\end{subfigure}
    \begin{subfigure}{0.32\textwidth}
		\caption{DGP: MIS-REL}
		\includegraphics[width=\textwidth]{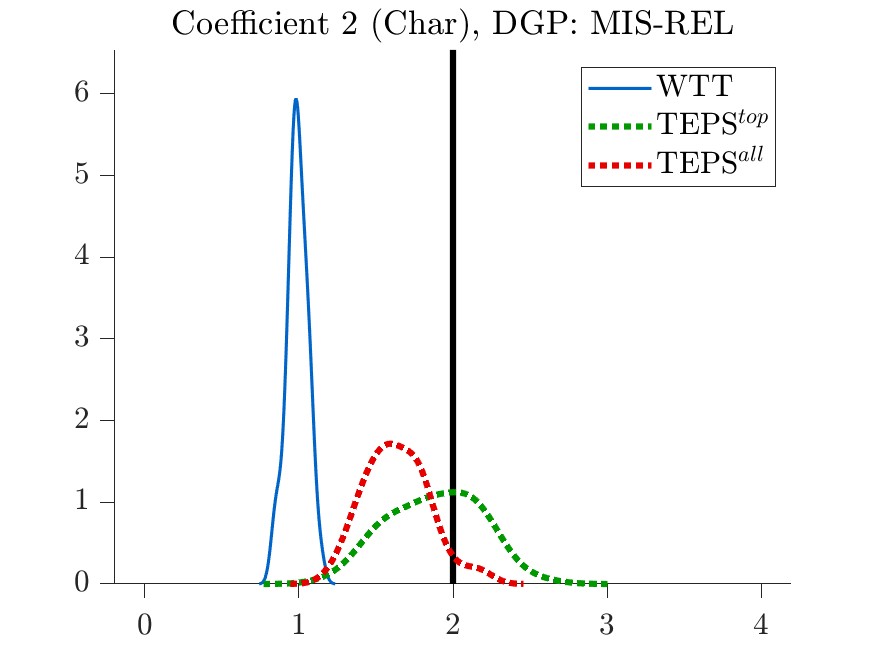}\\
		\includegraphics[width=\textwidth]{figures/MIS-REL_coeff_2_select.jpg}
	\end{subfigure}
	\caption{Distribution of Estimates based on WTT and TEPS ($\beta_2 = 2$)\label{fig:estplots}}
    \begin{tabnotes}
		We plot the kernel density plots of the estimates of $\beta_2$ from 100 Monte Carlo samples. The first column corresponds to when the data-generating process is TT, the second column corresponds to when the data-generating process is MIS-IRR, and the third column corresponds to when the data-generating process is MIS-REL. The black vertical line at 2 denotes the true value of the parameter.
	\end{tabnotes}
\end{figure}

\paragraph{Bias-Variance Tradeoff}
Figure \ref{fig:estplots} plots the distributions of the estimates of $\beta_2$ given each DGP; the true value of $\beta_2$ is 2. The figures plot WTT, \teps{top}, and \teps{all}. 

There are a few notable patterns. First, when the DGP is TT, all estimates are consistent, while the WTT-based estimator has the smallest variance as shown in Panel~(a). This is expected since no students make strategic mistakes violating the WTT or stability assumptions. Furthermore, the WTT-based estimator having the smallest variance is not only true under DGP TT but also under other DGPs. Intuitively, this is due to the fact that WTT uses the maximal (but possibly unreliable) information that one can infer from observed ROLs.

Second, Panel (b) shows the results from MIS-IRR in which some students make (almost) payoff-irrelevant mistakes. The WTT-based estimator is susceptible to strategic mistakes. For example, the WTT-based estimates have a mean 1.21 (standard deviation 0.08). That is, the WTT-based estimator is no longer consistent and the bias is sizable. On the other hand, the estimators based on TEPS are robust to payoff-irrelevant mistakes. 

Finally, Panel (c) show the results from MIS-REL in which some students make payoff-relevant mistakes. Recall that when students make payoff-relevant mistakes, first, WTT is not satisfied and thus WTT-based estimator is inconsistent, and second, stability is also not 100\% satisfied, and thus \teps{all} which does not take care of payoff relevant mistakes is also inconsistent. However, the bias of \teps{all} remains a lot smaller than that of the WTT-based estimator since the violation of WTT (74.4\%) is much more severe than the violation of stability (3.8\%) (see Table~\ref{table:skipsMC_app}). 

The bias-variance tradeoffs are summarized by the square root of the mean squared errors in Table \ref{table:MCresults}. In DGP TT, where all estimators are consistent, the WTT-based estimator attains the minimum mean squared error, as it has the minimum variance where the biases of all estimators are close to zero. However, in DGP MIS-IRR with payoff-irrelevant mistakes, the mean squared error of the WTT-based estimator is larger than that of any other estimator. This is due to the fact that even though the WTT-based estimator has the smallest variance, it is significantly inconsistent. The mean squared errors of the WTT-based estimator are even larger with DGP MIS-REL with payoff-relevant mistakes.

\paragraph{Performance of the TEPS estimator.}
We now compare the relative performance among TEPS-based estimators. As discussed earlier, to the extent that stability is satisfied, all TEPS-based estimators are consistent. However, the procedures by which the TEPS estimators are constructed imply that \teps{top} uses less (but potentially more reliable) information contained in the observed ROLs compared to other TEPS-based estimators. Therefore, \teps{\tau}, $\tau=20, 40, 60, 80, 100$, should have higher precision compared to \teps{top}. Similarly, \teps{\tau} is expected to have higher precision compared to \teps{\tau'} for all $\tau'<\tau$.

Figure~\ref{fig:estplots} shows the results, in which we only present \teps{top} and \teps{all} for clear comparison. \teps{all} is more precise than \teps{top}, having a higher level of concentration around the true value $\beta_2=2$. As reported in Table \ref{table:MCresults}, \teps{\tau}, $\tau=20, 40, 60, 80, 100$ always have (weakly) smaller standard deviations compared to \teps{top} and \teps{\tau'} for all $\tau'>\tau$. For example, in MIS-IRR in which approximately 74.4\%  students make payoff-irrelevant mistakes, the standard deviation of \teps{top} is 0.20, while that of \teps{all} is 0.13. 

Next, in DGP MIS-REL where students skip some schools with a positive admission probability and hence make payoff-relevant mistakes, the choice of TEPS threshold (i.e., how much payoff irrelevant mistakes are to be tolerated) becomes important. The bias of \teps{all} is larger than \teps{top}, \teps{20}, \teps{40}, \teps{60}, and \teps{80}. For example, for $\beta_2=2$, the bias of \teps{all} estimator is 0.36 where those of \teps{top} is 0.12 on average (Table \ref{table:MCresults}). As expected, the bias decreases as we tolerate fewer payoff-relevant mistakes (i.e., as we decrease $\tau$ in \teps{\tau}).

\begin{table}[h!]	
	\centering \footnotesize
	\caption{Determining the Selected Estimator: Test Results (at the $5\%$ significance level)\label{table:testresults}}
	\begin{tabular}{crcccc}
		\toprule
		\multicolumn{2}{r}{{Data Generating Process: }}& & \textsf{TT}   & \textsf{MIS-IRR} & \textsf{MIS-REL}  \\ \midrule
		%
		
		
		\multirow{7}{*}{Estimation method} & &WTT                  			& \textbf{0.94} & 0 & 0   \\
        &&\teps{top}& 0.01 & 0 & 0\\
        &&\teps{20}& 0& 0 & 0\\
        &&\teps{40}& 0& 0 & 0.01\\
        &&\teps{60}& 0& 0.01 & 0.14\\
        &&\teps{80}& 0.03& 0.12 & \textbf{0.68}\\
        &&\teps{all}& 0.01&\textbf{0.87} & 0.17\\
		\bottomrule
	\end{tabular}
	\begin{tabnotes}
		The results are from the 100 estimation samples. 
	\end{tabnotes}
\end{table}

\paragraph{Choosing among the estimation methods.}
Recall the main motivation that led us to introduce the TEPS-based estimator is to ``correct'' the strategic mistakes that students might make in a school choice environment in order to obtain robust estimates. Without the information on how students make mistakes, we follow the procedure described in Section~\ref{sec:tests} to determine the selected estimator that corrects (almost) all the mistakes (thus consistent) and uses the maximum information from the data (thus has the highest precision among all consistent estimators we consider.) Table~\ref{table:testresults} reports the test results based on the testing procedure with a size equal to 0.05, and the resulting `chosen' estimates are reported in the second row of Figure~\ref{fig:estplots} and `Selected' rows of Table~\ref{table:MCresults}. As expected, WTT is chosen 94\% out of 100 estimation samples when the DGP is TT since all estimators are consistent while WTT uses the maximum information (\Cref{prop:nest}). When the DGP is MIS-IRR, \teps{all} is chosen 87\%. Note that \teps{top}, \teps{20}, \teps{40} are never chosen since all TEPS-based estimators are consistent, and \teps{\tau}'s with larger $\tau$ use more information relative to that with smaller $\tau$'s. Finally, when the DGP is MIS-REL, \teps{80} is chosen 68\%. Note that the testing procedure selects \teps{\tau} with $\tau<100$ most of the time since students make payoff-relevant mistakes in MIS-REL and \teps{all} cannot handle payoff-relevant mistakes and thus is inconsistent.


\section{Markov Chain Monte Carlo Procedure for Preference Estimation \label{appendix:MCMC}} 
\appcaption{Appendix \ref{appendix:MCMC}: Markov Chain Monte Carlo Procedure for Preference Estimation}
\subsection{Setup}
There are $k$ students competing for admissions to $C$ schools/programs. Each school $c$ has a type $\tau(c)\in\{1,\cdots,\bar{T}\}$ where $\bar{T}\le C$. Denote the number of schools with type $\tau$ by $C_{\tau}$. WLOG, let schools be ordered in increasing order of type.  Student~$i$'s utility when being admitted to school~$c$ is given by,\vspace{-0cm}
\begin{eqnarray}\label{eq:mc_stu_u}
	U_{i,c}  = X_{i,c} \beta  + \epsilon_{i,c},
    \vspace{-0cm}
\end{eqnarray}
where $\epsilon_{i,c}$ is i.i.d.\ $N(0, \sigma^2_{\tau(c)})$.\footnote{Note that $\sigma_\tau^2$ for some $\tau$ has to be normalized to 1 for identification. WLOG, we set $\sigma_{\bar{T}}^2=1$ for the following.} Then $\Sigma\equiv Var(\epsilon_i)$ is a diagonal matrix with
$(\sigma_1^2,\cdots,\sigma_1^2,\cdots,\sigma_{\bar{T}}^2,$ $\cdots,\sigma_{\bar{T}}^2)$ on the diagonal.  We use $\Sigma$ wherever possible for notational simplicity.

Let $\pref{method}_i$ be the set of all preference relations inferred by some method (for example, WTT, \teps{top}, or \teps{\tau} for some $\tau\in(0,100]$) for student~$i$. 
The procedure we describe below applies to any method for inferring preference relation from choice data. When there is no ambiguity, we simply use $\pref{}_i$. 
Denote the schools that are inferred to be less preferred than $c$ as $\mathcal{L}_{i,c}$ and those are inferred to be preferred to $c$ as $\mathcal{M}_{i,c}$:\vspace{-0cm}
\begin{align*}
    \mathcal{L}_{i,c}&=\{c':(c,c')\in\pref{}_i\text{ i.e., }c'\text{ is revealed to be less preferred than }c\}\\
	\mathcal{M}_{i,c}&=\{c':(c',c)\in\pref{}_i\text{ i.e., }c'\text{ is revealed preferred to }c\}
    \vspace{-0cm}
\end{align*}

\subsection{A Gibbs sampling procedure}

We specify the following diffuse priors:\vspace{-0cm}
\begin{align*}
	\beta&\sim N(0,A^{-1})\\
	\sigma_{\tau}^2&\sim IW(\nu_{\tau},V_{0\tau}),~~\tau=1,2,\cdots,\bar{T}-1
    \vspace{-0cm}
\end{align*}
where\vspace{-0cm}
\begin{align*}
	A^{-1}&=100\cdot I_{dim(\beta)},\\
	\nu_{\tau}&=3+C_{\tau},~~V_{0\tau}=3+C_{\tau},~~\forall\tau
    \vspace{-0cm}
\end{align*}
We go through the following iterative process, a Gibbs sampler. 
\paragraph{Initialization.}  
\begin{enumerate}
	\item Draw $\Sigma^0$ from its prior distribution. 
	\item Draw $\beta^0$ from its prior distribution. 
	\item Draw $U^0$: 
	Following the index of students, $i=1,\ldots, k$, we draw $\{U_{i,c}^0\}_c$ sequentially as follows: for a given $i$, 
	\begin{enumerate}
		\item start with $c=1$. Draw $U_{i,1}^0$ from $N(X_{i,1}\beta^{0}, \Sigma_{1,1}^{0})$;
		\item for each $c=2,\ldots,C$, draw $U_{i,c}^0$ from $N(X_{i,c}\beta^{0}, \Sigma_{c,c}^{0})$ with truncations imposed by $\pref{}_i$. 
		To be specific, let\vspace{-0cm}
		\begin{align*}
			\widetilde{\mathcal{L}}_{i,c}&=\{c':c'<c\text{ and }(c,c')\in \pref{}_i\text{ i.e., }c'\text{ is revealed to be less preferred than }c\}\subseteq \mathcal{L}_{i,c}\\
			\widetilde{\mathcal{M}}_{i,c}&=\{c':c'<c\text{ and }(c',c)\in\pref{}_i\text{ i.e., }c' \text{ is revealed preferred to }c\}\subseteq \mathcal{M}_{i,c}
            \vspace{-0cm}
		\end{align*}
		That is, $\widetilde{\mathcal{L}}_{i,c}$ ($\widetilde{\mathcal{M}}_{i,c}$) is the set of schools whose utility is already drawn,\footnote{Note that we draw $c=1,\cdots,C$ sequentially so that $U_{i,c'}^0,c'<c$ is already drawn in this step when drawing $U_{i,c}^0$. } and at the same time less (more) preferred than school $c$ in $\pref{}_i$ respectively. Then, we draw $U_{i,c}^0$ with truncations from below at $\underline{u}^0_c$, and from above at $\bar{u}^0_c$ where
        \vspace{-0cm}
		\begin{align*}
			\underline{u}^0_c&=\begin{cases}\max\{U_{i,c'}:c'\in\widetilde{\mathcal{L}}_{i,c}\}&\text{if }\widetilde{\mathcal{L}}_{i,c}\neq\emptyset\\
				-\infty&\text{if }\widetilde{\mathcal{L}}_{i,c}=\emptyset\end{cases},~~
			\bar{u}^0_c&=\begin{cases}\min\{U_{i,c'}:c'\in\widetilde{\mathcal{M}}_{i,c}\}&\text{if }\widetilde{\mathcal{M}}_{i,c}\neq\emptyset\\
				\infty&\text{if }\widetilde{\mathcal{M}}_{i,c}=\emptyset\end{cases}
            \vspace{-0cm}
		\end{align*}
		Note that if both $\underline{u}^0_c$ and $\bar{u}^0_c$ are finite, we draw from a two-sided truncated distribution, and if only either one of them is finite, we draw from a one-sided truncated distribution, and if none of them is finite, we draw from the untruncated distribution.
	\end{enumerate}
	
\end{enumerate}

\paragraph{Iteration $r \geq 1$.}  
\begin{enumerate}
	\item Following the index of students, $i=1,\ldots, k$, we draw $\{U_{i,c}^r\}_c$ sequentially as follows.\\
	For a given $i$, for $c=1,\ldots,C$, draw $U_{i,c}^r$ from $N(X_{i,c}\beta^{r-1}, \Sigma_{c,c}^{r-1})$ with truncations imposed by $\pref{}_i$.	To be specific, we draw $U_{i,c}^r$ with truncations from below at $\underline{u}^r_c$, and from above at $\bar{u}^r_c$ where\vspace{-0cm}
	\begin{align*}
		\underline{u}^r_c&=\begin{cases}\max\Big\{\{U^r_{i,c'}:c'\in\widetilde{\mathcal{L}}_{i,c}\}\cup\{U^{r-1}_{i,c'}:c'\in\mathcal{L}_{i,c}\setminus\widetilde{\mathcal{L}}_{i,c}\}\Big\}&\text{if }\mathcal{L}_{i,c}\neq\emptyset\\
			-\infty&\text{if }\mathcal{L}_{i,c}=\emptyset\end{cases}\\
		\bar{u}^r_c&=\begin{cases}\min\Big\{\{U^r_{i,c'}:c'\in\widetilde{\mathcal{M}}_{i,c}\}\cup\{U^{r-1}_{i,c'}:c'\in\mathcal{M}_{i,c}\setminus\widetilde{\mathcal{M}}_{i,c}\}\Big\}&\text{if }\mathcal{M}_{i,c}\neq\emptyset\\
			\infty&\text{if }\mathcal{M}_{i,c}=\emptyset\end{cases}
    \vspace{-0cm}
	\end{align*}
    where $\mathcal{L}_{i,c},\widetilde{\mathcal{L}}_{i,c},\mathcal{M}_{i,c},\widetilde{\mathcal{M}}_{i,c}$ are defined above. In words, for example, if $c$ is revealed preferred to some school by $\pref{}_i$ (i.e., $\mathcal L_{i,c}\neq \emptyset$), the lower bound $\underline{u}^r_c$ is given by the maximum among the $r$-th draws of  utilities of schools that are less preferred than $c$ and  are already drawn in the current step (i.e., $\{U^r_{i,c'}:c'\in\widetilde{\mathcal{L}}_{i,c}\}$) and the $(r-1)$-th draws of utilities of schools that are less preferred than $c$ but not drawn in the current step yet (i.e., $\{U^{r-1}_{i,c'}:c'\in\mathcal{L}_{i,c}\setminus\widetilde{\mathcal{L}}_{i,c}\}$.) $\bar{u}^r_c$ is defined analogously but with schools that are more preferred than $c$ by $\pref{}_i$.
	
	\item Draw $\beta^r$ from the posterior distribution $N(\tilde{\beta},V)$\footnote{Note that $\tilde{\beta}$ is specific to the $r$-th iteration where we omit the dependence on $r$ for notational simplicity.} where
	\begin{align*}
		V&=(X^{*\top}X^*+A)^{-1}\\
		\tilde{\beta}&=VX^{*\top}U^*\\
		(\Sigma^{r-1})^{-1}&=\Lambda^{\top}\Lambda\\
		X_i^*&=\Lambda^{\top}X_i\\
		U_i^*&=\Lambda^{\top}U_i^r\\
		X&=\begin{bmatrix}
			X_1 \\ \vdots \\ X_k
		\end{bmatrix},~~
		U^r=\begin{bmatrix}
			U_1^r \\ \vdots \\ U_k^r
		\end{bmatrix}
	\end{align*}
	and $X_i$ is $C$ by $dim(\beta)$ matrix and $U_i\in\mathbb{R}^C$.
	
	This is a standard result for Bayesian regression with normal errors.
	\item For each $\tau=1,\cdots,\bar{T}-1$, draw $(\sigma_{\tau}^2)^r$ from the posterior distributions $IW(\nu_{\tau}+kC(\tau),V_{0\tau}+S_{\tau})$ where\vspace{-0cm}
	\begin{align*}
		S_{\tau}&=\sum_{c:\tau(c)=\tau}\sum_{i=1}^k \varepsilon_{i,c}^r(\varepsilon_{i,c}^{r})^{\top}\\
		\varepsilon_{i,c}^r &= U_{i,c}^r-X_{i,c}\beta^r
        \vspace{-0cm}
	\end{align*}
	where $\varepsilon_{i,c}^r$, $U^r_{i,c}$ and $X_{i,c}$ denote the part of $\varepsilon_i^r$, $U_i^r$ and $X_i$ corresponding to school $c$. 
	\item Save and pass $(\beta^r,\Sigma^r,U^r)$ to the next iteration.
\end{enumerate}

\section{Data on NYC High School Choice \label{appendix:data}}
\appcaption{Appendix \ref{appendix:data}: Data on NYC High School Choice}

\subsection{Institutional Background}

NYC public high school system consists of two sectors: specialized high schools and regular high schools. There are nine specialized high schools in NYC.\footnote{They are Stuyvesant High School; Brooklyn Technical High School; Bronx High School of Science; High School of American Studies at Lehman College; The Brooklyn Latin School; High School for Mathematics, Science and Engineering at City College; Queens High School for the Sciences at York College; Staten Island Technical High School; and Fiorello H. LaGuardia High School of Music \& Art and Performing Arts. These schools, except for Fiorello H. LaGuardia High School, use Specialized High School Admission Test (SHSAT) as the sole criterion of admission, which is a required exam for students wanting to attend any of the specialized high schools. Fiorello H. LaGuardia High School uses audition as its admission criterion.} We do not consider nine specialized high schools in our analysis because they use different admission methods from the regular high schools, and students  submit a separate ROL of specialized high schools.

Regular high schools are traditional public schools and have six types of programs differentiated in their admission method: \textit{Unscreened}, \textit{Limited unscreened}, \textit{Screened}, \textit{Audition}, \textit{Educational option}, and \textit{Zoned} programs. Multiple programs of different types may be offered by a single school. \textit{Unscreened} programs admit students by a random lottery number attached to each student. \textit{Limited unscreened} programs operate similarly as \textit{Unscreened} programs but give higher priority to students who attended an information session or open houses. \textit{Screened} programs as well as \textit{Audition} programs rank students by individual assortment of criteria. For example, \textit{Screened} programs use several criteria such as final report-card grade, statewide standardized test scores, and attendance and punctuality. \textit{Audition} programs hold school/program-specific auditions to admit students. \textit{Educational option} is a mixture of unscreened and screened programs. They have the purpose of serving students at diverse academic performance levels and divide students into high (the highest 16\%), middle (the middle 68\%), and low (the lowest 16\%) levels in terms of English Language Arts (ELA) scores. 50\% of the seats in each group are filled using school-specific criteria similarly as a \textit{Screened} program, and the other 50\% are filled randomly similarly as an \textit{Unscreened} program. \textit{Zoned} programs give priority or guarantee admission to students who apply and live in the zoned area of the school. 

Each program has its own eligibility and priority group criteria. For example, in the academic year 2016--17, Young Women's Leadership School in Astoria opened its seats only to female students, i.e., being a female student was the eligibility criterion. Besides, they gave the highest admission priority to continuing eighth graders, then to students or residents in Queens, and then to other NYC residents.

The number of priority groups is a lot smaller than the number of applicants to each program. Hence students who apply to programs that do not actively rank students ---\textit{Unscreened}, \textit{Limited Unscreened}, \textit{Zoned} and the unscreened part of \textit{Educational option}---are often in the same priority group. Hence the ties need to be broken for the SPDA algorithm to run. For this purpose, a random lottery number is drawn and attached to each student, which is used to break ties at all programs that require tie-breaking in the same fashion---single tie-breaking (STB) rule. The lottery number is unknown to the student at the moment of application.

The timeline of the admission process is as follows (\cite{corcoran/levin:11}). 
In October and November, students may apply to specialized high schools for which they should take SHSAT or audition at LaGuardia High School.  In December, they are required to submit up to 12 ranked non-specialized high school programs regardless of their application status to specialized high schools.  In March, the SPDA algorithms for specialized high schools (SHSAT takers) and non-specialized high schools (all students) are separately run, which is called Round 1. The Department of Education sends each student a letter with an offer from regular schools and an offer from specialized high schools, if any.  If a student receives offers from both regular and specialized high schools, he/she must choose one. All students who accept a Round 1 offer have a finalized admission decision.  If a student did not submit an application in Round 1, did not receive an offer in Round 1, or wants to apply to a program with availability, he/she can participate in Round 2. Round 2 takes place in March and operates with students who submitted Round 2 applications and programs that still have seat availability. Round 2 offers automatically replace Round 1 offers, if any. Students who are unassigned in Rounds 1 and 2 or reject the assignment go to the administrative round in which students are administratively assigned a school on a case-by-case basis.

\subsection{Data Source}

The main data that we use is the administrative data from the New York City Department of Education (NYC DOE) for the academic year 2016--17.
There are four sets of data used to construct information on the applicants. First, high school application (HSAP) data contains the submitted ROLs and student information such as ELA and math standardized test scores, English-language Learner (ELL) status, and student priorities at programs (including priority rank, priority criteria, and eligibility). Second, June biographic data provides comprehensive student biographic information, including ethnicity, gender, disability status, as well as information on attendance and punctuality. Third, standardized test data contains more detailed information on statewide standardized exams. Fourth, zoned DBN data provides information on the zoned school of each student, the census tract, and the school district of each student's residence. Finally, Middle School Course and Grade data contains all of the courses and information on credits and grades for each student in a given year. There exists a unique scrambled student ID variable that enables merging all NYC DOE datasets while personally identifying a student is impossible. Lastly, we use the information on each census tract and zip code in NYC obtained from the 2016 American Community Survey 5-Year Estimates from the US Census database.

School information is constructed using the NYC High School Directory which is published every year before the application process starts. This includes each program's capacity in the previous year, the number of students who applied in the previous year, eligibility and priority criteria, accountability data such as progress reports, graduation rate and college enrollment rate, and types of language classes provided. Other variables about the current 9th graders, such as ethnicity composition and the fraction of high-performing students, are constructed using the high school application data from the previous year, the academic year 2015--2016.

\subsection{Sample Restrictions}
We focus on students from Staten Island. There are two different samples---one for tracing out the uncertainties, the other for estimation of student preferences: \textit{Priority Construction} sample and \textit{Estimation} sample. They differ because of missing values in some variables. \textit{Priority Construction} sample is used to reconstruct the priority scores of each student at each school/program. \textit{Estimation} sample is a strict subset of \textit{Priority Construction} sample and is used for preference estimation and counterfactual simulations.

First, \textit{Priority Construction} sample consists of students who applied to at least one Staten Island school/program. Among such 4,824 students, 785 did not have information on variables needed to reconstruct priority and were dropped leaving us 4,039 students. Next, \textit{Estimation} sample consists of students who applied to at least one Staten Island school/program, went to a middle school in Staten Island, and resided in Staten Island at the point of application. Among such 4,480 students, 741 did not have information on variables needed for priority reconstruction or estimation and were dropped. Finally, 8 students had invalid ROLs such as containing an invalid program code on their ROLs, and were dropped,  leaving us 3,731 students.

Finally, we adjust the capacities of each school/program whenever we restrict our sample.  Specifically, we treat the Round 1 assignment in the data (or the Round 2 assignment if applicable) of each dropped student as fixed whenever we resimulate DA using the restricted sample. This is in order to ensure that we do not overestimate the probability of school/programs being feasible for each student in our TEPS procedure.

\section{TEPS for NYC High School Choice \label{appendix:SITE}}
\appcaption{Appendix \ref{appendix:SITE}: TEPS for NYC High School Choice}

\subsection{Constructing Priority Scores}

Before describing how to simulate uncertainty, let us first specify the procedure we take to prepare ingredients for the procedure.  The main inputs for the SPDA algorithm are the capacities of programs, students' preferences, and programs' preferences (priority scores). First, we use adjusted programs' capacities as described in Appendix \ref{appendix:data}. Next, we use students' submitted ROL in the data as students' preferences. 

School preferences are a bit more involved than the other two inputs. As described in Appendix \ref{appendix:data}, NYC public high schools have coarse priority rules. First, for eligibility criteria and admission priority groups which are publicly available before the admission process starts, we use the information listed on the High School Directory. 

Next, we estimate the priority ranks for \textit{Screened}, \textit{Audition}, and the screened part of \textit{Educational option} programs that actively rank students.  While there is information on the priority rank of students in the data set provided by NYC DOE, it is limited only to students who ranked that program. Furthermore, how each program ranks its applicants is not public information, and each program has its individual assortment of criteria. For our purposes to calculate the probability of each program being feasible to a student regardless of whether she ranked it or not, we need to construct a priority rank for all students at each program that actively ranks students.  To do so, we assume that there exists a program-specific latent variable $v_{ij}$ for student $i$ at actively ranking program $j$ which determines the priority ranks:
\vspace{-0cm}
\begin{align*}
	v_{ij} = \beta_j X_i + \varepsilon_{ij} \quad \text{ and }  \quad	i\succ_j i'\text{ if and only if }v_{ij}>v_{i'j},   
    \vspace{-0.5cm}
\end{align*}
where $X_i$ is a vector of student characteristics including 7th Standardized Math and ELA scores, middle school Math, Social Studies, English, Science GPA, days absent and days late, and $\varepsilon_{ij}$ is independent and identically distributed as extreme value type I conditional on $X_i$. We form a log-likelihood by considering all possible pairs of applicants to each program $j$ and estimate via MLE separately for each program. That is for $\mathcal{I}_j$, the set of applicants to program $j$,
\vspace{-0cm}
\begin{align*}
	\bhat_j=\underset{\beta_j}{\arg\max}~l(\beta_j) \equiv \sum_{i>i':i,i'\in\mathcal{I}_j}\log\Big(	\frac{\exp(v_{ij})1\{i\succ_j i'\} + \exp(v_{i'j})1\{i'\succ_j i\}}{\exp(v_{ij})+\exp(v_{i'j})}	\Big)
    \vspace{-0cm}
\end{align*}
With estimates $\bhat_j$, we predict $\hat{v}_{ij}=\bhat_j X_i$ for all students (not limited to applicants) and reconstruct priority ranks based on $\hat{v}_{ij}$. 

\subsection{Simulation of Uncertainties}
We describe how we simulate uncertainties present in the matching environment, the first step of TEPS. 

	After reconstructing priority scores for each student at each program, we simulate $B_L=10,000$ lotteries from a uniform distribution to break ties at non-actively ranking programs and run the SPDA 10,000 times. 
	This procedure would give us an empirical distribution of cutoffs of all programs $P=(P_1,\cdots, P_C)$ where the cutoff of a program is defined to be the lowest priority score of admitted students if the seats are filled, and zero if the seats are not filled. 
	
	Next, we draw $L$ number of lotteries 
	($L=5,000$) 
	from a uniform distribution in order to account for the fact that a student's own ex-post score is uncertain due to the random tie-breaking rules. For each lottery draw, we use the cutoff distribution simulated above to figure out the set of feasible schools and its probability for each student in each realization of uncertainties. Together with the submitted ROLs, we can also compute the assigned program in each realization of uncertainties.

\section{Analysis of the NYC High School Choice Data \label{appendix:nycresults}}
\appcaption{Appendix \ref{appendix:nycresults}: Analysis of the NYC High School Choice Data}

\subsection{Preference Estimates \label{appendix:preferenceestimates}}
Tables \ref{tab:SIpref1}--\ref{tab:SIpref4} present preference estimates for each covariate cell.  We report the mean and standard deviation of the posterior distribution as the point estimate and the standard error. We iterate through the MCMC 1 million times and discard the first 90\% to ensure mixing. We calculate the Potential Scale Reduction Factor (PSRF) using the draws that we keep following \cite{gelman1992inference}. For those that did not converge, we additionally iterate 1 million times and keep the last 0.1 million. The resulting PSRFs for all parameters for all cells are below 1.1 which ensures convergence.

\subsection{Additional Figures\label{app:add}}
\subsubsection{Program Characteristics by Feasibility Status}
\begin{figure}[h!]
	\centering
	\includegraphics[width=1\textwidth]{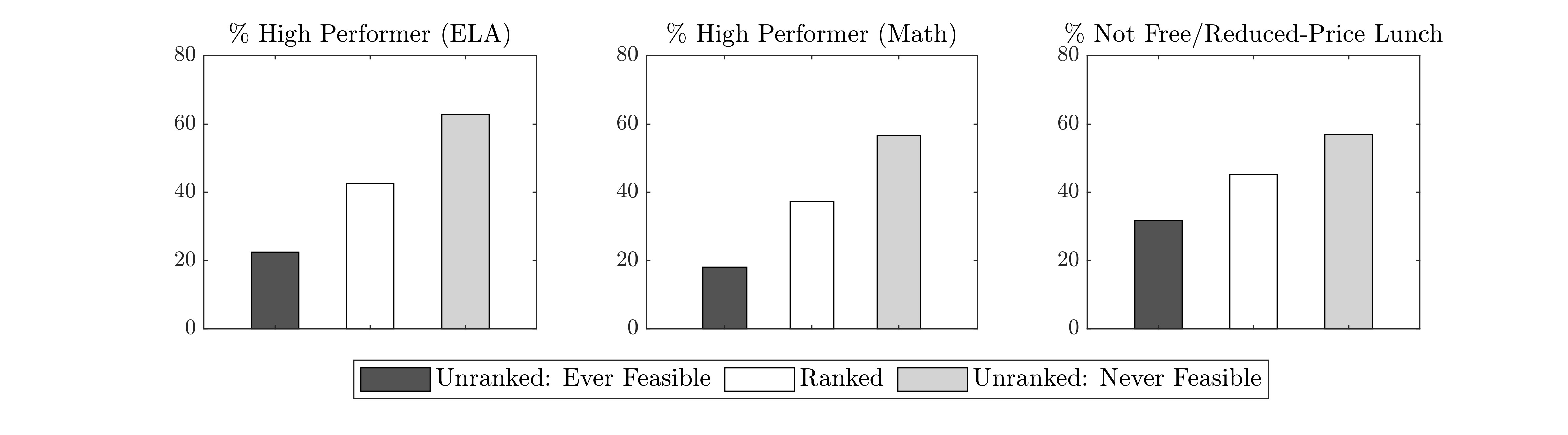}
	\vskip -0.3cm
	\caption{Characteristics of Ranked and Unranked Programs by Feasibility Status\label{fig:progcharbyfeasibility_more}}
	\begin{flushleft}\scriptsize
	\vskip -0.3cm
		Notes: 
		For each student, we classify the programs into three types---ever-feasible-unranked,  never-feasible-unranked, and ranked. Ranked programs are those included in the student's ROL and Stages 1 and 2 of the TEPS procedure  determine the feasibility of each unranked program.   
        We use the fraction of high-performing students (measured by ELA and math scores) and the fraction of those who are not eligible for free/reduced-price lunch in each program. 
        The figure reports the average across all students for each type of program. 
	\end{flushleft}
\end{figure}

\begin{landscape}
\begin{table}[h]
\centering
\footnotesize
\caption{Preference Estimates: Cell 1--2\label{tab:SIpref1}}
\resizebox{1.35\textwidth}{!}{\begin{tabular}{lccccccccccccccccccccccc}
\toprule 
& \multicolumn{11}{c}{Cell 1}&& \multicolumn{11}{c}{Cell 2}\\\cline{2-12}\cline{14-24}&\multicolumn{2}{c}{WTT}&&\multicolumn{2}{c}{\teps{all}}&&\multicolumn{2}{c}{\teps{top}}&&\multicolumn{2}{c}{Selected}&&\multicolumn{2}{c}{WTT}&&\multicolumn{2}{c}{\teps{all}}&&\multicolumn{2}{c}{\teps{top}}&&\multicolumn{2}{c}{Selected}\\\cline{2-3}\cline{5-6}\cline{8-9}\cline{11-12}\cline{14-15}\cline{17-18}\cline{20-21}\cline{23-24} & Coef. & S.E. &  & Coef. & S.E. &  & Coef. & S.E. &  & Coef. & S.E.& & Coef. & S.E. &  & Coef. & S.E. &  & Coef. & S.E. &  & Coef. & S.E. \\ 
\midrule 
School FE 4 & -0.26 & 0.05 &  & -0.45 & 0.08 &  & -0.86 & 0.11 &  & -0.55 & 0.08 &  & 0.11  & 0.05 &  & -0.12 & 0.08 &  & -0.38 & 0.13 &  & -0.12 & 0.08 \\ 
School FE 5 & -0.65 & 0.08 &  & -1.05 & 0.12 &  & -1.45 & 0.17 &  & -1.18 & 0.13 &  & -0.45 & 0.09 &  & -1.3  & 0.15 &  & -1.55 & 0.21 &  & -1.3  & 0.15 \\ 
School FE 6 & -0.24 & 0.08 &  & -0.48 & 0.13 &  & -0.64 & 0.18 &  & -0.53 & 0.14 &  & 0.18  & 0.09 &  & -0.44 & 0.15 &  & -0.61 & 0.22 &  & -0.44 & 0.15 \\ 
School FE 7 & 0     & 0.09 &  & -0.7  & 0.15 &  & -0.87 & 0.21 &  & -0.73 & 0.16 &  & -0.4  & 0.1  &  & -1.43 & 0.18 &  & -1.58 & 0.25 &  & -1.43 & 0.18 \\ 
School FE 8 & -0.29 & 0.06 &  & -0.39 & 0.09 &  & -0.47 & 0.12 &  & -0.43 & 0.09 &  & -0.2  & 0.06 &  & -0.48 & 0.09 &  & -0.58 & 0.13 &  & -0.48 & 0.09 \\ 
School FE 9 & -0.03 & 0.08 &  & -0.27 & 0.13 &  & -0.59 & 0.18 &  & -0.39 & 0.14 &  & -0.69 & 0.11 &  & -1.41 & 0.21 &  & -1.51 & 0.3  &  & -1.41 & 0.21 \\ 
$\mathds{1}$(STEM) &       &      &  &       &      &  &       &      &  &       &      &  &       &      &  &       &      &  &       &      &  &       &      \\ 
\multicolumn{1}{r}{Main Effect} & 0.22  & 0.06 &  & 0.33  & 0.11 &  & 0.44  & 0.16 &  & 0.3   & 0.13 &  & -0.09 & 0.07 &  & 0.07  & 0.14 &  & -0.26 & 0.25 &  & 0.07  & 0.14 \\ 
\multicolumn{1}{r}{$\times$ELA Score} & -0.22 & 0.05 &  & -0.19 & 0.08 &  & -0.14 & 0.1  &  & -0.14 & 0.08 &  & -0.06 & 0.06 &  & -0.13 & 0.1  &  & -0.2  & 0.15 &  & -0.13 & 0.1  \\ 
\multicolumn{1}{r}{$\times$Math Score} & 0.27  & 0.06 &  & 0.24  & 0.09 &  & 0.29  & 0.12 &  & 0.21  & 0.1  &  & 0.16  & 0.06 &  & 0.08  & 0.1  &  & 0.4   & 0.16 &  & 0.08  & 0.1  \\ 
\% High Perf. (ELA) &       &      &  &       &      &  &       &      &  &       &      &  &       &      &  &       &      &  &       &      &  &       &      \\ 
\multicolumn{1}{r}{Main Effect} & -0.13 & 0.15 &  & -0.86 & 0.23 &  & -1.07 & 0.34 &  & -0.73 & 0.26 &  & 0.16  & 0.16 &  & -0.37 & 0.25 &  & -0.48 & 0.35 &  & -0.37 & 0.25 \\ 
\multicolumn{1}{r}{$\times$ELA Score} & 0.32  & 0.07 &  & 0.4   & 0.15 &  & 0.49  & 0.2  &  & 0.38  & 0.16 &  & 0.42  & 0.08 &  & 0.35  & 0.18 &  & 0.56  & 0.25 &  & 0.35  & 0.18 \\ 
\% High Perf. (Math) &       &      &  &       &      &  &       &      &  &       &      &  &       &      &  &       &      &  &       &      &  &       &      \\ 
\multicolumn{1}{r}{Main Effect} & 0.46  & 0.2  &  & 1.49  & 0.33 &  & 1.58  & 0.51 &  & 1.36  & 0.36 &  & 0.7   & 0.22 &  & 2.31  & 0.4  &  & 2.96  & 0.6  &  & 2.31  & 0.4  \\ 
\multicolumn{1}{r}{$\times$Math Score} & 0.91  & 0.08 &  & 0.71  & 0.17 &  & 0.84  & 0.25 &  & 0.77  & 0.19 &  & 1     & 0.09 &  & 1.12  & 0.21 &  & 0.7   & 0.27 &  & 1.12  & 0.21 \\ 
\% FRPL (program) &       &      &  &       &      &  &       &      &  &       &      &  &       &      &  &       &      &  &       &      &  &       &      \\ 
\multicolumn{1}{r}{Main Effect} & 1.07  & 0.18 &  & -0.59 & 0.35 &  & -0.23 & 0.52 &  & -0.64 & 0.39 &  & 0.08  & 0.2  &  & -2    & 0.37 &  & -1.89 & 0.52 &  & -2    & 0.37 \\ 
\multicolumn{1}{r}{$\times$Med Income} & -0.06 & 0.15 &  & 0.17  & 0.28 &  & 0.17  & 0.4  &  & 0.32  & 0.31 &  & 0.06  & 0.15 &  & -0.03 & 0.3  &  & -0.23 & 0.42 &  & -0.03 & 0.3  \\ 
\multicolumn{1}{r}{$\times$Avg Score} & -0.62 & 0.14 &  & -0.22 & 0.28 &  & -0.03 & 0.44 &  & -0.18 & 0.31 &  & 0.04  & 0.14 &  & 0.19  & 0.32 &  & 0.04  & 0.46 &  & 0.19  & 0.32 \\ 
\% FRPL (school) &       &      &  &       &      &  &       &      &  &       &      &  &       &      &  &       &      &  &       &      &  &       &      \\ 
\multicolumn{1}{r}{$\times$Med Income} & 0.08  & 0.22 &  & 0.35  & 0.35 &  & 0.44  & 0.48 &  & 0.25  & 0.39 &  & 0.58  & 0.24 &  & 0.53  & 0.4  &  & 1.05  & 0.56 &  & 0.53  & 0.4  \\ 
\multicolumn{1}{r}{$\times$Avg Score} & 0.39  & 0.18 &  & -0.42 & 0.32 &  & -0.83 & 0.45 &  & -0.38 & 0.34 &  & -0.43 & 0.2  &  & -0.92 & 0.39 &  & -1.03 & 0.51 &  & -0.92 & 0.39 \\ 
9th Grade Size (100s) & 0.16  & 0.01 &  & 0.19  & 0.02 &  & 0.23  & 0.02 &  & 0.19  & 0.02 &  & 0.17  & 0.01 &  & 0.25  & 0.02 &  & 0.29  & 0.03 &  & 0.25  & 0.02 \\ 
\% Asian & 0.68  & 0.31 &  & 0.75  & 0.51 &  & 0.64  & 0.71 &  & 0.75  & 0.55 &  & -2.27 & 0.35 &  & -3.84 & 0.63 &  & -3.5  & 0.89 &  & -3.84 & 0.63 \\ 
\% Black & -1.25 & 0.21 &  & -0.62 & 0.33 &  & -0.49 & 0.46 &  & -0.6  & 0.36 &  & -0.81 & 0.23 &  & 1.17  & 0.41 &  & 1.13  & 0.62 &  & 1.17  & 0.41 \\ 
\% Hispanic & -0.75 & 0.21 &  & -0.66 & 0.33 &  & -0.71 & 0.45 &  & -0.51 & 0.36 &  & -1.05 & 0.22 &  & 0.25  & 0.38 &  & 0.68  & 0.53 &  & 0.25  & 0.38 \\ 
1(Nearest School) & 0.28  & 0.04 &  & 0.26  & 0.07 &  & 0.26  & 0.09 &  & 0.27  & 0.07 &  & 0.2   & 0.05 &  & 0.26  & 0.07 &  & 0.24  & 0.11 &  & 0.26  & 0.07 \\ 
Distance & -0.26 & 0.01 &  & -0.3  & 0.02 &  & -0.3  & 0.02 &  & -0.3  & 0.02 &  & -0.34 & 0.01 &  & -0.35 & 0.02 &  & -0.42 & 0.04 &  & -0.35 & 0.02 \\ 
$\sigma^2_{STEM}$ & 0.86  & 0.07 &  & 0.75  & 0.11 &  & 0.73  & 0.15 &  & 0.85  & 0.15 &  & 0.99  & 0.08 &  & 1.06  & 0.17 &  & 1.53  & 0.34 &  & 1.06  & 0.17 \\ 
\\ 
\bottomrule 
\end{tabular}}
\begin{flushleft}
\scriptsize
Notes: We report the full estimates of the parameters in \eqref{eqn:utility_SI}.  All \% variables are within $[0,1]$. For conciseness, we report only the estimates based on WTT, \teps{all}, \teps{top}, and the selected estimates obtained by following the procedure described in Section~\ref{sec:tests}. We report the mean and standard deviation of the posterior distribution as the point estimate and the standard error.
\end{flushleft}
\end{table}
\clearpage
\begin{table}[h]
\centering
\footnotesize
\caption{Preference Estimates: Cell 3--4\label{tab:SIpref2}}
\resizebox{1.35\textwidth}{!}{\begin{tabular}{lccccccccccccccccccccccc}
\toprule 
& \multicolumn{11}{c}{Cell 3}&& \multicolumn{11}{c}{Cell 4}\\\cline{2-12}\cline{14-24}&\multicolumn{2}{c}{WTT}&&\multicolumn{2}{c}{\teps{all}}&&\multicolumn{2}{c}{\teps{top}}&&\multicolumn{2}{c}{Selected}&&\multicolumn{2}{c}{WTT}&&\multicolumn{2}{c}{\teps{all}}&&\multicolumn{2}{c}{\teps{top}}&&\multicolumn{2}{c}{Selected}\\\cline{2-3}\cline{5-6}\cline{8-9}\cline{11-12}\cline{14-15}\cline{17-18}\cline{20-21}\cline{23-24} & Coef. & S.E. &  & Coef. & S.E. &  & Coef. & S.E. &  & Coef. & S.E.& & Coef. & S.E. &  & Coef. & S.E. &  & Coef. & S.E. &  & Coef. & S.E. \\ 
\midrule 
School FE 4 & -0.35 & 0.06 &  & -0.44 & 0.08 &  & -0.55 & 0.13 &  & -0.49 & 0.1  &  & -0.01 & 0.06 &  & -0.12 & 0.08 &  & -0.42 & 0.14 &  & -0.39 & 0.13 \\ 
School FE 5 & -0.53 & 0.09 &  & -1.06 & 0.13 &  & -1.18 & 0.19 &  & -1.21 & 0.16 &  & -0.42 & 0.09 &  & -1.17 & 0.14 &  & -1.36 & 0.21 &  & -1.38 & 0.2  \\ 
School FE 6 & -0.5  & 0.09 &  & -0.61 & 0.14 &  & -0.72 & 0.2  &  & -0.8  & 0.17 &  & -0.08 & 0.09 &  & -0.72 & 0.15 &  & -0.86 & 0.22 &  & -0.95 & 0.22 \\ 
School FE 7 & 0.18  & 0.11 &  & -0.47 & 0.17 &  & -0.43 & 0.24 &  & -0.51 & 0.21 &  & -0.27 & 0.11 &  & -1.03 & 0.2  &  & -1.3  & 0.29 &  & -1.19 & 0.27 \\ 
School FE 8 & -0.34 & 0.07 &  & -0.42 & 0.1  &  & -0.38 & 0.14 &  & -0.44 & 0.12 &  & -0.15 & 0.06 &  & -0.54 & 0.1  &  & -0.58 & 0.15 &  & -0.62 & 0.15 \\ 
School FE 9 & -0.42 & 0.09 &  & -0.57 & 0.14 &  & -0.53 & 0.2  &  & -0.64 & 0.17 &  & -0.76 & 0.11 &  & -1.63 & 0.21 &  & -1.7  & 0.31 &  & -1.79 & 0.3  \\ 
$\mathds{1}$(STEM) &       &      &  &       &      &  &       &      &  &       &      &  &       &      &  &       &      &  &       &      &  &       &      \\ 
\multicolumn{1}{r}{Main Effect} & 0.19  & 0.09 &  & 0.27  & 0.14 &  & -0.02 & 0.28 &  & -0.07 & 0.24 &  & -0.26 & 0.1  &  & -0.29 & 0.2  &  & -0.54 & 0.35 &  & -0.41 & 0.32 \\ 
\multicolumn{1}{r}{$\times$ELA Score} & -0.04 & 0.07 &  & 0.01  & 0.09 &  & 0.09  & 0.15 &  & 0.13  & 0.13 &  & -0.02 & 0.08 &  & -0.01 & 0.14 &  & -0.22 & 0.19 &  & -0.1  & 0.18 \\ 
\multicolumn{1}{r}{$\times$Math Score} & 0.13  & 0.07 &  & 0.01  & 0.1  &  & 0.07  & 0.16 &  & -0.05 & 0.13 &  & 0.1   & 0.08 &  & -0.04 & 0.13 &  & 0.38  & 0.2  &  & 0.12  & 0.18 \\ 
\% High Perf. (ELA) &       &      &  &       &      &  &       &      &  &       &      &  &       &      &  &       &      &  &       &      &  &       &      \\ 
\multicolumn{1}{r}{Main Effect} & 0.21  & 0.17 &  & -0.26 & 0.26 &  & -0.2  & 0.36 &  & 0     & 0.3  &  & 0.13  & 0.17 &  & -0.2  & 0.26 &  & -0.47 & 0.38 &  & -0.26 & 0.35 \\ 
\multicolumn{1}{r}{$\times$ELA Score} & 0.43  & 0.09 &  & 0.69  & 0.17 &  & 0.76  & 0.24 &  & 0.72  & 0.21 &  & 0.31  & 0.09 &  & 0.14  & 0.19 &  & 0.41  & 0.27 &  & 0.3   & 0.25 \\ 
\% High Perf. (Math) &       &      &  &       &      &  &       &      &  &       &      &  &       &      &  &       &      &  &       &      &  &       &      \\ 
\multicolumn{1}{r}{Main Effect} & -0.34 & 0.24 &  & 0.65  & 0.37 &  & 0.65  & 0.56 &  & 0.69  & 0.44 &  & 0.8   & 0.23 &  & 2.27  & 0.41 &  & 3.12  & 0.63 &  & 2.95  & 0.58 \\ 
\multicolumn{1}{r}{$\times$Math Score} & 0.79  & 0.09 &  & 0.73  & 0.18 &  & 0.97  & 0.27 &  & 0.92  & 0.23 &  & 0.59  & 0.09 &  & 0.9   & 0.21 &  & 0.23  & 0.29 &  & 0.43  & 0.28 \\ 
\% FRPL (program) &       &      &  &       &      &  &       &      &  &       &      &  &       &      &  &       &      &  &       &      &  &       &      \\ 
\multicolumn{1}{r}{Main Effect} & 1.17  & 0.21 &  & -0.39 & 0.35 &  & -1.13 & 0.48 &  & -0.66 & 0.43 &  & 0.29  & 0.2  &  & -0.94 & 0.38 &  & -1.01 & 0.55 &  & -0.52 & 0.52 \\ 
\multicolumn{1}{r}{$\times$Med Income} & 0     & 0.16 &  & 0.14  & 0.27 &  & 0.17  & 0.39 &  & 0.1   & 0.32 &  & 0.03  & 0.14 &  & -0.04 & 0.29 &  & 0.04  & 0.4  &  & 0.03  & 0.4  \\ 
\multicolumn{1}{r}{$\times$Avg Score} & -0.25 & 0.15 &  & 0.08  & 0.27 &  & 0.08  & 0.37 &  & 0.19  & 0.31 &  & -0.15 & 0.15 &  & 0.05  & 0.33 &  & -0.23 & 0.49 &  & -0.5  & 0.48 \\ 
\% FRPL (school) &       &      &  &       &      &  &       &      &  &       &      &  &       &      &  &       &      &  &       &      &  &       &      \\ 
\multicolumn{1}{r}{$\times$Med Income} & -0.4  & 0.27 &  & -0.38 & 0.37 &  & -0.32 & 0.52 &  & -0.29 & 0.43 &  & 0.06  & 0.24 &  & 0.17  & 0.36 &  & 0.27  & 0.53 &  & -0.03 & 0.57 \\ 
\multicolumn{1}{r}{$\times$Avg Score} & 0.29  & 0.21 &  & -0.3  & 0.31 &  & -0.06 & 0.43 &  & -0.37 & 0.38 &  & 0.12  & 0.21 &  & 0.01  & 0.4  &  & -0.27 & 0.59 &  & 0.24  & 0.56 \\ 
9th Grade Size (100s) & 0.14  & 0.01 &  & 0.19  & 0.02 &  & 0.25  & 0.03 &  & 0.23  & 0.03 &  & 0.15  & 0.01 &  & 0.23  & 0.02 &  & 0.29  & 0.04 &  & 0.28  & 0.03 \\ 
\% Asian & 2.3   & 0.38 &  & 1.7   & 0.6  &  & 2.41  & 0.87 &  & 1.88  & 0.72 &  & -1.31 & 0.38 &  & -3.27 & 0.67 &  & -2.99 & 0.95 &  & -3.36 & 0.9  \\ 
\% Black & -1.29 & 0.26 &  & -0.59 & 0.38 &  & 0.08  & 0.59 &  & 0.03  & 0.47 &  & -0.48 & 0.26 &  & 1.18  & 0.43 &  & 1.85  & 0.64 &  & 1.66  & 0.62 \\ 
\% Hispanic & -0.7  & 0.25 &  & -0.08 & 0.38 &  & 0.73  & 0.54 &  & 0.46  & 0.45 &  & -0.51 & 0.25 &  & 0.45  & 0.4  &  & 0.71  & 0.56 &  & 0.91  & 0.55 \\ 
1(Nearest School) & -0.01 & 0.06 &  & 0.06  & 0.08 &  & 0.01  & 0.11 &  & 0.03  & 0.09 &  & 0.26  & 0.06 &  & 0.22  & 0.08 &  & 0.22  & 0.12 &  & 0.3   & 0.11 \\ 
Distance & -0.31 & 0.02 &  & -0.33 & 0.02 &  & -0.35 & 0.03 &  & -0.35 & 0.03 &  & -0.26 & 0.01 &  & -0.29 & 0.02 &  & -0.3  & 0.03 &  & -0.28 & 0.03 \\ 
$\sigma^2_{STEM}$ & 1.17  & 0.12 &  & 0.97  & 0.17 &  & 1.28  & 0.36 &  & 1.42  & 0.33 &  & 1.21  & 0.14 &  & 1.53  & 0.3  &  & 1.66  & 0.47 &  & 1.7   & 0.47 \\ 
\\ 
\bottomrule 
\end{tabular}}
\begin{flushleft}
\scriptsize
Notes: We report the full estimates of the parameters in \eqref{eqn:utility_SI}.  All \% variables are within $[0,1]$. For conciseness, we report only the estimates based on WTT, \teps{all}, \teps{top}, and the selected estimates obtained by following the procedure described in Section~\ref{sec:tests}. We report the mean and standard deviation of the posterior distribution as the point estimate and the standard error.
\end{flushleft}
\end{table}
\clearpage
\begin{table}[h]
\centering
\footnotesize
\caption{Preference Estimates: Cell 5--6\label{tab:SIpref3}}
\resizebox{1.35\textwidth}{!}{\begin{tabular}{lccccccccccccccccccccccc}
\toprule 
& \multicolumn{11}{c}{Cell 5}&& \multicolumn{11}{c}{Cell 6}\\\cline{2-12}\cline{14-24}&\multicolumn{2}{c}{WTT}&&\multicolumn{2}{c}{\teps{all}}&&\multicolumn{2}{c}{\teps{top}}&&\multicolumn{2}{c}{Selected}&&\multicolumn{2}{c}{WTT}&&\multicolumn{2}{c}{\teps{all}}&&\multicolumn{2}{c}{\teps{top}}&&\multicolumn{2}{c}{Selected}\\\cline{2-3}\cline{5-6}\cline{8-9}\cline{11-12}\cline{14-15}\cline{17-18}\cline{20-21}\cline{23-24} & Coef. & S.E. &  & Coef. & S.E. &  & Coef. & S.E. &  & Coef. & S.E.& & Coef. & S.E. &  & Coef. & S.E. &  & Coef. & S.E. &  & Coef. & S.E. \\ 
\midrule 
School FE 4 & -0.29 & 0.1  &  & -0.5  & 0.12 &  & -0.5  & 0.24 &  & -0.64 & 0.13 &  & -0.13 & 0.11 &  & -0.23 & 0.15 &  & -0.41 & 0.25 &  & -0.32 & 0.16 \\ 
School FE 5 & -0.8  & 0.12 &  & -1.47 & 0.19 &  & -1.72 & 0.33 &  & -1.61 & 0.21 &  & -0.58 & 0.14 &  & -1.19 & 0.24 &  & -0.98 & 0.34 &  & -1.2  & 0.25 \\ 
School FE 6 & -0.45 & 0.12 &  & -0.83 & 0.19 &  & -0.54 & 0.33 &  & -0.85 & 0.2  &  & -0.07 & 0.14 &  & -0.56 & 0.23 &  & -0.22 & 0.37 &  & -0.54 & 0.24 \\ 
School FE 7 & 0.22  & 0.18 &  & -0.62 & 0.29 &  & -1.37 & 0.55 &  & -0.74 & 0.31 &  & 0.14  & 0.2  &  & -0.55 & 0.37 &  & -1.16 & 0.53 &  & -0.62 & 0.39 \\ 
School FE 8 & -0.3  & 0.1  &  & -0.44 & 0.15 &  & -0.33 & 0.26 &  & -0.48 & 0.16 &  & -0.07 & 0.11 &  & -0.48 & 0.19 &  & -0.7  & 0.29 &  & -0.56 & 0.2  \\ 
School FE 9 & -0.3  & 0.12 &  & -0.64 & 0.17 &  & -0.78 & 0.33 &  & -0.8  & 0.19 &  & -0.85 & 0.16 &  & -1.22 & 0.27 &  & -0.88 & 0.39 &  & -1.14 & 0.27 \\ 
$\mathds{1}$(STEM) &       &      &  &       &      &  &       &      &  &       &      &  &       &      &  &       &      &  &       &      &  &       &      \\ 
\multicolumn{1}{r}{Main Effect} & 0.28  & 0.11 &  & 0.46  & 0.18 &  & 0.02  & 0.43 &  & 0.41  & 0.22 &  & -0.15 & 0.16 &  & -0.48 & 0.34 &  & -0.85 & 0.54 &  & -0.45 & 0.37 \\ 
\multicolumn{1}{r}{$\times$ELA Score} & -0.02 & 0.07 &  & 0.07  & 0.09 &  & 0.17  & 0.2  &  & 0.08  & 0.1  &  & -0.17 & 0.12 &  & -0.43 & 0.21 &  & -0.4  & 0.33 &  & -0.36 & 0.22 \\ 
\multicolumn{1}{r}{$\times$Math Score} & 0     & 0.07 &  & -0.08 & 0.1  &  & 0.11  & 0.23 &  & -0.09 & 0.11 &  & 0.14  & 0.11 &  & 0.38  & 0.23 &  & 0.64  & 0.43 &  & 0.3   & 0.24 \\ 
\% High Perf. (ELA) &       &      &  &       &      &  &       &      &  &       &      &  &       &      &  &       &      &  &       &      &  &       &      \\ 
\multicolumn{1}{r}{Main Effect} & 0.4   & 0.25 &  & -0.12 & 0.37 &  & -0.64 & 0.6  &  & -0.14 & 0.4  &  & 0.74  & 0.3  &  & 0.24  & 0.48 &  & -0.09 & 0.66 &  & 0.38  & 0.51 \\ 
\multicolumn{1}{r}{$\times$ELA Score} & 0.52  & 0.12 &  & 0.61  & 0.23 &  & 1.19  & 0.42 &  & 0.7   & 0.25 &  & 0.37  & 0.15 &  & 0.33  & 0.31 &  & 0.38  & 0.41 &  & 0.29  & 0.31 \\ 
\% High Perf. (Math) &       &      &  &       &      &  &       &      &  &       &      &  &       &      &  &       &      &  &       &      &  &       &      \\ 
\multicolumn{1}{r}{Main Effect} & -0.75 & 0.34 &  & 0.85  & 0.5  &  & 1.56  & 0.88 &  & 0.87  & 0.55 &  & -0.36 & 0.39 &  & 1.48  & 0.69 &  & 2.91  & 0.96 &  & 1.38  & 0.74 \\ 
\multicolumn{1}{r}{$\times$Math Score} & 0.51  & 0.13 &  & 0.61  & 0.25 &  & 0.44  & 0.47 &  & 0.57  & 0.27 &  & 0.35  & 0.14 &  & 0.69  & 0.37 &  & 0.45  & 0.51 &  & 0.81  & 0.39 \\ 
\% FRPL (program) &       &      &  &       &      &  &       &      &  &       &      &  &       &      &  &       &      &  &       &      &  &       &      \\ 
\multicolumn{1}{r}{Main Effect} & 0.78  & 0.3  &  & -1.02 & 0.51 &  & -2.79 & 0.81 &  & -1.12 & 0.56 &  & 0.6   & 0.35 &  & 0.19  & 0.68 &  & -0.18 & 0.95 &  & 0.09  & 0.73 \\ 
\multicolumn{1}{r}{$\times$Med Income} & -0.14 & 0.18 &  & -0.18 & 0.33 &  & -0.82 & 0.56 &  & -0.16 & 0.36 &  & -0.09 & 0.24 &  & -0.36 & 0.48 &  & -0.69 & 0.69 &  & -0.56 & 0.5  \\ 
\multicolumn{1}{r}{$\times$Avg Score} & -0.05 & 0.21 &  & 0.46  & 0.36 &  & 1.02  & 0.54 &  & 0.42  & 0.39 &  & -0.51 & 0.26 &  & -0.86 & 0.56 &  & -1.46 & 0.88 &  & -0.59 & 0.6  \\ 
\% FRPL (school) &       &      &  &       &      &  &       &      &  &       &      &  &       &      &  &       &      &  &       &      &  &       &      \\ 
\multicolumn{1}{r}{$\times$Med Income} & 0.44  & 0.26 &  & 0.12  & 0.44 &  & 1.23  & 0.75 &  & 0.18  & 0.46 &  & -0.54 & 0.37 &  & -0.21 & 0.59 &  & 1.24  & 0.93 &  & 0.09  & 0.64 \\ 
\multicolumn{1}{r}{$\times$Avg Score} & -0.07 & 0.31 &  & -0.66 & 0.46 &  & -1.14 & 0.75 &  & -0.44 & 0.48 &  & 0.84  & 0.36 &  & 1.42  & 0.65 &  & 1.98  & 0.96 &  & 1.16  & 0.66 \\ 
9th Grade Size (100s) & 0.1   & 0.02 &  & 0.16  & 0.04 &  & 0.33  & 0.07 &  & 0.17  & 0.04 &  & 0.09  & 0.03 &  & 0.18  & 0.04 &  & 0.33  & 0.07 &  & 0.19  & 0.05 \\ 
\% Asian & 2.34  & 0.51 &  & 1.28  & 0.75 &  & 0.45  & 1.34 &  & 1.1   & 0.82 &  & -0.41 & 0.63 &  & -2.23 & 1.08 &  & -3.27 & 1.6  &  & -1.81 & 1.16 \\ 
\% Black & -0.96 & 0.34 &  & 0.11  & 0.47 &  & 0.67  & 0.84 &  & -0.06 & 0.52 &  & -0.46 & 0.41 &  & 0.12  & 0.64 &  & 0.35  & 0.96 &  & 0.16  & 0.67 \\ 
\% Hispanic & -0.24 & 0.32 &  & 0.68  & 0.48 &  & 1.84  & 0.82 &  & 0.84  & 0.54 &  & -0.98 & 0.39 &  & -0.32 & 0.65 &  & -0.16 & 0.88 &  & -0.22 & 0.68 \\ 
1(Nearest School) & -0.12 & 0.08 &  & -0.22 & 0.11 &  & -0.07 & 0.18 &  & -0.15 & 0.12 &  & 0.13  & 0.09 &  & -0.01 & 0.15 &  & -0.02 & 0.21 &  & -0.01 & 0.16 \\ 
Distance & -0.3  & 0.02 &  & -0.29 & 0.03 &  & -0.35 & 0.05 &  & -0.31 & 0.03 &  & -0.23 & 0.02 &  & -0.25 & 0.04 &  & -0.3  & 0.06 &  & -0.25 & 0.04 \\ 
$\sigma^2_{STEM}$ & 1.18  & 0.18 &  & 0.86  & 0.21 &  & 2.17  & 0.88 &  & 1.03  & 0.29 &  & 1.1   & 0.2  &  & 1.31  & 0.42 &  & 1.39  & 0.58 &  & 1.3   & 0.45 \\ 
\\ 
\bottomrule 
\end{tabular}}
\begin{flushleft}
\scriptsize
Notes: We report the full estimates of the parameters in \eqref{eqn:utility_SI}.  All \% variables are within $[0,1]$. For conciseness, we report only the estimates based on WTT, \teps{all}, \teps{top}, and the selected estimates obtained by following the procedure described in Section~\ref{sec:tests}. We report the mean and standard deviation of the posterior distribution as the point estimate and the standard error.
\end{flushleft}
\end{table}
\clearpage
\begin{table}[h]
\centering
\footnotesize
\caption{Preference Estimates: Cell 7--8\label{tab:SIpref4}}
\resizebox{1.35\textwidth}{!}{\begin{tabular}{lccccccccccccccccccccccc}
\toprule 
& \multicolumn{11}{c}{Cell 7}&& \multicolumn{11}{c}{Cell 8}\\\cline{2-12}\cline{14-24}&\multicolumn{2}{c}{WTT}&&\multicolumn{2}{c}{\teps{all}}&&\multicolumn{2}{c}{\teps{top}}&&\multicolumn{2}{c}{Selected}&&\multicolumn{2}{c}{WTT}&&\multicolumn{2}{c}{\teps{all}}&&\multicolumn{2}{c}{\teps{top}}&&\multicolumn{2}{c}{Selected}\\\cline{2-3}\cline{5-6}\cline{8-9}\cline{11-12}\cline{14-15}\cline{17-18}\cline{20-21}\cline{23-24} & Coef. & S.E. &  & Coef. & S.E. &  & Coef. & S.E. &  & Coef. & S.E.& & Coef. & S.E. &  & Coef. & S.E. &  & Coef. & S.E. &  & Coef. & S.E. \\ 
\midrule 
School FE 4 & -0.11 & 0.06 &  & -0.25 & 0.08 &  & -0.28 & 0.13 &  & -0.31 & 0.09 &  & 0.32  & 0.06 &  & 0.13  & 0.08 &  & 0.16  & 0.14 &  & 0.06  & 0.12 \\ 
School FE 5 & -0.5  & 0.07 &  & -0.98 & 0.1  &  & -1.03 & 0.17 &  & -1.03 & 0.12 &  & -0.22 & 0.07 &  & -0.85 & 0.11 &  & -0.74 & 0.17 &  & -0.79 & 0.16 \\ 
School FE 6 & -0.17 & 0.08 &  & -0.44 & 0.12 &  & -0.42 & 0.19 &  & -0.48 & 0.14 &  & 0.09  & 0.08 &  & -0.51 & 0.12 &  & -0.43 & 0.19 &  & -0.48 & 0.18 \\ 
School FE 7 & 0.66  & 0.1  &  & 0.24  & 0.16 &  & -0.33 & 0.28 &  & 0.08  & 0.18 &  & 0.06  & 0.11 &  & -0.6  & 0.19 &  & -1.1  & 0.37 &  & -1.04 & 0.35 \\ 
School FE 8 & 0.19  & 0.06 &  & 0.07  & 0.09 &  & 0.03  & 0.15 &  & 0.08  & 0.1  &  & 0.17  & 0.06 &  & -0.17 & 0.1  &  & 0.04  & 0.17 &  & -0.04 & 0.15 \\ 
School FE 9 & -0.04 & 0.07 &  & -0.24 & 0.1  &  & -0.25 & 0.17 &  & -0.33 & 0.12 &  & -0.48 & 0.08 &  & -0.89 & 0.12 &  & -0.73 & 0.2  &  & -0.79 & 0.18 \\ 
$\mathds{1}$(STEM) &       &      &  &       &      &  &       &      &  &       &      &  &       &      &  &       &      &  &       &      &  &       &      \\ 
\multicolumn{1}{r}{Main Effect} & 0.18  & 0.07 &  & -0.19 & 0.18 &  & -0.99 & 0.55 &  & -0.15 & 0.2  &  & -0.39 & 0.1  &  & -0.92 & 0.29 &  & -0.93 & 0.55 &  & -0.7  & 0.41 \\ 
\multicolumn{1}{r}{$\times$ELA Score} & -0.04 & 0.04 &  & -0.08 & 0.07 &  & -0.09 & 0.12 &  & -0.03 & 0.07 &  & -0.03 & 0.06 &  & 0.02  & 0.1  &  & -0.1  & 0.16 &  & -0.09 & 0.14 \\ 
\multicolumn{1}{r}{$\times$Math Score} & -0.01 & 0.04 &  & -0.1  & 0.07 &  & -0.11 & 0.14 &  & -0.16 & 0.08 &  & -0.07 & 0.06 &  & -0.27 & 0.11 &  & -0.1  & 0.19 &  & -0.2  & 0.17 \\ 
\% High Perf. (ELA) &       &      &  &       &      &  &       &      &  &       &      &  &       &      &  &       &      &  &       &      &  &       &      \\ 
\multicolumn{1}{r}{Main Effect} & 0.31  & 0.15 &  & 0.23  & 0.23 &  & 0.39  & 0.34 &  & 0.38  & 0.26 &  & 0.33  & 0.14 &  & 0.17  & 0.22 &  & -0.28 & 0.31 &  & -0.08 & 0.29 \\ 
\multicolumn{1}{r}{$\times$ELA Score} & 0.44  & 0.07 &  & 0.57  & 0.15 &  & 0.66  & 0.2  &  & 0.54  & 0.15 &  & 0.4   & 0.08 &  & 0.5   & 0.17 &  & 0.27  & 0.23 &  & 0.34  & 0.21 \\ 
\% High Perf. (Math) &       &      &  &       &      &  &       &      &  &       &      &  &       &      &  &       &      &  &       &      &  &       &      \\ 
\multicolumn{1}{r}{Main Effect} & -0.57 & 0.19 &  & 0.18  & 0.3  &  & 0.61  & 0.47 &  & 0.24  & 0.33 &  & 0.62  & 0.19 &  & 1.34  & 0.34 &  & 1.68  & 0.48 &  & 1.43  & 0.45 \\ 
\multicolumn{1}{r}{$\times$Math Score} & 0.3   & 0.07 &  & 0.41  & 0.15 &  & 0.49  & 0.22 &  & 0.54  & 0.16 &  & 0.66  & 0.08 &  & 1.05  & 0.2  &  & 1.08  & 0.28 &  & 1.25  & 0.28 \\ 
\% FRPL (program) &       &      &  &       &      &  &       &      &  &       &      &  &       &      &  &       &      &  &       &      &  &       &      \\ 
\multicolumn{1}{r}{Main Effect} & 1.29  & 0.18 &  & 0.5   & 0.34 &  & 0.18  & 0.47 &  & 0.53  & 0.36 &  & 0.76  & 0.18 &  & 0.07  & 0.34 &  & -0.48 & 0.49 &  & -0.3  & 0.49 \\ 
\multicolumn{1}{r}{$\times$Med Income} & -0.14 & 0.1  &  & -0.3  & 0.18 &  & -0.67 & 0.3  &  & -0.46 & 0.19 &  & 0.07  & 0.1  &  & 0.15  & 0.2  &  & 0.03  & 0.3  &  & -0.01 & 0.29 \\ 
\multicolumn{1}{r}{$\times$Avg Score} & 0.01  & 0.12 &  & 0.12  & 0.21 &  & 0.17  & 0.31 &  & 0.24  & 0.23 &  & -0.12 & 0.13 &  & -0.26 & 0.27 &  & -0.5  & 0.38 &  & -0.26 & 0.35 \\ 
\% FRPL (school) &       &      &  &       &      &  &       &      &  &       &      &  &       &      &  &       &      &  &       &      &  &       &      \\ 
\multicolumn{1}{r}{$\times$Med Income} & 0.27  & 0.15 &  & 0.26  & 0.23 &  & 0.89  & 0.42 &  & 0.37  & 0.27 &  & -0.5  & 0.16 &  & -0.83 & 0.25 &  & -0.71 & 0.42 &  & -0.67 & 0.41 \\ 
\multicolumn{1}{r}{$\times$Avg Score} & 0.22  & 0.18 &  & 0.13  & 0.25 &  & 0.45  & 0.39 &  & 0.25  & 0.28 &  & 0.71  & 0.18 &  & 1.02  & 0.32 &  & 0.94  & 0.49 &  & 0.89  & 0.44 \\ 
9th Grade Size (100s) & 0.01  & 0.01 &  & 0.05  & 0.02 &  & 0.16  & 0.03 &  & 0.08  & 0.02 &  & 0.04  & 0.02 &  & 0.11  & 0.03 &  & 0.24  & 0.04 &  & 0.22  & 0.04 \\ 
\% Asian & 1.24  & 0.28 &  & 0.78  & 0.45 &  & 0.2   & 0.73 &  & 0.59  & 0.5  &  & -1.78 & 0.31 &  & -2.92 & 0.52 &  & -1.76 & 0.76 &  & -1.97 & 0.73 \\ 
\% Black & -1.01 & 0.2  &  & -0.78 & 0.28 &  & -0.42 & 0.46 &  & -0.65 & 0.31 &  & 0.19  & 0.2  &  & 0.71  & 0.31 &  & 1.74  & 0.48 &  & 1.49  & 0.44 \\ 
\% Hispanic & -0.58 & 0.17 &  & -0.17 & 0.26 &  & 0.27  & 0.41 &  & 0.14  & 0.3  &  & -0.28 & 0.18 &  & 0.23  & 0.28 &  & 0.5   & 0.39 &  & 0.37  & 0.37 \\ 
1(Nearest School) & 0.13  & 0.05 &  & 0.12  & 0.06 &  & 0.15  & 0.1  &  & 0.12  & 0.07 &  & 0.14  & 0.05 &  & 0.15  & 0.07 &  & 0.29  & 0.1  &  & 0.28  & 0.1  \\ 
Distance & -0.19 & 0.01 &  & -0.22 & 0.02 &  & -0.25 & 0.03 &  & -0.23 & 0.02 &  & -0.14 & 0.01 &  & -0.15 & 0.02 &  & -0.14 & 0.03 &  & -0.12 & 0.02 \\ 
$\sigma^2_{STEM}$ & 1.14  & 0.11 &  & 1.43  & 0.25 &  & 2.75  & 1.03 &  & 1.43  & 0.29 &  & 1.13  & 0.13 &  & 1.6   & 0.36 &  & 1.68  & 0.69 &  & 1.42  & 0.49 \\ 
\\ 
\bottomrule 
\end{tabular}}
\begin{flushleft}
\scriptsize
Notes: We report the full estimates of the parameters in \eqref{eqn:utility_SI}.  All \% variables are within $[0,1]$. For conciseness, we report only the estimates based on WTT, \teps{all}, \teps{top}, and the selected estimates obtained by following the procedure described in Section~\ref{sec:tests}. We report the mean and standard deviation of the posterior distribution as the point estimate and the standard error.
\end{flushleft}
\end{table}
\end{landscape}

\end{appendix}
\end{spacing}

\end{document}